\documentclass[epj]{svjour}

\usepackage{graphics}
\usepackage{mathtools}
\usepackage{longtable}
\usepackage{lscape}
\usepackage{caption}
\usepackage{lineno}

\begin{document}

\title{Experimental searches for rare alpha and beta decays}

\author{P. Belli\inst{1,2,}\thanks{\email{pierluigi.belli@roma2.infn.it}} \and 
R. Bernabei\inst{1,2} \and 
F.A. Danevich\inst{3} \and 
A. Incicchitti\inst{4,5} \and
V.I. Tretyak\inst{3}
}

\institute{Dip. Fisica, Universit\`a di Roma ``Tor Vergata'', I-00133 Rome, Italy  \and
INFN sezione Roma ``Tor Vergata'', I-00133 Rome, Italy \and
Institute for Nuclear Research, 03028 Kyiv, Ukraine \and
Dip. Fisica, Universit\`a di Roma ``La Sapienza'', I-00185 Rome, Italy \and
INFN sezione Roma, I-00185 Rome, Italy}

\date{}
\vspace{0.3cm}

\abstract{
The current status of the experimental searches for rare alpha and beta decays
is reviewed. Several interesting observations of alpha and beta
decays, previously unseen due to their large half-lives ($10^{15} - 
10^{20}$ yr),
have been achieved during the last years thanks to the improvements in
the experimental techniques and to the underground locations of 
experiments that
allows to suppress backgrounds. In particular, the list includes first
observations of alpha decays of $^{151}$Eu, $^{180}$W (both to the ground state
of the daughter nuclei), $^{190}$Pt (to excited state of the daughter 
nucleus), $^{209}$Bi (to the
ground and excited states of the daughter nucleus). The isotope $^{209}$Bi has the longest
known half-life of $T_{1/2} \approx 10^{19}$ yr relatively to alpha decay. The beta
decay of $^{115}$In to the first excited state of $^{115}$Sn (E$_{exc} = 497.334$
keV), recently observed for the first time, has the $Q_\beta$ value of
only $(147 \pm 10)$ eV, which is the lowest $Q_\beta$ value known to-date.
Searches and investigations of other rare alpha and beta decays ($^{48}$Ca,
$^{50}$V, $^{96}$Zr, $^{113}$Cd, $^{123}$Te, $^{178m2}$Hf, $^{180m}$Ta and others) are also
discussed.
\PACS{
      {}{Keywords.} \and
      {}{$\alpha$ decay} \and
      {}{$\beta$ decay} \and
      {}{Scintillation detectors} \and
      {}{Radiation detectors} \and
      {}{Low temperature detectors} \and
      {}{HPGe detectors} 
     } 
}

\maketitle

\tableofcontents

\vspace{0.2cm}
\section{Introduction}
\label{intro}

From the theoretical and experimental points of view, a very interesting topic of nuclear physics is the 
study of nuclear instability, which played so far an important role in the foundation and development 
of nuclear physics. In fact, it can offer information about the nuclear structure, the nuclear levels 
and the properties of nuclei. In the last hundred years a lot of experimental and theoretical studies have
been performed and the knowledge of the nuclear matter becomes more and more determined. In particular, 
the first two processes ever studied were $\alpha$ and $\beta$ decays (actually, the generic word $\beta$ decay 
consists of $\beta^-$, $\beta^+$ and electron capture, $\varepsilon$). Generally, $\alpha$ and $\beta$ decays
are also accompanied by emission of $\gamma$ quanta, conversion electrons and $e^+e^-$ pairs
and/or by emission of subsequent atomic radiation, X-rays and/or Auger and Coster-Kronig electrons.

More recently, other forms of radioactive decays have been considered, as: 
\begin{itemize}
\item {\bf the cluster decays}, where a nuclide decays by emitting nuclides heavier than $\alpha$ particle (from $^{14}$C to $^{34}$Si);
   about 40 nuclides from $^{221}$Fr to $^{242}$Cm can decay in such a way to isotopes close to $^{208}$Pb (``lead radioactivity'')
   with partial half-lives in the range $10^3 - 10^{20}$ yr \cite{Bonet:1999,Santhosh:2007}. The cluster decay was theoretically predicted in 1980 \cite{Sandul:1984} 
   and experimentally observed for the first time in 1984 \cite{Rose:1984,Aleks:1984}.
\item {\bf the $2\beta$ decay}, which is a transformation of nucleus $(A,Z)$ into $(A,Z+2)$ with simultaneous emission of two electrons.
   The two-neutrino $2\beta$ decay is the rarest nuclear decay ever observed in several nuclei with the half-lives
   $T_{1/2}^{2\nu2\beta^-} \approx 10^{18} - 10^{24}$ yr \cite{Tretyak:2002,Saak:2013,Barabash:2015}. On the contrary, the neutrinoless $2\beta$ decay
   is forbidden in the Standard Model of particle
   physics (SM) since the process violates the lepton number by two units and is possible if the neutrino is a massive Majorana particle.
   Typically the most sensitive experiments give only half-life limits on the decay at level of lim $T_{1/2}^{0\nu2\beta^-} 
\approx 10^{24} - 10^{26}$ yr.
   Investigations of the double beta plus processes, such as double
   electron capture ($2\varepsilon$), electron capture with positron emission ($\varepsilon\beta^+$) and double positron decay ($2\beta^+$) are also on-going \cite{Tretyak:2002,Malam:2013}.
\item {\bf the spontaneous fission} occurs when a heavy nucleus fragments itself in two (or, with lower probabilities, in more than two) nuclides. It is theoretically possible for many nuclei with $A>100$,
   but it has been observed for
   several isotopes starting from $^{232}$Th. The measured partial half-lives range from $10^{-3}$ s of $^{264}$Hs to $10^{19}$ yr of $^{235}$U.
\item {\bf nucleon, di-nucleon, tri-nucleon ... emission} occurs in short living isotopes in about forty isotopes with half-lives from ps to s.
\item {\bf higher-order elec\-tro\-mag\-netic phe\-nom\-ena}, ac\-c\-ompanying $\beta$ decay, shell-electron capture, and $\alpha$ decay, are \cite{Petters:1966,Ljub:1973}: 
   i) internal bremsstrahlung; ii) ionization and excitation of the electron cloud; iii) internal $e^+e^-$ pair production (IPP). They are much less intense than 
   the first-order processes; for example, in the case of the internal bremsstrahlung about one photon is produced per 100 $\beta$ decays. 
   The first estimates of the contribution of the IPP process was reported 
   in Ref. \cite{Arley:1938,Tisza:1937,Greens:1956} and in Ref. \cite{Ljub:1973} for the $\beta$ and the $\alpha$ decays, respectively.
   The IPP in the $\alpha$ decay has been measured for $^{210}$Po, $^{239}$Pu, and $^{241}$Am (see Ref. \cite{Bernabei:2013} for a recent measurement and references therein).
\end{itemize}

{\bf $\alpha$ decay} corresponds to a very asymmetric spontaneous fission, where a nucleus $(A,Z)$ transforms into $(A-4,Z-2)$
with the ejection of a $^4$He nucleus ($\alpha$ particle).
Most nuclei with $A > 140$ are potential $\alpha-$emitters. 
The lightest $\alpha-$emitter is $^{105}$Te ($T_{1/2}$ = 0.6 $\mu$s, $Q_\alpha$ = 5.1 MeV)\footnote{There are also few light nuclei which decay with emission of $\alpha$ particle: $^5$He, $^5$Li, $^6$Be, $^8$Be, 
with half-lives $10^{-22} - 10^{-16}$ s.}.
However, among the naturally occurring $\alpha-$emit\-ting nuclides only those with 
either $A > 208$ or $A \approx 145$ have $\alpha-$half-lives short enough to be detected; the lightest one is the $^{142}$Ce nuclide.
Considering all the $\alpha$ unstable nuclei, the partial $\alpha-$half-lives range from $10^{-8}$ s of $^{217}$Ac to $10^{19}$ yr of $^{209}$Bi.
The $\alpha$ decay rate is an exponentially increasing function of $Q_\alpha$. Thus, the $\alpha$ decay is preferentially to the ground state of the daughter nucleus, 
since decays to excited states necessarily have smaller values of $Q_\alpha$
(on the contrary, for $\beta$ decay the $Q_\beta$ dependence is weaker).

{\bf $\beta$ decay} can occur in nuclei where the neutron-to-proton ratio is not optimal. In this process the parent and the
daughter nuclei have the same atomic mass:
an electron (positron) plus an electron anti-neutrino (neutrino) are emitted; the nucleus transforms from $(A,Z)$ into $(A,Z+1)$ or $(A,Z-1)$.
In first approximation the $\beta$ decay rate is proportional to the fifth power of $Q_\beta$.
The $\beta-$decay half-lives range from $10^{-3}$ s of e.g. $^{35}$Na to $10^{16}$ yr of $^{113}$Cd (the longest to date observed $\beta$ decay).

The purpose of this paper is to review the current status of the experimental searches for rare alpha and beta decays.
Several interesting observations of alpha and beta
decays, previously unseen due to their large half-lives ($10^{15} - 10^{20}$ yr)
or/and very low branching ratios ($10^{-3}-10^{-8}$),
have been achieved just during the last years thanks to the improvements in
the experimental techniques and to the underground locations of the experiments that
allows to suppress backgrounds.

\vspace{0.2cm}
\section{The rare $\alpha$ decays}
\label{intro_alpha}

The interest to the more than 100 years old phenomenon of $\alpha$ decay \cite{rut99} is still great, 
both from the theoretical and the experimental sides. 
In fact, it can offer information about both the nuclear structure and the fusion--fission reactions since 
the $\alpha$ decay process involves sub-barrier penetration of $\alpha$ particles through the barrier, 
caused by the interaction between the $\alpha$ and the nucleus. 

The description of $\alpha$ decay as a process of quantum tunnelling through an energy barrier (a process not possible from the point of view of classical physics) in 1928 by Gamow \cite{Gam1928} and Gurney and Condon \cite{Gur1928} was a great success of the quantum mechanics. It naturally explained the exponential dependence of the probability 
of the $\alpha$ decay on the released energy $Q_\alpha$, already known empirically since 1911 as the Geiger-Nuttall law \cite{Gei1911}. The energy releases in naturally occurring $\alpha$-decaying nuclides are in the range 
from 1.9 MeV ($^{144}$Nd) to 4.9 MeV ($^{234}$U), while more energetic $\alpha$ particles are present in chains of $^{232}$Th, $^{235}$U, $^{238}$U decay (with the highest value of $Q_\alpha=8.954$ MeV for $^{212}$Po in the $^{232}$Th chain). The released energy is shared between the $\alpha$ particle, $E_\alpha$, and the daughter nuclear recoil, $E_d$, in good approximation as:
$E_\alpha=Q_\alpha \times (A_m-4)/A_m$,
$E_d=Q_\alpha \times 4/A_m$ 
($A_m$ is the mass number of the mother nucleus).
It means that for typical values $A_m \simeq 200$ and $Q_\alpha \simeq 5$ MeV, the recoiled daughter has energy $\simeq 100$ keV. Such recoils are well observed e.g. in experiments to search for dark matter with bolometers where U/Th chains are still present in tiny amounts creating background but also serving as a calibration points (see e.g. \cite{Ang2005}).

Various theoretical models are continuously
developed or improved, see e.g. \cite{den09,poe12,qia14,Den15,san15,ash16,sah16,zha17,akr17,qia11,qia11b} and refs. therein, in particular
motivated by the searches for stable or long-lived superheavy isotopes \cite{hof15,oga17} and the prediction of
their half-lives. 
One can find in the list of approaches:
cluster model \cite{Buck:1991}, 
generalized density-dependent cluster model \cite{Ni:2010,Ni:2012}, 
modified two-potential approach for deformed nuclei by constructing the microscopic double-folding potential \cite{qia11},
density-dependent cluster model using a two-potential approach \cite{qia11b},
density-dependent M3Y effective interaction \cite{Chow:2008,Sama:2007}, 
generalized liquid drop model \cite{Wang:2010}, unified model for $\alpha$ decay and $\alpha$ capture \cite{Denis:2005,Denis:2010}. 
In particular, the inclusion of nuclear deformation in the $\alpha$ decay calculations is important since most of the $\alpha$ emitters 
have deformed shapes; this has been accounted in various way 
\cite{Denis:2005,Cob:2012,Pahl:2013,Ghod:2016,Ism:2017,Bohr:1998,Saxe:1986,Ishk:2005,Scam:2013,Adam:2014}. 
Recently, the effective radial potential between the axially symmetric deformed daughter nucleus and the $\alpha$ particle has been considered 
as sum of the deformed Woods--Saxon nuclear potential, of the deformed Coulomb potential, of the centrifugal potential, and also taking into account 
the quadrupole and the hexade-capole deformation parameters of the daughter nucleus \cite{Dahmar:2017}.
The effects of the orientation and of the deformed surface diffuseness on the effective potential, transmission coefficient, 
and assault frequency were significant, suggesting that the inclusion of a more accurate description of the deformed surface 
diffuseness instead of 
an average form, and analysis of its effect on a wide range of $\alpha$ decays within different models can be interesting. On the 
other hand, 
experimentalists require the possibility of evaluating the expected half-lives during the
planning and the design of an experiment, especially when $\alpha$ 
decays of super-heavy elements or of nuclei very far from the stability line are investigated; both processes are very rare and difficult to observe. 
Thus, simple and accurate empirical formulas are still developed and used. 
Presently, more accurate atomic mass values lead to a more exact definition of the $Q_\alpha$ values of the ground-state-to-ground-state $\alpha$ transitions;
in addition, more precise definitions of the value of the angular momentum of the emitted $\alpha$ particle (related to factual values of the ground-state spins) 
lead to a more rigorous determination and description of the $\alpha$ decays half-lives.
See Ref. \cite{Qi2019} for a recent review on various microscopic and phenomenological models of $\alpha$ decay which successfully reproduce experimentally measured half-lives.

Recently the detection of naturally long-lived $\alpha-$ de\-cay\-ing nuclides has become a hot subject.
Improvements in the experimental sensitivity, especially related with the use of
super-low-background set-ups located in underground laboratories, have led during the last
decade to discovering new $\alpha$ decays which were not observed previously because of their extremely
long half-lives or small branching ratios. 

The process of $\alpha$ decay can be accompanied by emission of $\gamma$ quanta when the decay goes to excited
levels of a daughter nucleus or/and the daughter in the ground state is also unstable. Thus,
a possible strategy to detect rare $\alpha$ decay is to 
look for the $\gamma$ rays following the $\alpha$ decay. This can be accomplished
by using e.g. an ultra-low background germanium detector in an underground laboratory,
as the Gran Sasso National Laboratory [LNGS, 3600 m water equivalent (w.e.)] 
of Istituto Nazionale di Fisica Nucleare (INFN), Italy 
(for an ultra-pure Os sample with natural isotopic composition see for example Ref. \cite{bel18}). 

A collection of half-lives (or limits) on $\alpha$ decays of all the naturally occurring nuclides with $Q_\alpha > 0$ is
reported in Table \ref{tab:alpha}.
The Table also shows the theoretical half-life values (from the ground state to the ground state, g.s. to g.s., transitions)
calculated with the cluster model of Ref. \cite{Buc91,Buc92} and with the semiempirical
formulae \cite{Poe83} based on the liquid drop model and the description of the
$\alpha$ decay as a very asymmetric fission process. The approaches
\cite{Buc91,Buc92,Poe83} were tested with a set of experimental half-lives of
almost four hundred $\alpha$ emitters and demonstrated good agreement between
calculated and experimental $T_{1/2}$ values, mainly inside a factor of 2--3.
In some cases, when there is a difference between spins and parities of the parent and the daughter
nuclei, and the emitted $\alpha$ particle has non-zero angular momentum $l$,
one has to take into account the additional hindrance factor, calculated in accordance with Ref. \cite{Hey99} 
(for the lowest possible $l$ value).
In addition, we also report in Table \ref{tab:alpha} the results obtained by
the semiempirical formulae of Ref.~\cite{Den15} which also were successfully tested
with about four hundred experimental $\alpha$ decays and which take into account 
non-zero $l$ explicitly.

In the following a systematic discussion on long--living isotopes is carried out.

\vspace{0.2cm}
\section{Investigations of rare $\alpha$ decays}
\label{inv_alpha}

A systematic discussion on long-living isotopes is given in the following.

Note that the four naturally occurring isotopes, $^{232}$Th, $^{234}$U, $^{235}$U, and $^{238}$U, are
well known to experimentalists working in the low background field, because they are present 
at different levels in many materials. Moreover, they are the progenitors of
long radioactive decay chains, which supply the bulk of the heat production in our planet \cite{geoneu}.
Their decay parameters and their half-lives are well established (see Table \ref{tab:alpha}); 
thus, they will not be considered in the following.

\subsection{$^{209}$Bi $\alpha$ decay}
\label{bi209}

The attempts to detect $\alpha$ decay of $^{209}$Bi began in 1949. The
used experimental technique exploited bismuth-loaded

\onecolumn
\begin{landscape}
\begin{longtable}{l|c|r|c|cr|lll}
\caption{Collection of partial half-lives (or limits) on $\alpha$ decays. 
Only nuclides with natural abundance ($\delta$) greater than zero 
(i.e. naturally present in nature) and with $Q_\alpha > 0$ are considered. 
The $\alpha$ decays to excited levels of the daughter nucleus are also reported if they were observed.
The nuclide with the lowest experimental $T_{1/2}$ reported in this table 
is $^{234}$U with the measured total half-life of $(3.44 \pm 0.01) \times 10^{5}$ yr. 
The limits, where not specified, are at 90\% C.L.
The ground state to the ground state transition is supposed if not stated explicitly.}
\label{tab:alpha} \\
\hline\noalign{\smallskip}
Nuclide                      & $J^{\pi}$               & $\delta$ (\%)     & $Q_\alpha$ (keV)   & \multicolumn{5}{c}{Partial $T_{1/2}$ (yr)}             \\
transition                   & parent, daughter        & \cite{Meija:2016} & \cite{Wang:2017}   & \multicolumn{2}{c|}{experimental}                & \multicolumn{3}{c}{theoretical} \\
~                            & nuclei                  & ~                 & ~                  & \multicolumn{2}{c|}{~}                           & \cite{Buc91}       & \cite{Poe83}       & \cite{Den15} \\
\hline
\noalign{\smallskip}
\endfirsthead

\multicolumn{9}{c}%
{{\bfseries \tablename\ \thetable{} -- continued from previous page}} \\
\hline\noalign{\smallskip}
Nuclide                      & $J^{\pi}$               & $\delta$ (\%)     & $Q_\alpha$ (keV)   & \multicolumn{5}{c}{Partial $T_{1/2}$ (yr)}             \\
transition                   & parent, daughter        & \cite{Meija:2016} & \cite{Wang:2017}   & \multicolumn{2}{c|}{experimental}                & \multicolumn{3}{c}{theoretical} \\
~                            & nuclei                  & ~                 & ~                  & \multicolumn{2}{c|}{~}                           & \cite{Buc91}       & \cite{Poe83}       & \cite{Den15} \\
\hline
\noalign{\smallskip}
\endhead

\hline \multicolumn{9}{r}{{Continued on next page}} \\ \hline
\endfoot

\noalign{\smallskip}\hline\noalign{\smallskip}
\noalign{\smallskip}\hline
\endlastfoot

$^{142}$Ce $\to$ $^{138}$Ba  & $0^+\to0^+$     & 11.114(51)        & $ 1303.5 \pm  2.5$ & $> 2.9 \times 10^{18}$      & \cite{bel03}       & $3.4\times10^{27}$ & $7.0\times10^{27}$ & $4.2\times10^{27}$ \\
$^{143}$Nd $\to$ $^{139}$Ce  & $7/2^-\to3/2^+$ & 12.173(26)        & $  521   \pm  7  $ & --                          &                    & $3.8\times10^{80}$ & $1.8\times10^{94}$ & $3.3\times10^{89}$ \\
$^{144}$Nd $\to$ $^{140}$Ce  & $0^+\to0^+$     & 23.798(19)        & $ 1903.2 \pm  1.6$ & $2.29(16) \times 10^{15}$   & \cite{Sonz:2001}   & $2.1\times10^{15}$ & $4.6\times10^{15}$ & $3.6\times10^{15}$ \\
$^{145}$Nd $\to$ $^{141}$Ce  & $7/2^-\to7/2^-$ & 8.293(12)         & $ 1576.0 \pm  1.6$ & $> 6 \times 10^{16}$        & \cite{nd144_04}    & $1.9\times10^{22}$ & $4.3\times10^{23}$ & $7.5\times10^{22}$ \\
$^{146}$Nd $\to$ $^{142}$Ce  & $0^+\to0^+$     & 17.189(32)        & $ 1182.4 \pm  2.2$ & $> 1.6 \times 10^{18}$ $^a$ & \cite{Stengl:2015} & $2.0\times10^{34}$ & $3.9\times10^{34}$ & $1.7\times10^{34}$ \\
$^{148}$Nd $\to$ $^{144}$Ce  & $0^+\to0^+$     & 5.756(21)         & $  599   \pm  3  $ & --                          &                    & $6.1\times10^{70}$ & $1.1\times10^{71}$ & $2.2\times10^{69}$ \\
$^{147}$Sm $\to$ $^{143}$Nd  & $7/2^-\to7/2^-$ & 15.00(14)         & $ 2311.0 \pm  0.4$ & $1.079(26) \times 10^{11}$  & \cite{wil17}       & $3.7\times10^{10}$ & $4.1\times10^{11}$ & $1.7\times10^{11}$ \\
$^{148}$Sm $\to$ $^{144}$Nd  & $0^+\to0^+$     & 11.25(9)          & $ 1986.8 \pm  0.4$ & $6.4(13) \times 10^{15}$    & \cite{cas16}       & $5.6\times10^{15}$ & $1.2\times10^{16}$ & $9.2\times10^{15}$ \\
$^{149}$Sm $\to$ $^{145}$Nd  & $7/2^-\to7/2^-$ & 13.82(10)         & $ 1871.3 \pm  1.0$ & $> 2 \times 10^{15}$        & \cite{aud17}       & $8.3\times10^{17}$ & $9.7\times10^{18}$ & $3.1\times10^{18}$ \\
$^{150}$Sm $\to$ $^{146}$Nd  & $0^+\to0^+$     & 7.37(9)           & $ 1449.8 \pm  1.0$ & --                          &                    & $8.7\times10^{27}$ & $1.8\times10^{28}$ & $9.0\times10^{27}$ \\
$^{152}$Sm $\to$ $^{148}$Nd  & $0^+\to0^+$     & 26.74(9)          & $  220.5 \pm  1.9$ & --                          &                    & $7.3\times10^{159}$& $1.5\times10^{160}$& $1.9\times10^{150}$\\
$^{151}$Eu $\to$ $^{147}$Pm  & $5/2^+\to7/2^+$ & 47.81(6)          & $ 1964.5 \pm  1.1$ & $4.6(12) \times 10^{18}$    & \cite{cas14}       & $2.9\times10^{17}$ & $3.6\times10^{18}$ & $7.1\times10^{18}$ \\
$^{153}$Eu $\to$ $^{149}$Pm  & $5/2^+\to7/2^+$ & 52.19(6)          & $  272.1 \pm  2.0$ & $> 5.5 \times 10^{17}$ $^b$ & \cite{dan12}       & $4.0\times10^{141}$& $7.3\times10^{144}$& $4.6\times10^{140}$\\
$^{152}$Gd $\to$ $^{148}$Sm  & $0^+\to0^+$     & 0.20(3)           & $ 2204.4 \pm  1.0$ & $1.08(8) \times 10^{14}$    & \cite{Macfarlane:1961,aud17} & $9.0\times10^{13}$ & $1.8\times10^{14}$ & $1.5\times10^{14}$\\
$^{154}$Gd $\to$ $^{150}$Sm  & $0^+\to0^+$     & 2.18(2)           & $  920.3 \pm  0.7$ & --                          &                    & $2.7\times10^{52}$ & $6.0\times10^{52}$ & $5.6\times10^{21}$ \\
$^{155}$Gd $\to$ $^{151}$Sm  & $3/2^-\to5/2^-$ & 14.80(9)          & $   81.5 \pm  0.7$ & --                          &                    & $2.6\times10^{311}$& $1.8\times10^{316}$& $2.9\times10^{305}$\\
$^{156}$Dy $\to$ $^{152}$Gd  & $0^+\to0^+$     & 0.056(3)          & $ 1753.0 \pm  0.3$ & $ > 10^{18}$                & \cite{Riezler:1958}& $3.4\times10^{24}$ & $6.8\times10^{24}$ & $4.2\times10^{24}$ \\
$^{158}$Dy $\to$ $^{154}$Gd  & $0^+\to0^+$     & 0.095(3)          & $  873.7 \pm  2.4$ & --                          &                    & $3.1\times10^{58}$ & $7.6\times10^{58}$ & $3.3\times10^{57}$ \\
$^{160}$Dy $\to$ $^{156}$Gd  & $0^+\to0^+$     & 2.329(18)         & $  437.3 \pm  1.1$ & --                          &                    & $2.8\times10^{106}$& $1.2\times10^{107}$& $2.8\times10^{102}$\\
$^{161}$Dy $\to$ $^{157}$Gd  & $5/2^+\to3/2^-$ & 18.889(42)        & $  342.8 \pm  1.1$ & --                          &                    & $6.6\times10^{127}$& $5.1\times10^{130}$& $1.6\times10^{129}$\\
$^{162}$Dy $\to$ $^{158}$Gd  & $0^+\to0^+$     & 25.475(36)        & $   83.2 \pm  1.1$ & --                          &                    & $6.2\times10^{318}$& $3.4\times10^{320}$& $1.0\times10^{273}$\\
$^{165}$Ho $\to$ $^{161}$Tb  & $7/2^-\to3/2^+$ & 100               & $  137.7 \pm  1.5$ & --                          &                    & $8.8\times10^{239}$& $1.2\times10^{246}$& $4.7\times10^{235}$\\
$^{162}$Er $\to$ $^{158}$Dy  & $0^+\to0^+$     & 0.139(5)          & $ 1647.9 \pm  2.3$ & $> 1.4 \times 10^{14}$      &\cite{Porschen:1956}& $1.9\times10^{29}$ & $4.1\times10^{29}$ & $1.6\times10^{29}$ \\
$^{164}$Er $\to$ $^{160}$Dy  & $0^+\to0^+$     & 1.601(3)          & $1304.92 \pm 0.17$ & --                          &                    & $1.0\times10^{40}$ & $2.8\times10^{40}$ & $4.0\times10^{39}$ \\
$^{166}$Er $\to$ $^{162}$Dy  & $0^+\to0^+$     & 33.503(36)        & $  830.5 \pm  1.1$ & --                          &                    & $5.8\times10^{64}$ & $2.8\times10^{65}$ & $2.2\times10^{63}$ \\
$^{167}$Er $\to$ $^{163}$Dy  & $7/2^+\to5/2^-$ & 22.869(9)         & $  665.1 \pm  1.1$ & --                          &                    & $1.9\times10^{79}$ & $1.2\times10^{82}$ & $3.4\times10^{81}$ \\
$^{168}$Er $\to$ $^{164}$Dy  & $0^+\to0^+$     & 26.978(18)        & $  551.9 \pm  1.1$ & --                          &                    & $4.0\times10^{92}$ & $4.2\times10^{93}$ & $2.9\times10^{89}$ \\
$^{170}$Er $\to$ $^{166}$Dy  & $0^+\to0^+$     & 14.910(36)        & $   51.2 \pm  1.7$ & --                          &                    & $1.3\times10^{437}$& $5.2\times10^{441}$& $1.7\times10^{345}$\\
$^{169}$Tm $\to$ $^{165}$Ho  & $1/2^+\to7/2^-$ & 100               & $ 1198.9 \pm  1.1$ & --                          &                    & $1.5\times10^{46}$ & $8.0\times10^{47}$ & $6.3\times10^{48}$ \\
$^{168}$Yb $\to$ $^{164}$Er  & $0^+\to0^+$     & 0.123(3)          & $ 1936.1 \pm  1.2$ & $> 1.3 \times 10^{14}$      &\cite{Porschen:1956}& $3.8\times10^{24}$ & $8.7\times10^{24}$ & $3.4\times10^{24}$ \\
$^{170}$Yb $\to$ $^{166}$Er  & $0^+\to0^+$     & 2.982(39)         & $ 1737.2 \pm  1.2$ & --                          &                    & $1.4\times10^{29}$ & $4.1\times10^{29}$ & $9.4\times10^{28}$ \\
$^{171}$Yb $\to$ $^{167}$Er  & $1/2^-\to7/2^+$ & 14.086(140)       & $ 1559.5 \pm  1.2$ & --                          &                    & $4.0\times10^{34}$ & $3.2\times10^{36}$ & $2.5\times10^{38}$ \\
$^{172}$Yb $\to$ $^{168}$Er  & $0^+\to0^+$     & 21.686(130)       & $ 1310.8 \pm  1.2$ & --                          &                    & $2.1\times10^{42}$ & $9.2\times10^{42}$ & $5.3\times10^{41}$ \\
$^{173}$Yb $\to$ $^{169}$Er  & $5/2^-\to1/2^-$ & 16.103(63)        & $  947.0 \pm  1.2$ & --                          &                    & $2.3\times10^{60}$ & $4.4\times10^{63}$ & $1.3\times10^{63}$ \\
$^{174}$Yb $\to$ $^{170}$Er  & $0^+\to0^+$     & 32.025(80)        & $  739.3 \pm  1.5$ & --                          &                    & $4.4\times10^{75}$ & $6.2\times10^{76}$ & $3.1\times10^{73}$ \\
$^{176}$Yb $\to$ $^{172}$Er  & $0^+\to0^+$     & 12.995(83)        & $  567   \pm  4  $ & --                          &                    & $4.9\times10^{94}$ & $1.9\times10^{96}$ & $1.7\times10^{91}$ \\
$^{175}$Lu $\to$ $^{171}$Tm  & $7/2^+\to1/2^+$ & 97.401(13)        & $ 1619.8 \pm  1.5$ & --                          &                    & $2.5\times10^{34}$ & $7.0\times10^{35}$ & $4.0\times10^{35}$ \\
$^{176}$Lu $\to$ $^{172}$Tm  & $7/2^-\to2^-$   & 2.599(13)         & $ 1567   \pm  6  $ & --                          &                    & $1.1\times10^{37}$ & $1.3\times10^{42}$ & $7.2\times10^{40}$ \\
$^{174}$Hf $\to$ $^{170}$Yb  & $0^+\to0^+$     & 0.16(12)          & $ 2494.5 \pm  2.3$ & $2.0(4) \times 10^{15}$     & \cite{Macfarlane:1961,aud17} & $3.5\times10^{16}$ & $7.4\times10^{16}$ & $3.5\times10^{16}$ \\
$^{176}$Hf $\to$ $^{172}$Yb  & $0^+\to0^+$     & 5.26(70)          & $ 2254.2 \pm  1.5$ & --                          &                    & $2.5\times10^{20}$ & $6.6\times10^{20}$ & $2.0\times10^{20}$ \\
$^{177}$Hf $\to$ $^{173}$Yb  & $7/2^-\to5/2^-$ & 18.60(16)         & $ 2245.7 \pm  1.4$ & --                          &                    & $4.5\times10^{20}$ & $5.2\times10^{22}$ & $4.4\times10^{22}$ \\
$^{178}$Hf $\to$ $^{174}$Yb  & $0^+\to0^+$     & 27.28(28)         & $ 2084.4 \pm  1.4$ & --                          &                    & $3.4\times10^{23}$ & $1.1\times10^{24}$ & $2.2\times10^{23}$ \\
$^{179}$Hf $\to$ $^{175}$Yb  & $9/2^+\to7/2^-$ & 13.62(11)         & $ 1807.7 \pm  1.4$ & --                          &                    & $4.5\times10^{29}$ & $4.0\times10^{32}$ & $4.7\times10^{31}$ \\
$^{180}$Hf $\to$ $^{176}$Yb  & $0^+\to0^+$     & 35.08(33)         & $ 1287.1 \pm  1.4$ & --                          &                    & $6.4\times10^{45}$ & $5.7\times10^{46}$ & $9.2\times10^{44}$ \\
$^{180m}$Ta $\to$ $^{176}$Lu & $9^-\to7^-$     & 0.01201(32)       & $ 2101.6 \pm  2.5$ & --                          &                    & $2.8\times10^{24}$ & $6.7\times10^{28}$ & $3.5\times10^{27}$ \\
$^{181}$Ta $\to$ $^{177}$Lu  & $7/2^+\to7/2^+$ & 99.98799(32)      & $ 1520.6 \pm  1.7$ & --                          &                    & $4.2\times10^{38}$ & $1.7\times10^{40}$ & $3.0\times10^{37}$ \\
$^{180}$W  $\to$ $^{176}$Hf  & $0^+\to0^+$     & 0.12(1)           & $ 2515.3 \pm  1.0$ & $1.59(5) \times 10^{18}$    &\cite{Munster:2014a}& $8.5\times10^{17}$ & $2.1\times10^{18}$ & $7.4\times10^{17}$ \\
$^{182}$W  $\to$ $^{178}$Hf  & $0^+\to0^+$     & 26.50(16)         & $ 1764.3 \pm  1.6$ & $> 7.7 \times 10^{21}$      &\cite{Cozzini:2004a}& $5.3\times10^{32}$ & $2.6\times10^{33}$ & $2.1\times10^{32}$ \\
$^{183}$W  $\to$ $^{179}$Hf  & $1/2^-\to9/2^+$ & 14.31(4)          & $ 1672.4 \pm  1.6$ & $> 4.1 \times 10^{21}$      &\cite{Cozzini:2004a}& $4.9\times10^{36}$ & $3.4\times10^{40}$ & $1.4\times10^{41}$ \\
$^{184}$W  $\to$ $^{180}$Hf  & $0^+\to0^+$     & 30.64(2)          & $ 1649.1 \pm  1.6$ & $> 8.9 \times 10^{21}$      &\cite{Cozzini:2004a}& $7.4\times10^{35}$ & $5.0\times10^{36}$ & $2.2\times10^{35}$ \\
$^{186}$W  $\to$ $^{182}$Hf  & $0^+\to0^+$     & 28.43(19)         & $ 1116   \pm  6  $ & $> 8.2 \times 10^{21}$      &\cite{Cozzini:2004a}& $2.0\times10^{56}$ & $4.6\times10^{57}$ & $9.2\times10^{54}$ \\
$^{185}$Re $\to$ $^{181}$Ta  & $5/2^+\to7/2^+$ & 37.40(5)          & $ 2194.4 \pm  1.5$ & --                          &                    & $3.4\times10^{24}$ & $4.4\times10^{25}$ & $7.8\times10^{24}$ \\
$^{187}$Re $\to$ $^{183}$Ta  & $5/2^+\to7/2^+$ & 62.60(5)          & $ 1651.4 \pm  1.5$ & --                          &                    & $1.6\times10^{37}$ & $6.1\times10^{38}$ & $2.2\times10^{37}$ \\
$^{184}$Os $\to$ $^{180}$W   & $0^+\to0^+$     & 0.02(2)           & $ 2958.7 \pm  1.6$ & $1.12(23) \times 10^{13}$   &\cite{pet14}        & $3.5\times10^{13}$ & $7.2\times10^{13}$ & $3.3\times10^{13}$ \\
$^{186}$Os $\to$ $^{182}$W   & $0^+\to0^+$     & 1.59(64)          & $ 2821.2 \pm  0.9$ & $2.0(11) \times 10^{15}$    &\cite{vio75}        & $1.9\times10^{15}$ & $4.7\times10^{15}$ & $1.6\times10^{15}$ \\
$^{187}$Os $\to$ $^{183}$W   & $1/2^-\to1/2^-$ & 1.96(17)          & $ 2721.7 \pm  0.9$ & --                          &                    & $4.1\times10^{16}$ & $4.5\times10^{19}$ & $5.1\times10^{16}$ \\
$^{188}$Os $\to$ $^{184}$W   & $0^+\to0^+$     & 13.24(27)         & $ 2143.2 \pm  0.9$ & --                          &                    & $1.4\times10^{26}$ & $6.8\times10^{26}$ & $7.2\times10^{25}$ \\
$^{189}$Os $\to$ $^{185}$W   & $3/2^-\to3/2^-$ & 16.15(23)         & $ 1976.1 \pm  0.9$ & --                          &                    & $4.8\times10^{29}$ & $2.4\times10^{34}$ & $3.1\times10^{29}$ \\
$^{190}$Os $\to$ $^{186}$W   & $0^+\to0^+$     & 26.26(20)         & $ 1375.8 \pm  1.2$ & --                          &                    & $2.0\times10^{47}$ & $3.6\times10^{48}$ & $2.1\times10^{46}$ \\
$^{192}$Os $\to$ $^{188}$W   & $0^+\to0^+$     & 40.78(32)         & $  361   \pm  4  $ & --                          &                    & $1.8\times10^{149}$& $1.7\times10^{153}$& $1.4\times10^{140}$\\
$^{191}$Ir $\to$ $^{187}$Re  & $3/2^+\to5/2^+$ & 37.3(2)           & $ 2082.8 \pm  1.2$ & --                          &                    & $4.2\times10^{28}$ & $7.2\times10^{29}$ & $6.0\times10^{28}$ \\
$^{193}$Ir $\to$ $^{189}$Re  & $3/2^+\to5/2^+$ & 62.7(2)           & $ 1018   \pm  8  $ & --                          &                    & $2.6\times10^{66}$ & $9.9\times10^{68}$ & $4.0\times10^{65}$ \\
$^{190}$Pt $\to$ $^{186}$Os  & $0^+\to0^+$     & 0.012(2)          & $ 3268.6 \pm  0.6$ & $4.97(16) \times 10^{11}$   & \cite{Brau17}      & $3.2\times10^{11}$ & $6.6\times10^{11}$ & $2.8\times10^{11}$ \\
\multicolumn{1}{r|}{$\to$ $^{186}$Os$^*$}&    $0^+\to2^+$, 137.2 keV &      &           & $2.2(6) \times 10^{14}$ $^c$& \cite{Belli:2011}  & $2.1\times10^{13}$ & $4.5\times10^{13}$ & $9.1\times10^{12}$ \\
$^{192}$Pt $\to$ $^{188}$Os  & $0^+\to0^+$     & 0.782(24)         & $ 2423.9 \pm  2.5$ & $> 6.0 \times 10^{16}$      & \cite{Kau66}       & $6.8\times10^{22}$ & $2.9\times10^{23}$ & $3.9\times10^{22}$ \\
$^{194}$Pt $\to$ $^{190}$Os  & $0^+\to0^+$     & 32.864(410)       & $ 1522.8 \pm  0.5$ & --                          &                    & $2.2\times10^{44}$ & $3.9\times10^{45}$ & $2.9\times10^{43}$ \\
$^{195}$Pt $\to$ $^{191}$Os  & $1/2^-\to9/2^-$ & 33.775(240)       & $ 1176.4 \pm  0.5$ & $> 6.3 \times 10^{18}$      & \cite{Belli:2011}  & $4.5\times10^{59}$ & $1.8\times10^{69}$ & $3.1\times10^{63}$ \\
$^{196}$Pt $\to$ $^{192}$Os  & $0^+\to0^+$     & 25.211(340)       & $  812.8 \pm  2.3$ & --                          &                    & $6.6\times10^{82}$ & $1.9\times10^{85}$ & $8.2\times10^{79}$ \\
$^{198}$Pt $\to$ $^{194}$Os  & $0^+\to0^+$     & 7.356(130)        & $  106   \pm  3  $ & $> 4.7 \times 10^{17}$      & \cite{Belli:2011}  & $2.8\times10^{336}$& $5.8\times10^{346}$& $9.2\times10^{284}$\\
$^{197}$Au $\to$ $^{193}$Ir  & $3/2^+\to3/2^+$ & 100               & $  971.6 \pm  1.4$ & --                          &                    & $1.3\times10^{72}$ & $6.0\times10^{74}$ & $2.8\times10^{68}$ \\
$^{196}$Hg $\to$ $^{192}$Pt  & $0^+\to0^+$     & 0.15(1)           & $ 2038   \pm  4  $ & --                          &                    & $1.6\times10^{32}$ & $1.2\times10^{33}$ & $5.5\times10^{31}$ \\
$^{198}$Hg $\to$ $^{194}$Pt  & $0^+\to0^+$     & 10.04(3)          & $ 1380.8 \pm  0.6$ & --                          &                    & $1.1\times10^{52}$ & $3.5\times10^{53}$ & $6.6\times10^{50}$ \\
$^{199}$Hg $\to$ $^{195}$Pt  & $1/2^-\to1/2^-$ & 16.94(12)         & $  822.9 \pm  0.7$ & --                          &                    & $1.8\times10^{85}$ & $5.6\times10^{99}$ & $2.4\times10^{82}$ \\
$^{200}$Hg $\to$ $^{196}$Pt  & $0^+\to0^+$     & 23.14(9)          & $  716.3 \pm  0.7$ & --                          &                    & $5.6\times10^{95}$ & $5.5\times10^{98}$ & $7.6\times10^{91}$ \\
$^{201}$Hg $\to$ $^{197}$Pt  & $3/2^-\to1/2^-$ & 13.17(9)          & $  332.3 \pm  0.8$ & --                          &                    & $2.4\times10^{169}$& $1.6\times10^{198}$& $3.4\times10^{168}$\\
$^{202}$Hg $\to$ $^{198}$Pt  & $0^+\to0^+$     & 29.74(13)         & $  133.8 \pm  2.2$ & --                          &                    & $5.1\times10^{301}$& $4.3\times10^{311}$& $1.2\times10^{261}$\\
$^{203}$Tl $\to$ $^{199}$Au  & $1/2^+\to3/2^+$ & 29.52(7)          & $  907.4 \pm  1.2$ & --                          &                    & $1.4\times10^{80}$ & $1.5\times10^{83}$ & $2.4\times10^{78}$ \\
$^{205}$Tl $\to$ $^{201}$Au  & $1/2^+\to3/2^+$ & 70.48(7)          & $  155   \pm  3  $ & --                          &                    & $2.5\times10^{280}$& $1.3\times10^{291}$& $7.8\times10^{254}$\\
$^{204}$Pb $\to$ $^{200}$Hg  & $0^+\to0^+$     & 1.4(6)            & $ 1968.5 \pm  1.1$ & $> 1.4 \times 10^{20}$      & \cite{bee13}       & $6.5\times10^{35}$ & $1.2\times10^{37}$ & $1.4\times10^{35}$ \\
$^{206}$Pb $\to$ $^{202}$Hg  & $0^+\to0^+$     & 24.1(30)          & $ 1134.8 \pm  1.1$ & $> 2.5 \times 10^{21}$      & \cite{bee13}       & $2.5\times10^{66}$ & $7.4\times10^{68}$ & $2.4\times10^{64}$ \\
$^{207}$Pb $\to$ $^{203}$Hg  & $1/2^-\to5/2^-$ & 22.1(50)          & $  392.3 \pm  1.3$ & $> 1.9 \times 10^{21}$      & \cite{bee13}       & $1.2\times10^{156}$& $3.3\times10^{189}$& $2.4\times10^{155}$\\
$^{208}$Pb $\to$ $^{204}$Hg  & $0^+\to0^+$     & 52.4(70)          & $  516.6 \pm  1.2$ & $> 2.6 \times 10^{21}$      & \cite{bee13}       & $7.8\times10^{127}$& $8.4\times10^{132}$& $9.3\times10^{120}$\\
$^{209}$Bi $\to$ $^{205}$Tl  & $9/2^-\to1/2^+$ & 100               & $ 3137.3 \pm  0.8$ & $2.04(8) \times 10^{19}$    & \cite{Beeman:2012a}& $1.1\times10^{18}$ & $6.5\times10^{18}$ & $2.0\times10^{19}$ \\
\multicolumn{1}{r|}{$\to$ $^{205}$Tl$^*$}& $9/2^-\to3/2^+$, 203.7 keV &     &           & $1.4(2) \times 10^{21}$     & \cite{Beeman:2012a}& $6.8\times10^{19}$ & $4.9\times10^{20}$ & $4.9\times10^{20}$ \\
$^{232}$Th $\to$ $^{228}$Ra  & $0^+\to0^+$     & 100               & $ 4081.6 \pm  1.4$ & $1.79(3) \times10^{10}$ $^d$& \cite{aud17}       & $3.7\times10^{10}$ & $3.0\times10^{10}$ & $1.5\times10^{10}$ \\
$^{234}$U  $\to$ $^{230}$Th  & $0^+\to0^+$     & 0.0054(5)         & $ 4857.5 \pm  0.7$ & $3.44(1) \times 10^{5}$ $^e$& \cite{aud17}       & $4.8\times10^{5}$  & $3.8\times10^{5}$  & $3.1\times10^{5}$  \\
$^{235}$U  $\to$ $^{231}$Th  & $7/2^-\to5/2^+$ & 0.7204(6)         & $ 4678.0 \pm  0.7$ & $1.48(2) \times10^{10}$ $^f$& \cite{aud17}       & $1.1\times10^{7}$  & $2.4\times10^{7}$  & $2.3\times10^{10}$ \\
$^{238}$U  $\to$ $^{234}$Th  & $0^+\to0^+$     & 99.2742(10)       & $ 4269.9 \pm  2.1$ & $5.66(2) \times 10^{9}$ $^g$& \cite{aud17}       & $1.5\times10^{10}$ & $1.3\times10^{10}$ & $5.9\times10^{9}$  \\
\end{longtable}
\noindent
$^a$ $\alpha$ decay to the first excited level of the daughter nucleus, $^{142}$Ce; $E_\gamma = 641.3$ keV. \\
$^b$ the limit is at 68\% C.L. \\
$^c$ the value of Ref. \cite{Belli:2011} has been re-calculated 
using the new recommended value of the natural abundance of $^{190}$Pt 0.012\% \cite{Meija:2016}, instead of 0.014\% used at that time. \\
$^d$ Partial $T_{1/2}$ for $^{232}$Th decay to g.s. of $^{228}$Ra from the total half-life $1.40(1) \times 10^{10}$ yr \cite{aud17} and branching $b=78.2\pm1.3\%$. \\
$^e$ Partial $T_{1/2}$ for decay to g.s. from the total half-life $2.455(6) \times 10^{5}$ yr \cite{aud17} and branching $b=71.38\pm0.16\%$. \\
$^f$ Partial $T_{1/2}$ for decay to g.s. from the total half-life $7.04(1) \times 10^{8}$ yr \cite{aud17} and branching $b=4.77\pm0.07\%$. \\
$^g$ Partial $T_{1/2}$ for decay to g.s. from the total half-life $4.468(6) \times 10^{9}$ yr \cite{aud17} and branching $b=79.0\pm0.27\%$. \\
\end{landscape}
\twocolumn

\noindent nuclear
emulsions. The alpha activity of $^{209}$Bi was not observed in the
first experiment of Ref. \cite{Jenkner:1949}, and only a lower limit on the
half-life $T_{1/2} > 3 \times 10^{15}$ yr was set. Further
experiments claimed detection of this decay with half-life at
level of $T_{1/2} \sim 2 \times 10^{17}$ yr
\cite{Faraggi:1951,Riezler:1952}. However, as shown in Fig. \ref{fig:bu-a-emulsion}, the 
performances (first of all, a rather poor energy resolution) of the emulsion detection 
technique was rather poor at that time.

\begin{figure}[!h]
\begin{center}
\vspace{-0.4cm}
\resizebox{0.38\textwidth}{!}{\includegraphics{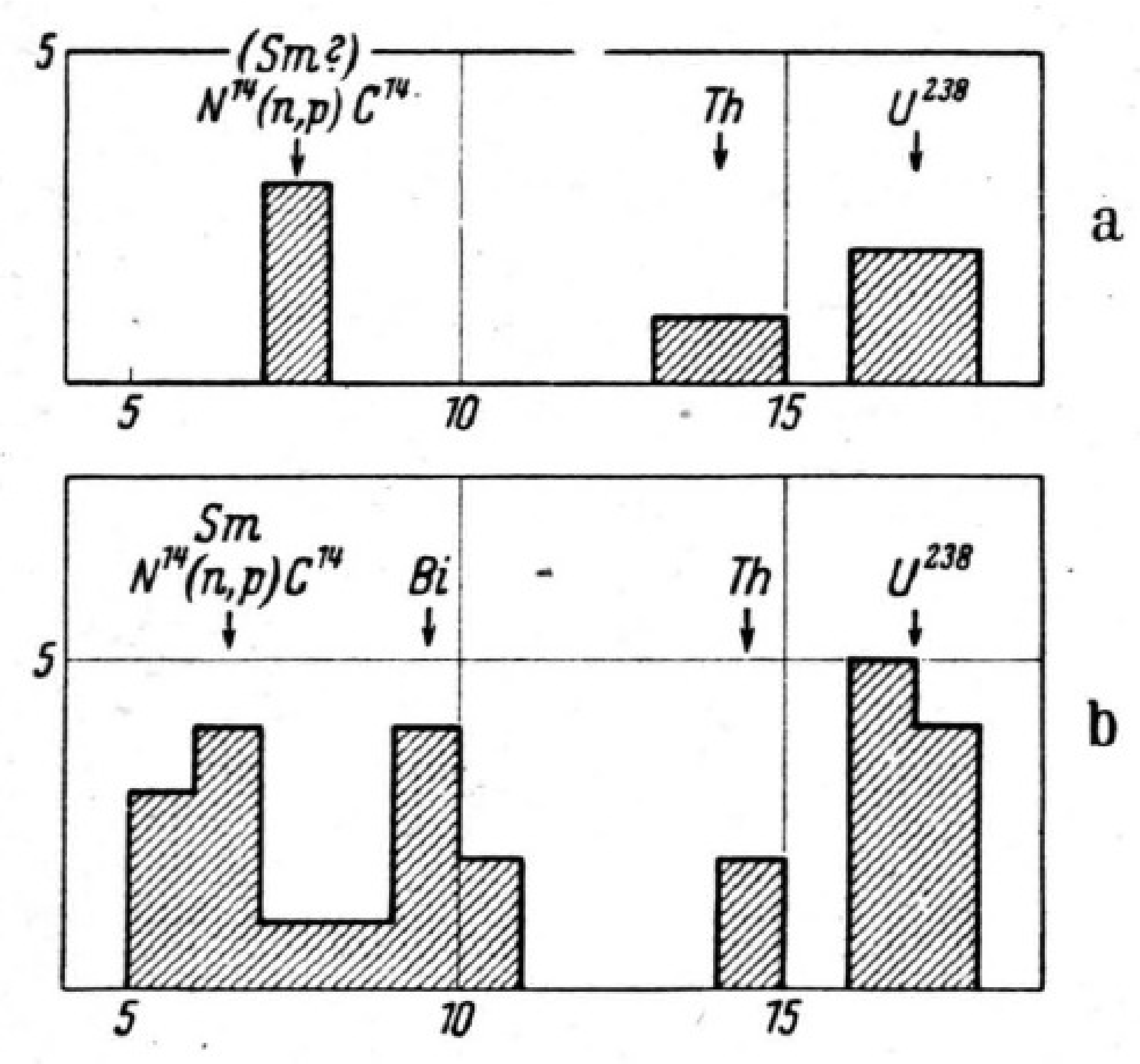}}
\end{center}
\vspace{-0.4cm}
\caption{Distributions of track lengths (on the X-axis) from 0 $\mu$m to 18 $\mu$m
in a blank emulsion plate with area 4.2 cm$^2$ over 100 days
(upper panel), and in the Bi-loaded emulsion with area 4.5 cm$^2$
over 100 days (lower panel). On the Y-axis is given the number of tracks per
interval of track lengths 1 $\mu$m. There is a peculiarity in the
data accumulated with the Bi-loaded emulsion which was 
interpreted by authors at that time as $\alpha$ decay of $^{209}$Bi with a
half-life $T_{1/2}=2\times10^{17}$ yr. Figure taken from Ref.
\cite{Riezler:1952}.}
\label{fig:bu-a-emulsion}
\end{figure}

This interpretation was not confirmed by other 
investigations where half-life limits at the level
of $T_{1/2} > 2 \times 10^{18}$ yr \cite{Hincks:1958} and
$T_{1/2}>10^{19}$ yr \cite{Carvalho:1972} were set. The alpha decay to the
first $3/2^+$ excited level of the daughter $^{205}$Tl ($E_{exc} =
203.7$ keV) was searched by using a low-background 
high-purity germanium (HPGe)
$\gamma$ spectrometry \cite{Norman:2000}; however, the
experimental sensitivity, $T_{1/2}>3\times10^{19}$ yr, was not
enough to detect the process.

\begin{figure}[!ht]
\begin{center}
\vspace{-0.3cm}
\resizebox{0.4\textwidth}{!}{\includegraphics{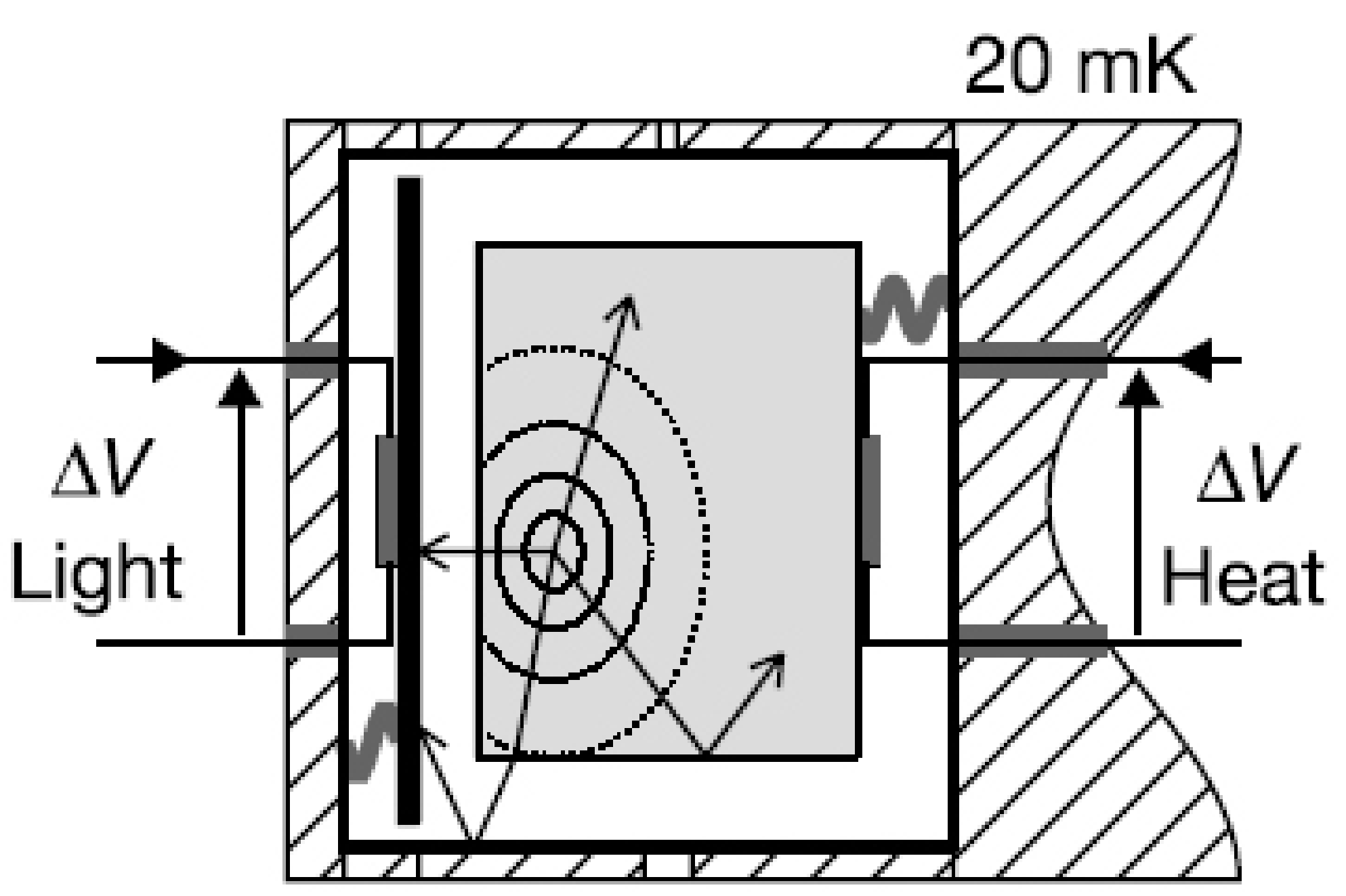}}
\end{center}
\caption{Simplified scheme of the low temperature scintillating
bolometer. The scintillating bolometer (BGO crystal, light-grey box) and the optical
bolometer (germanium disk, bold-grey bar) face each other in an Ag-coated
light-reflecting copper cavity cooled to milli-Kelvin temperature.
Heat sensors made from neutron transmuted doped germanium (NTD-Ge, small grey boxes 
in contact with the bolometers)
convert the temperature rise in the BGO crystal and germanium disk
into voltage signals. Figure taken from Ref. \cite{Marcillac:2003a}.}
\label{fig:sc-bol}
\end{figure}

\begin{figure*}[!ht]
\begin{center}
\resizebox{0.6\textwidth}{!}{\includegraphics{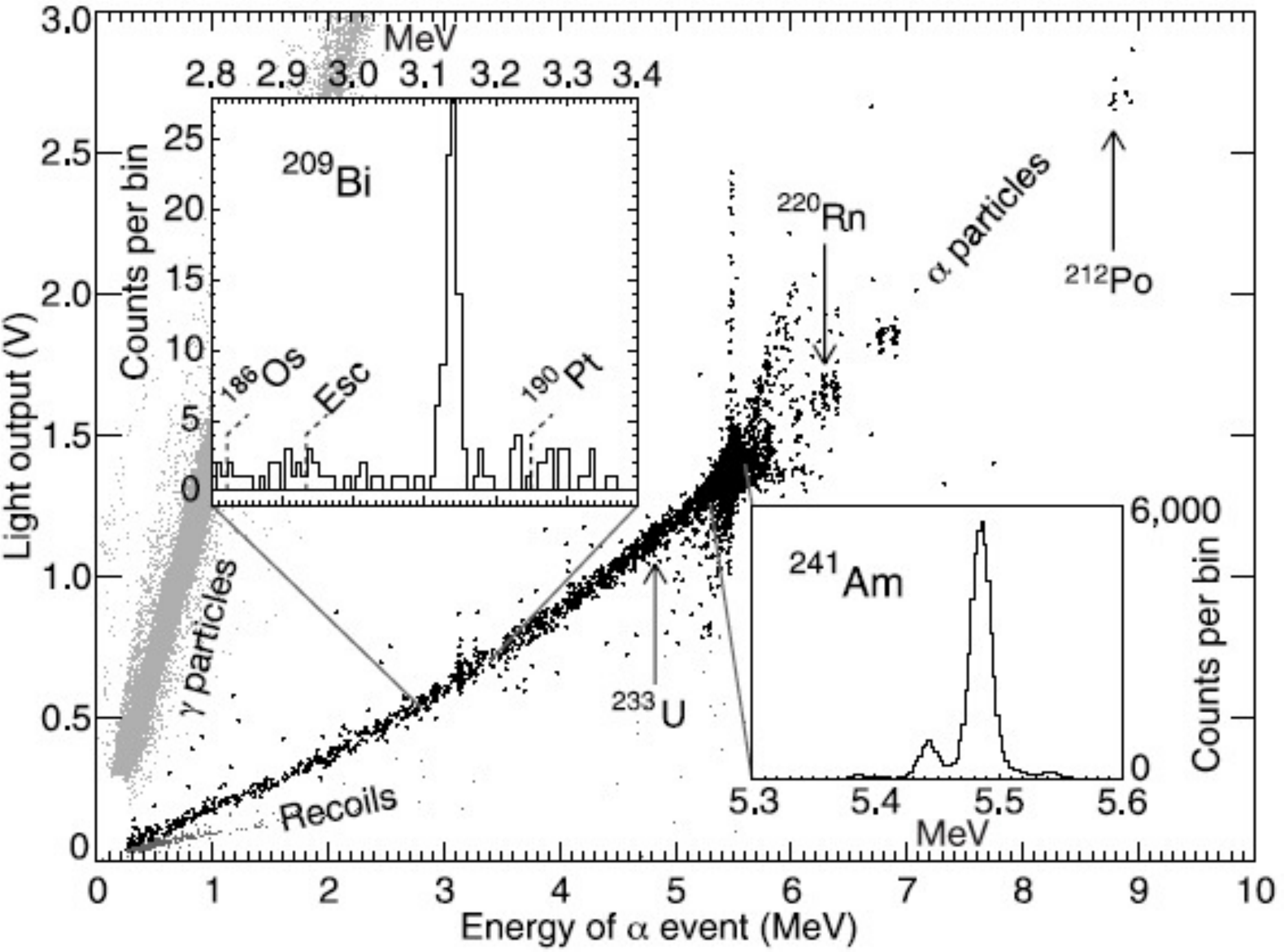}}
\end{center}
\vspace{-0.2cm}
\caption{Scatter plot of light-versus-heat signals measured with a
91-g BGO scintillating bolometer during a $^{241}$Am calibration run
over 114 h. The $\alpha$ events (black dots) are separated from $\beta$
and $\gamma$ events (gray dots) thanks to a much lower
scintillation light yield of the BGO scintillator to $\alpha$
particles. The insets show energy spectra of the $^{241}$Am
calibration source and the energy region where the peak of
$^{209}$Bi is observed. Figure taken from Ref. \cite{Marcillac:2003a}.}
\label{fig:bi-a-ground}
\end{figure*}

\begin{figure*}[!ht]
\begin{center}
\resizebox{0.55\textwidth}{!}{\includegraphics{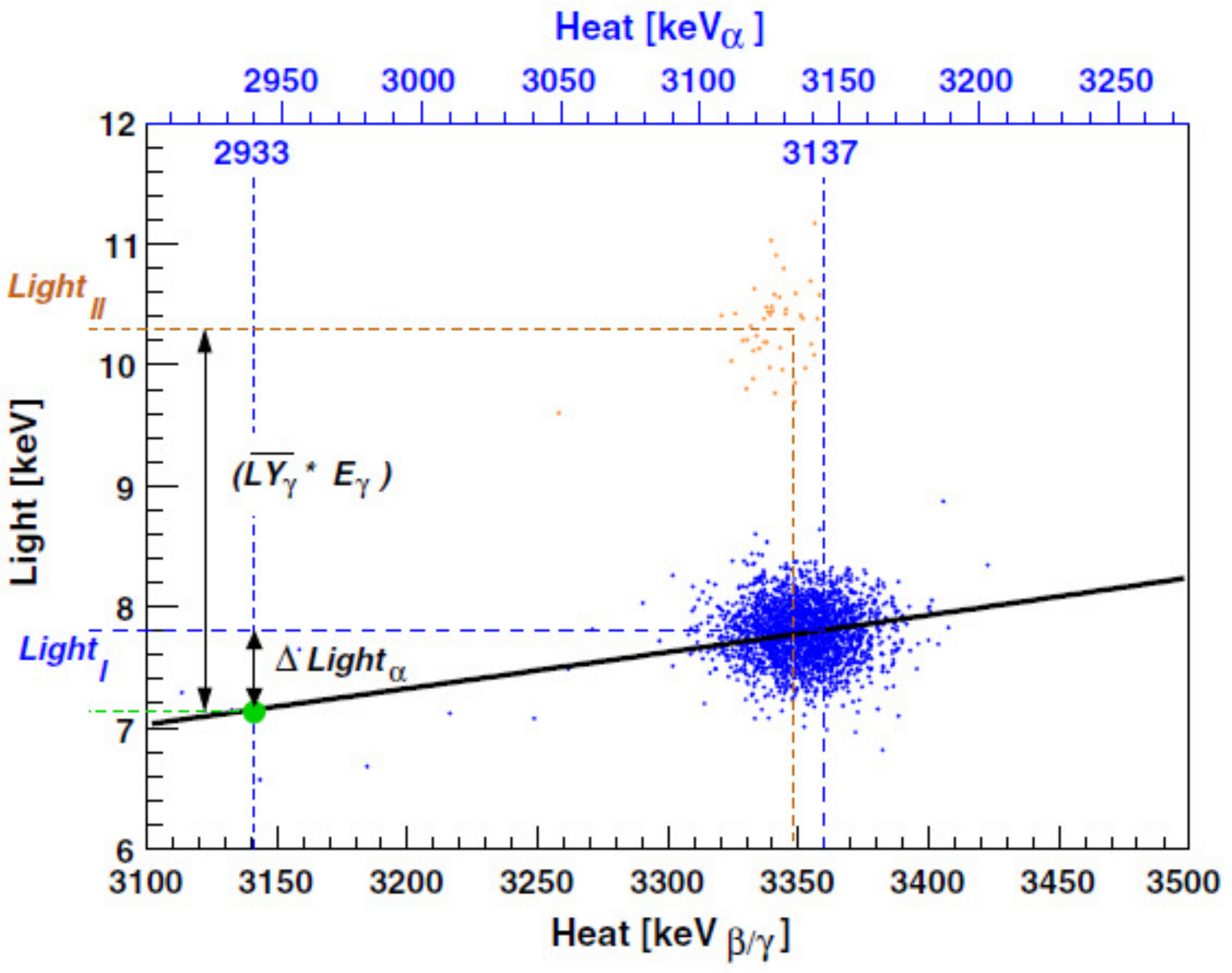}}
\end{center}
\vspace{-0.2cm}
\caption{(Color online) Scatter plot of light-versus-heat signals
measured with a 889-g BGO scintillating bolometer over 374.6 h. The
population of events in the right lower corner (blue dots) is due
to detection of $\alpha$ particles plus recoils after the $\alpha$
decay of $^{209}$Bi to the ground state of the daughter nuclei.
The events above the main population (red dots) are due to the
mixed $\alpha~+$~recoil$~+~\gamma$(conversion electron) events
corresponding to the $\alpha$ decay of $^{209}$Bi to the first
$3/2^+$ 204 keV excited level of $^{205}$Tl. Figure taken from
Ref. \cite{Beeman:2012a}.}
\label{fig:bi-a-excited}
\end{figure*}

\begin{table*}[!ht]
\caption{Limits and half-lives from experiments on the $\alpha$ decay of $^{209}$Bi. The abbreviation ``LT'' denotes low temperature.}
\begin{center}
\begin{tabular}{l|l|l|l}
 \hline
  Level of          & $T_{1/2}$ (yr)                & Experimental technique                        & Reference (year) \\
  daughter nucleus  & ~                             & ~                                             & \\
  ~                 & ~                             & ~                                             & \\
 \hline
  g.s. $1/2^+$      & $>3\times10^{15}$             & Bi-loaded nuclear emulsion                    & \cite{Jenkner:1949} (1949) \\
  g.s. $1/2^+$      & $=2.7\times10^{17}$           & Bi-loaded nuclear emulsion                    & \cite{Faraggi:1951} (1951) \\
  g.s. $1/2^+$      & $=2\times10^{17}$             & Bi-loaded nuclear emulsion                    & \cite{Riezler:1952} (1952) \\
  g.s. $1/2^+$      & $>2\times10^{18}$             & Bi-loaded nuclear emulsion                    & \cite{Hincks:1958} (1958) \\
  g.s. $1/2^+$      & $>10^{19}$                    & Bi-loaded nuclear emulsion                    & \cite{Carvalho:1972} (1972) \\
  g.s. $1/2^+$      & $=(1.9\pm0.2)\times10^{19}$   & Bi$_4$Ge$_3$O$_{12}$ LT scintillating bolometers & \cite{Marcillac:2003a} (2003) \\
  g.s. $1/2^+$      & $=(2.04\pm0.08)\times10^{19}$ & Bi$_4$Ge$_3$O$_{12}$ LT scintillating bolometer  & \cite{Beeman:2012a} (2012) \\
$203.7$ keV $3/2^+$ & $>3\times10^{19}$             & HPGe $\gamma$ spectrometry     & \cite{Norman:2000} (2000) \\
$203.7$ keV $3/2^+$ & $=(1.4\pm0.2)\times10^{21}$   & Bi$_4$Ge$_3$O$_{12}$ LT scintillating bolometer & \cite{Beeman:2012a} (2012) \\
 \hline
\end{tabular}
\end{center}
\label{tab:Bi-a}
\end{table*}

The $\alpha$ decay of $^{209}$Bi was detected for the first time
in 2003 using Bi$_4$Ge$_3$O$_{12}$ (BGO) low temperature
scintillating bolometers \cite{Marcillac:2003a}. The principle of the
low temperature scintillating bolometer operation is schematically
shown in Fig. \ref{fig:sc-bol}.

Two crystals with mass 45.7 g and 91.2 g cooled down to 20 mK were
utilized in the experiment carried out in a surface laboratory.
The $\alpha$ particles were separated from the events induced by cosmic
rays thanks to the high quenching of the scintillation signal for
$\alpha$ particles (we refer reader to work \cite{Tretyak:2010}
where quenching in scintillation detectors is reviewed). 
Fig. \ref{fig:bi-a-ground} shows the scatter plots of light-versus-heat 
signals measured during a $^{241}$Am calibration run.
The energy release of $[3137\pm 1(\mathrm{stat})\pm 2(\mathrm{syst})]$
keV was measured that is in agreement with the $Q_{\alpha}$ of
$^{209}$Bi: 3137.3(0.8) keV \cite{Wang:2017}. The half-life of
$^{209}$Bi was determined as $T_{1/2}=(1.9\pm0.2)\times10^{19}$ yr.

The observation was confirmed using a much more massive 889-g BGO
scintillating bolometer operated at the LNGS underground
laboratory \cite{Beeman:2012a}. The experiment not only measured
the half-life to the ground state transition with higher accuracy
$T_{1/2}=(2.04\pm0.08)\times10^{19}$ yr, but also observed the
$\alpha$ decay of $^{209}$Bi to the first excited level of
$^{205}$Tl with a half-life $T_{1/2} = (1.4\pm0.2)\times10^{21}$
yr. The pure $\alpha+~$recoil and $\alpha
+~$recoil$~+\gamma$(conversion electron) events were clearly
separated in the light channel of the detector thanks to the large
enough energy of the $\gamma$ rays and conversion electrons emitted in
the de-excitation of the 204 keV level, and comparatively high energy
resolution in the light channel (see Fig. \ref{fig:bi-a-excited}).

The experiments where $\alpha$ decay of $^{209}$Bi was searched
for (observed) are summarized in Table \ref{tab:Bi-a}. It should
be stressed that $^{209}$Bi has 100\% isotopic abundance
\cite{Meija:2016}, which facilitates the success of the experiments.

\subsection{The $\alpha$ decay of $^{190}$Pt and other platinum isotopes}
\label{pt190}

The $^{190}$Pt--$^{186}$Os system is an interesting tool to investigate the chronology and evolution of geo- and cosmo-chemistry samples (see e.g. Refs. \cite{Coo04,Bra06}).
All the six naturally occurring isotopes of platinum are potentially unstable in relation to $\alpha$
decay (see Table \ref{tab:alpha}). However, only for one of them, the $^{190}$Pt (with the largest energy release: $Q_\alpha = 3268.6(6)$ keV), this process was experimentally observed to-date, 
with the first successful
measurement in 1921 \cite{Hof21}. 
The natural abundance of $^{190}$Pt is very low: $\delta$ = (0.012 $\pm$ 0.002)\% \cite{Meija:2016}.
This aspect and the rather high cost of platinum, probably explains the lack of experimental results.

\subsubsection{$^{190}$Pt $\alpha$ decay}

In Ref. \cite{Hof21} and in subsequent works the half-life of $^{190}$Pt was determined to be in
the range of $(3.2-10)\times10^{11}$ yr for the decay to the ground state of $^{186}$Os (see review \cite{Tav06} and Refs. therein); the recommended
half-life value for this transition was:  $T^\alpha_{1/2}=(6.5\pm0.3)\times10^{11}$ yr \cite{Bag03},
but recently a laboratory measurement of the $\alpha$ decay half-life of $^{190}$Pt has been performed using a low background Frisch grid ionization chamber. The resulting half-life is (4.97 $\pm$ 0.16) $\times10^{11}$ yr, with a total uncertainty of 
3.2\% \cite{Brau17}. This result is in good agreement with the half-life obtained using the geological comparison method of (4.78 $\pm$ 0.05) $\times10^{11}$ yr.
In such methods the Pt--Os system is considered and the isotopic abundances are measured inside either ores or meteorites of known age (calibrated by different techniques,
mainly by $^{238}$U); thus, the standard method of the radiometric dating allows the evaluation of the half-life of the $^{190}$Pt parent. 

\begin{figure}[ht]
\begin{center}
\resizebox{0.48\textwidth}{!}{ 
  \includegraphics{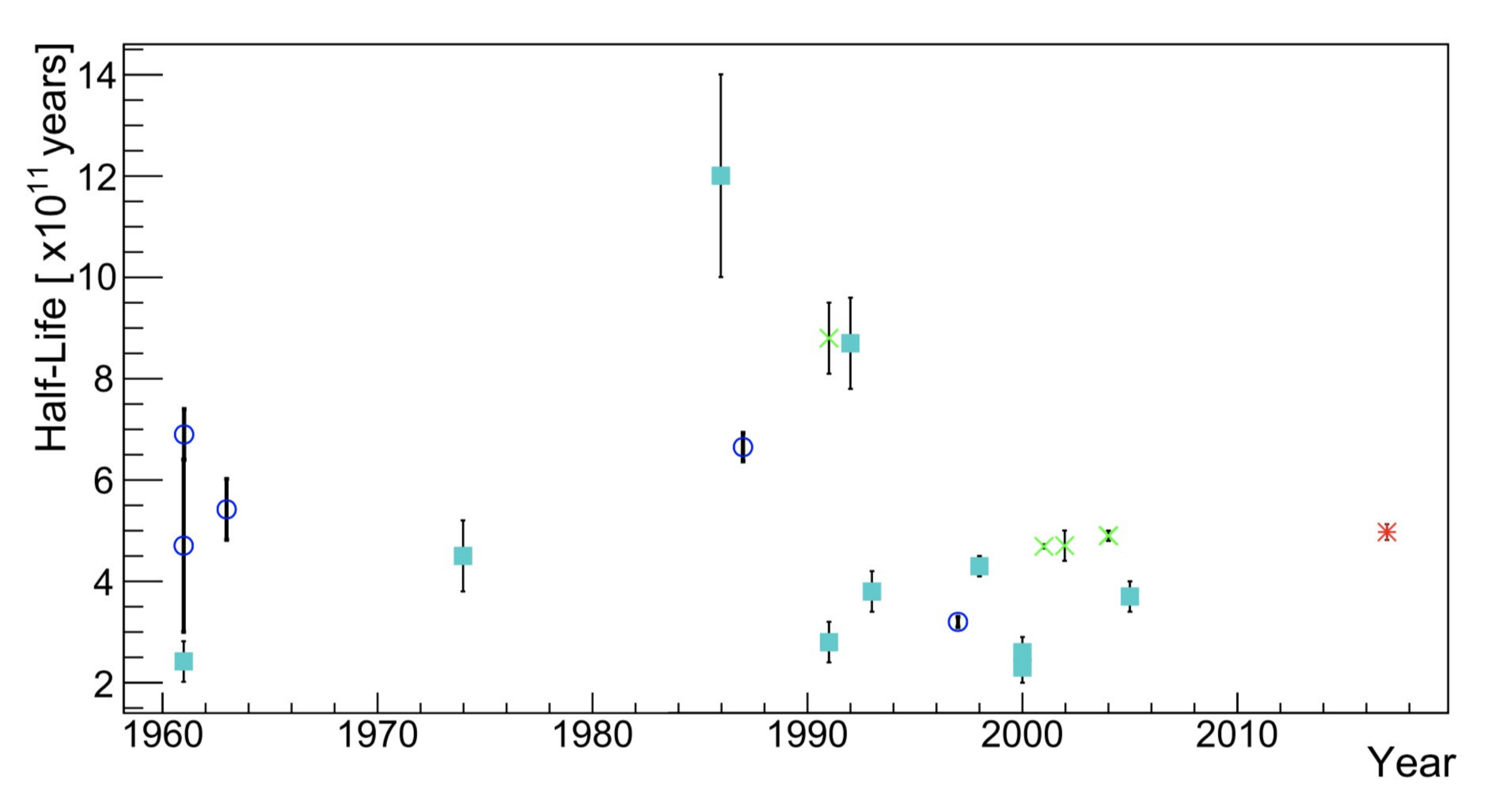}
}
\end{center}
\vspace{-0.4cm}
\caption{(Color online) $^{190}$Pt half-life measurements as a function of time. For the interpretation of the references to colour in this figure legend see Ref. 
\cite{Brau17}. Direct counting method (blue circle), geological comparison method (green cross), semiempirical calculation model (teal square) and the result from Ref. \cite{Brau17} (red cross) are shown. Figure taken from Ref. \cite{Brau17}.}
\label{Pt2}
\end{figure}

In Fig.~\ref{Pt2} the $^{190}$Pt half-life measurements as a function of time are shown. In general the most recent direct counting method results do not agree well with each other, whereas the 
most recent half-lives determined using the geological comparison method are consistent with each other. The source of the discrepancy is still unclear and further direct counting measurements
could be very useful for the interpretation of the results.

The $^{190}$Pt $\alpha$ decay was
observed only to the g.s. of $^{186}$Os and to the first excited state. In fact, the first excited level of the daughter
nuclide $^{186}$Os ($J^\pi = 2^+$) has a quite low energy: $E_{exc} = 137.2$ keV \cite{ToI98}, 
and the energy available to the
$\alpha$ particle in the decay to this level: $Q_\alpha^* = 3131.4(6)$ keV, is not much lower than that 
in the g.s. to g.s. transition. Theoretical estimates of the corresponding half-life \cite{Belli:2011} gave values in the range of $T^\alpha_{1/2}=10^{13}-10^{14}$ yr. 
This  $^{190}$Pt $\to$ $^{186}$Os($2^+_1$) decay was  in fact discovered
through the observation of the 137.2 keV $\gamma$ quantum emitted in the deexcitation of
the $^{186}$Os$^*$ nucleus with a well shielded ultra-low background HPGe detector even using a Pt sample of 42.5 g
with natural isotopic composition \cite{Belli:2011};
the measurements were performed at the LNGS.

In addition to the $^{190}$Pt decay, the $\gamma$ quanta can
be emitted in the $\alpha$ decays of other Pt isotopes: 
(1) when excited levels of the daughter Os nuclei are populated; 
(2) when the daughter Os isotopes are unstable and further decay with emission
of some $\gamma$ quanta, as in the case of $^{195}$Pt and $^{198}$Pt; in Ref.~\cite{Belli:2011} the
$T_{1/2}$ limits for these latter decays were determined for the first time.

\begin{figure}[ht]
\begin{center}
\resizebox{0.48\textwidth}{!}{ 
  \includegraphics{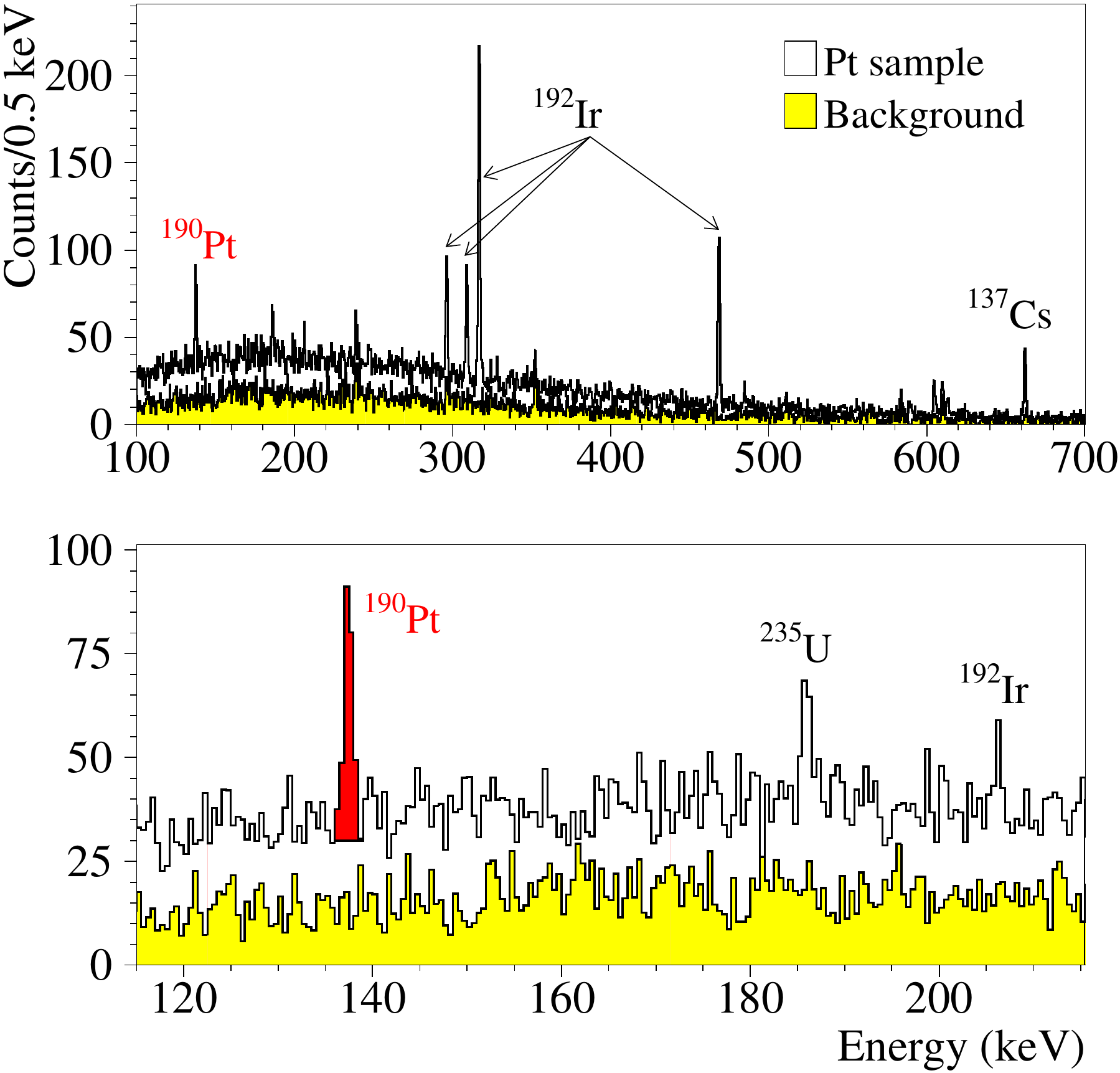}
}
\end{center}
\vspace{-0.4cm}
\caption{(Color online) Energy spectrum of the Pt sample with mass of 42.5 g
measured for 1815 h in the (100 -- 700) keV energy interval (upper part),
and in more detail around the 137 keV region (lower part).
The background spectrum (measured for 1046 h but normalized here to 1815 h) 
is also shown (filled histogram). 
Peak at 137 keV after $\alpha$ decay of $^{190}$Pt $\to$ $^{186}$Os$(2^+_1)$
is clearly visible in the Pt spectrum being absent in the background \cite{Belli:2011}.
Figure taken from Ref. \cite{Belli:2011}.}
\label{Pt3}
\end{figure}

Part of the spectrum accumulated with the Pt sample in comparison with the background in the energy range of 100--700 keV is shown in Fig.~\ref{Pt3}.

The 137.1 keV peak can be explained with the $\alpha$ decay of $^{190}$Pt to the first
excited level of $^{186}$Os, whose excitation energy is ($137.159\pm0.008$) keV \cite{Bag03}.
If populated, this level deexcites with the emission of a $\gamma$ quantum. Using the area of the 137 keV peak, the corresponding partial
half-life for the transition to the first excited level of $^{186}$Os can be calculated as:
\begin{equation}
T_{1/2}(^{190}\mbox{Pt} \to~ ^{186}\mbox{Os}(2^+_1, 137.2~\mbox{keV})) =
\frac{\ln 2 \cdot N_{190} \cdot \varepsilon \cdot t}{S \cdot (1+\alpha)},
\label{plat1}
\end{equation}
where $N_{190}=1.84\times10^{19}$ is the number of the $^{190}$Pt nuclei in the Pt sample, 
$\varepsilon$ is the efficiency to detect the full energy $\gamma$ ray with the HPGe detector (3.0\%), 
$t=1815.4$ h is the measurement time,
$S=(132\pm17)$ counts is the area of the peak and 
$\alpha$ is the coefficient of conversion of $\gamma$
quanta to electrons for the given nuclear transition.

The full peak efficiency at 137 keV was calculated with the EGS4 \cite{EGS4} and,
for crosscheck, also with GEANT4 \cite{GEANT4} simulation packages. 
It was validated through a voluminous ($ 7.0 \times 1.1$ cm) water source with several radioactive isotopes dissolved, 
among them $^{57}$Co with gamma quanta of 136.5 keV. The disagreement between the experimental and the calculated 
efficiencies was 6\% (0\%) with EGS4 (GEANT4). However, because of a more complicated geometry of the Pt sample 
in comparison with the simple cylindrical shape of the water source, the systematic uncertainty in the efficiency 
was conservatively estimated as 20\%.

Summing all the systematic uncertainties in squares, the following value for the half-life of
$\alpha$ decay $^{190}$Pt $\to$ $^{186}$Os$(2^+_1)$ was obtained \cite{Belli:2011}:
\begin{equation}
T_{1/2}(^{190}\mbox{Pt} \to~ ^{186}\mbox{Os}(2^+_1, 137.2~\mbox{keV})) =
\label{plat2}
\end{equation}
\[ 2.6_{-0.3}^{+0.4}\mbox{(stat.)}\pm0.6\mbox{(syst.)}\times10^{14} ~\mbox{yr}.\]
This value reported in Ref. \cite{Belli:2011} was calculated using the old value of the 
natural abundance of $^{190}$Pt equal to 0.014\%, from the previous recommendations \cite{Boh2005}.
Thus, using the new recommended value of 0.012\% \cite{Meija:2016}, the recalculated half-life is:
$T_{1/2} = (2.2 \pm 0.6) \times 10^{14}$ (see Table \ref{tab:alpha}).

It should be noted that $^{192}$Ir, present in the Pt sample, emits
also $\gamma$ rays with energy of 136.3 keV, however, with very low yield: $I=0.183\%$ \cite{ToI98}. 
Taking into account that the area of the most intensive ($I=82.80\%$) peak of $^{192}$Ir at
316.5 keV is ($619\pm32$) counts (and considering also the different efficiencies of
2.3\% at 137 keV\footnote{The efficiency for the 137 keV peak of $^{192}$Ir is lower than that
for the single 137 keV $\gamma$ quantum because of summing effects for $^{192}$Ir
$\gamma$ quanta emitted in cascade.}
and 5.5\% at 316 keV), contribution of $^{192}$Ir to the 137 keV peak is 
0.6 counts; this does not change the $T_{1/2}$ value presented in Eq.~(\ref{plat2}).
Gamma rays with energies close to 137 keV are emitted in some other nuclear 
processes (see e.g. \cite{WWW}), 
and this could give an alternative explanation of the peak observed in the Pt
experimental spectrum; however, usually additional $\gamma$ rays are
also emitted in such decays, and their absence allows us to exclude these effects. 
For example, $^{181}$Hf ($T_{1/2}=42.4$ d \cite{ToI98}), 
which could be created in result of cosmogenic activation of Pt,
emits 136.3 keV ($I=5.85\%$) and 136.9 keV ($I=0.86\%$) $\gamma$ quanta but
also 133.0 keV $\gamma$ quanta should be emitted with much higher yield of $I=43.3\%$
which, however, are absent in the experimental data.
Other example could be the 136.6 keV ($I=0.012\%$) $\gamma$ rays from $^{235}$U  
but peak at 143.8 keV ($I=10.96\%$) is absent. The contributions from the nuclides of the U/Th
natural radioactive chains: $^{214}$Pb ($E_\gamma=137.5$ keV, $I<0.006\%$) and
$^{228}$Ac ($E_\gamma=137.9$ keV, $I=0.024\%$) are calculated to be negligible
using the data on U/Th pollution in the Pt sample.

The $^{181}$W ($T_{1/2}=121.2$ d), other possible cosmogenic contamination of Pt, cannot be estimated in
the above-de\-scri\-bed way because its 136.3 keV $\gamma$ rays are the most intensive
with yield of $I=0.0311\%$. The $^{181}$W induced activity in Pt was calculated with the COSMO code 
\cite{Mar92}; the result was 1.5 decays per day per kg of $^{nat}$Pt. 
Taking into account the time of measurements (75.6 d), the small mass of the used Pt sample (42.5 g) and the 
low yield of these $\gamma$ rays, the contribution from $^{181}$W to the 137 keV peak
is estimated also to be negligible ($5\times10^{-5}$ counts).

\begin{figure*}[ht]
\begin{center}
\resizebox{0.70\textwidth}{!}{ 
  \includegraphics{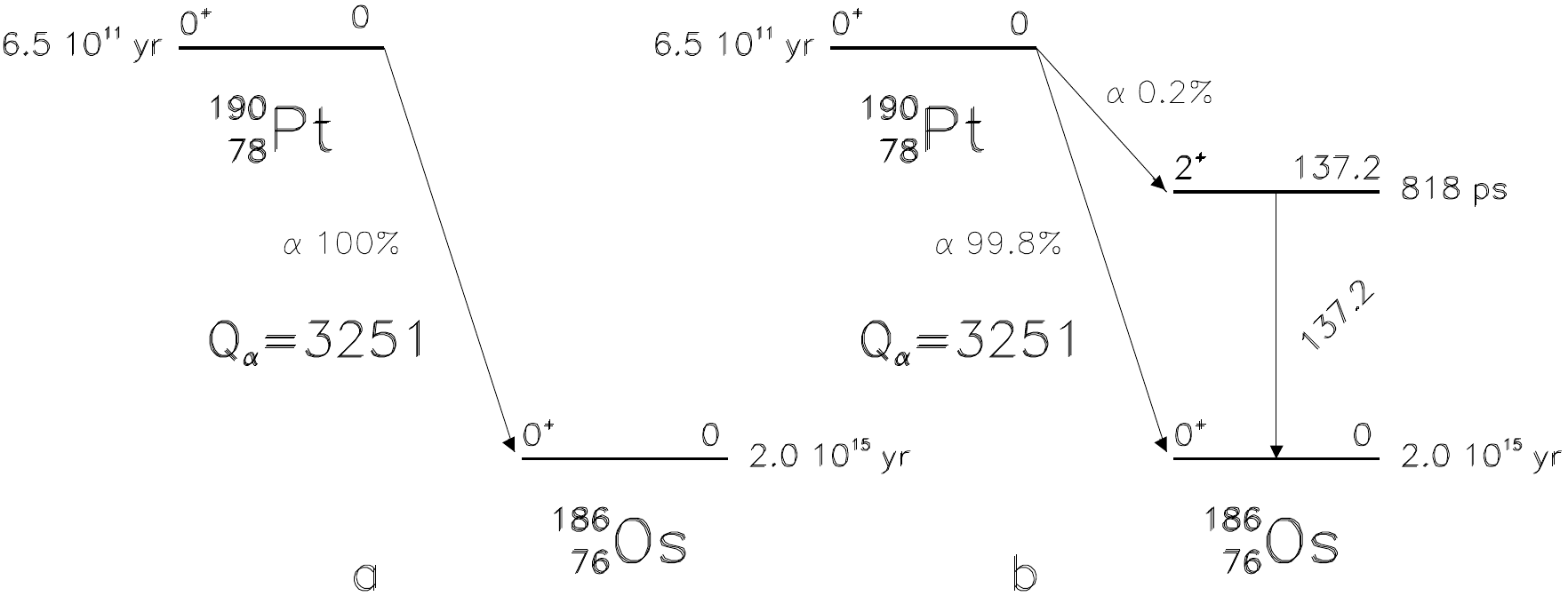}
}
\end{center}
\caption{Old (a) and new (b) schemes of the $\alpha$ decay of $^{190}$Pt.
The energies of the levels and the deexcitation $\gamma$ quantum are given 
in keV \cite{Bag03,ToI98}. Figure taken from Ref. \cite{Belli:2011}.}
\label{Pt4}
\end{figure*}

In conclusion, analysing other possible sources of the 137 keV peak, no 
real alternative that could mimic $\alpha$ decay 
$^{190}$Pt $\to$ $^{186}$Os$(2^+_1)$ was found. 
The old scheme of the $^{190}$Pt $\alpha$ decay \cite{Bag03,ToI98} and the updated scheme
which follows the observation of the $^{190}$Pt $\to$ $^{186}$Os$(2^+_1)$ transition
is presented in Fig.~\ref{Pt4}.

In the process of $^{190}$Pt $\alpha$ decay other excited levels of $^{186}$Os could be 
populated too. Because the probability of $\alpha$ decay exponentially decreases with the decrease
of the energy, we consider here only the possible transition to the next excited level of 
$^{186}$Os ($J^\pi=4^+$) with $E_{exc}=434.1$ keV. If this level is populated, two $\gamma$ quanta are
emitted in its deexcitation with energies of $E_{\gamma1}=296.9$ keV and $E_{\gamma2}=137.2$ keV.
Both peaks were present in the energy spectrum of Ref. \cite{Belli:2011}. 
While the 137 keV peak can be explained by the 
$^{190}$Pt $\to$ $^{186}$Os$(2^+_1)$ $\alpha$ decay, the 296 keV peak is related with 
$^{192}$Ir. 

The contribution to the 296 keV peak from $^{192}$Ir can be estimated using its nearby peak at 
316.5 keV. Taking into account that the area of the 316 keV peak is ($619\pm32$) counts, the
yields $I_{316}=82.80\%$ and $I_{296}=28.67\%$ \cite{ToI98}, one could expect ($193\pm10$) 
counts from $^{192}$Ir. At the same time, in the real Pt spectrum area of the 296 keV peak
is $S=200\pm23$ counts. The difference between these two values ($7\pm25$) counts is consistent 
with 0 and, in accordance with the Feldman-Cousins procedure \cite{Fel98}, results in  the
limit: $S<48$ counts at 90\% C.L. 
To be conservative, when deriving the $T_{1/2}$ limits for the Pt $\alpha$ decays, 
the used values for efficiencies were obtained 
with the code giving systematically lower efficiency.
Substituting in Eq.~(\ref{plat1}) the value of the
electron conversion coefficient $\alpha=0.095$ \cite{Bag03} and the
efficiency $\varepsilon=7.1\%$ for the 296 keV $\gamma$ quanta,
 the following half-life limit was obtained \cite{Belli:2011}:
\[T_{1/2}(^{190}\mbox{Pt} \to~ ^{186}\mbox{Os}(4^+_1, 434.1~\mbox{keV})) > \]
\[ > 3.6\times10^{15} ~\mbox{yr at 90\% C.L.}\]

\subsubsection{$T_{1/2}$ limits on $\alpha$ decays of other Pt isotopes}
 
The data collected in Ref.~\cite{Belli:2011} also allowed the search for the $\alpha$ decays of 
other Pt isotopes related with the emission of $\gamma$ quanta. 
Here only transitions to the lowest excited levels of the daughter Os isotopes  are considered as the
most probable. In the case of $^{195}$Pt and $^{198}$Pt, the daughter Os nuclei are unstable,
and the search for $\gamma$ quanta related with their decays gives the possibility to
look for the g.s. to g.s. transitions of Pt to Os. 
In general, no such decays were seen,
and only limits on the corresponding half-lives were determined. 

Following the procedure described above, limits on $\alpha$
decays of other primordial platinum isotopes were obtained by the analysis of the energy spectra 
in the regions of interest. A summary of all the
obtained results is given in Table~\ref{Ptsummary}.

\begin{table*}[!ht]
\caption{Summary of $T_{1/2}$ results on Pt $\alpha$ decays. The limits are at 90\% C.L.
Theoretical half-lives are calculated in accordance with the prescriptions of \cite{Buc91,Buc92} and \cite{Poe83}
taking into account additional hindrance factors.
For $^{195}$Pt and $^{198}$Pt (and for $^{192}$Pt all states), the experimental $T_{1/2}$ limits are valid for the $\alpha$ decays to
all the states of the daughter Os nuclei while the theoretical estimations are given for the g.s. to 
g.s. transitions.}
\begin{center}
\begin{tabular*}{\textwidth}{@{\extracolsep{\fill}}lrrrl@{}}
\hline
Alpha                       & Level of daughter  & Experimental                                  & Theoretical estimation  \\
transition                  & nucleus            & $T_{1/2}$ (yr)                                & $T_{1/2}$ (yr) \\
\hline
~ & ~ & ~ & ~ \\
$^{190}$Pt $\to$ $^{186}$Os & $0^+$, g.s.        & $(4.97\pm0.16)\times10^{11}$ \cite{Brau17}    & $3.2\times10^{11}$  \cite{Buc91}; $6.6\times10^{11}$ \cite{Poe83} \\
~                           &                    &                                               & $2.8\times10^{11}$  \cite{Den15}                                  \\
~                           & $2^+$, 137.2 keV   & $(2.2\pm0.6)\times10^{14}$ \cite{Belli:2011} $^a$ & $2.1\times10^{13}$  \cite{Buc91}; $4.5\times10^{13}$ \cite{Poe83} \\
~                           &                    &                                               & $9.1\times10^{12}$  \cite{Den15}; $1.6\times10^{13}$ \cite{Tav11}; $2.6 \times 10^{13}$ \cite{Ni011} \\
~                           & $4^+$, 434.1 keV   & $>3.6\times10^{15}$ \cite{Belli:2011}         & $7.4\times10^{17}$  \cite{Buc91,Buc92}    \\
~                           &                    &                                               & $2.0\times10^{18}$  \cite{Poe83}          \\
$^{192}$Pt $\to$ $^{188}$Os & all states         & $>6.0\times10^{16}$  \cite{Kau66}             & $6.8\times10^{22}$  \cite{Buc91}; $2.9\times10^{23}$ \cite{Poe83} \\
~                           &                    &                                               & $3.9\times10^{22}$  \cite{Den15}                                  \\
~                           & $2^+$, 155.0 keV   & $>1.3\times10^{17}$ \cite{Belli:2011}         & $9.1\times10^{25}$  \cite{Buc91,Buc92}    \\
~                           &                    &                                               & $4.6\times10^{26}$  \cite{Poe83} \\
~                           & $4^+$, 477.9 keV   & $>2.6\times10^{17}$ \cite{Belli:2011}         & $2.7\times10^{33}$  \cite{Buc91,Buc92} \\
~                           &                    &                                               & $1.9\times10^{34}$  \cite{Poe83} \\
$^{194}$Pt $\to$ $^{190}$Os & $2^+$, 186.7 keV   & $>2.0\times10^{18}$ \cite{Belli:2011}         & $7.0\times10^{51}$  \cite{Buc91,Buc92} \\
~                           &                    &                                               & $1.9\times10^{53}$  \cite{Poe83} \\
~                           & $4^+$, 547.9 keV   & $>6.8\times10^{19}$ \cite{Belli:2011}         & $4.3\times10^{71}$  \cite{Buc91,Buc92} \\
~                           &                    &                                               & $3.5\times10^{73}$  \cite{Poe83} \\
$^{195}$Pt $\to$ $^{191}$Os & all states         & $>6.3\times10^{18}$ \cite{Belli:2011}         & $4.5\times10^{59}$  \cite{Buc91}; $1.8\times10^{69}$ \cite{Poe83} \\
~                           &                    &                                               & $3.1\times10^{63}$  \cite{Den15}                                  \\
$^{196}$Pt $\to$ $^{192}$Os & $2^+$, 205.8 keV   & $>6.9\times10^{18}$ \cite{Belli:2011}         & $1.7\times10^{106}$ \cite{Buc91,Buc92} \\
~                           &                    &                                               & $2.1\times10^{109}$ \cite{Poe83} \\
$^{198}$Pt $\to$ $^{194}$Os & all states         & $>4.7\times10^{17}$ \cite{Belli:2011}         & $2.8\times10^{336}$ \cite{Buc91}; $5.8\times10^{346}$ \cite{Poe83}\\
~                           &                    &                                               & $9.2\times10^{284}$ \cite{Den15}                                  \\
\hline
\end{tabular*}
\end{center}
$^a$ The value of Ref. \cite{Belli:2011} has been re-calculated 
using the new recommended value of the natural abundance of $^{190}$Pt 0.012\% \cite{Meija:2016}, instead of 0.014\% used at that time.
\label{Ptsummary}
\end{table*}

The theoretical estimations for the half-lives of the $\alpha$ decay of the Pt isotopes
are reported in the last column of Table~\ref{Ptsummary}.
The approaches \cite{Buc91,Buc92,Poe83,Den15} have already been mentioned above.
The ground state of all Pt isotopes has spin and parity $J^\pi=0^+$, with the exception of
$^{195}$Pt with $J^\pi=1/2^-$. We consider here the transitions to the daughter levels with
$J^\pi=2^+$ and $4^+$ ($9/2^-$ for $^{195}$Pt $\to$ $^{191}$Os g.s.).
Because of the difference between the spins of the initial and of the final nuclear states,
the emitted $\alpha$ particle will have non-zero angular momentum: values of $l=2$
or $l=4$ are the most preferable among the allowed ones by the selection rules for the 
transitions with $\Delta J^{\Delta \pi}=2^+$ or $4^+$, respectively.
Thus, in the $T_{1/2}$ estimation with \cite{Buc91,Buc92}, we take into account the
hindrance factors of HF = 1.9 for $l=2$ and HF = 9.0 for $l=4$, 
calculated in accordance with Ref. \cite{Hey99}.
The results of the calculations are given in Table~\ref{Ptsummary}.
For the $\alpha$ decay $^{190}$Pt $\to$ $^{186}$Os($2^+$, 137.2 keV) 
we also report the results of Ref. \cite{Tav11} with $T_{1/2} = 1.6 \times 10^{13}$ yr
and of Ref. \cite{Ni011} with $T_{1/2} = 2.6 \times 10^{13}$ yr.

From the theoretical results, it is clear that the observation of other Pt $\alpha$ decays
presented in Table~\ref{Ptsummary} is out of reach of current experimental possibilities,
probably with the exception of the $^{190}$Pt $\to ^{186}$Os($4^+$, 434.1 keV) decay. The latter
will need 3 orders of magnitude higher sensitivity which, however, could be possible
by using enriched $^{190}$Pt instead of natural Pt with $\delta(^{190}$Pt) = 0.012\%.

\subsection{The $\alpha$ decay of osmium isotopes}
\label{osmium}

All the seven naturally occurring Os isotopes are potentially unstable in relation to $\alpha$ decay
(see Table \ref{tab:alpha} for energy releases and abundances). However, only for two of them (with the highest
$Q_\alpha$ values) the process was experimentally observed or indication of its existence obtained. 

For $^{184}$Os, only limits were known previously: $T_{1/2} > 2.0 \times 10^{13}$ yr \cite{Porschen:1956,bag10}
obtained with nuclear emulsions, 
and $T_{1/2} > 5.6 \times 10^{13}$ yr \cite{spe76} with proportional counter measurements of Os sample 
enriched in $^{184}$Os to 2.25\%. However, recently an indication on $\alpha$ decay of $^{184}$Os was found in
geochemical measurements \cite{pet14} where an excess of the daughter $^{180}$W was measured in meteorites
and terrestrial rocks; the half-life was determined to be $T_{1/2} = (1.12 \pm 0.23) \times 10^{13}$ yr. 
This experimental value is in good agreement with theoretical calculations:
$T_{1/2} = 2.6 \times 10^{13}$ yr \cite{Med06} and 
$T_{1/2} = 1.3 \times 10^{13}$ yr \cite{Gan09}
(see also Table \ref{tab:alpha}).

The decay of $^{186}$Os with $T_{1/2} = (2.0 \pm 1.1) \times 10^{15}$~yr was 
observed in a direct experiment with a semiconductor detector and an Os sample 
enriched in $^{186}$Os at 61.27\% \cite{vio75}.
Theoretical estimations, e.g. the recent ones 
$T_{1/2} = 2.7 \times 10^{15}$~yr \cite{Ism17} and
$T_{1/2} = (2.8 - 3.5) \times 10^{15}$~yr \cite{Wu18},
are in good agreement with this experimental value 
(see also Table \ref{tab:alpha}).

\begin{table*}
\caption{Naturally occurring Os isotopes (with $J^\pi$ value for the ground state), 
their abundances $\delta$ \cite{Meija:2016}, 
the expected energy release $Q_\alpha$ \cite{Wang:2017}, 
and the states of W daughters which could be populated in the $\alpha$ decays
accompanied by $\gamma$ quanta emission.}
\label{tab:os}  
\begin{center}     
\begin{tabular}{lllllllll}
\hline\noalign{\smallskip}
\multicolumn{2}{l}{Parent}  & $\delta$, \% & $Q_\alpha$, keV & \multicolumn{3}{l}{Populated level of W} & \multicolumn{2}{l}{Calculated $T_{1/2}$ (yr)}  \\
\multicolumn{2}{l}{isotope} &              &                 & \multicolumn{3}{l}{daughter nucleus}     & \cite{Den15}    &  \cite{Poe83}                \\
\noalign{\smallskip}
\hline
\noalign{\smallskip}
$^{184}_{76}$Os & $0^+$   &  0.02(2)$^a$ & 2958.7(16) & $^{180}_{74}$W & $2^+$   & 103.6 keV & $6.3\times 10^{14}$  & $2.9\times 10^{15}$  \\
                &         &              &            &                & $4^+$   & 337.6 keV & $9.2\times 10^{17}$  & $2.5\times 10^{19}$  \\
$^{186}_{76}$Os & $0^+$   &  1.59(64)    & 2821.2(9)  & $^{182}_{74}$W & $2^+$   & 100.1 keV & $3.3\times 10^{16}$  & $2.2\times 10^{17}$  \\
                &         &              &            &                & $4^+$   & 329.4 keV & $7.2\times 10^{19}$  & $2.8\times 10^{21}$  \\
$^{187}_{76}$Os & $1/2^-$ &  1.96(17)    & 2721.7(9)  & $^{183}_{74}$W & $3/2^-$ &  46.5 keV & $6.7\times 10^{18}$  & $4.4\times 10^{20}$  \\
                &         &              &            &                & $5/2^-$ &  99.1 keV & $4.0\times 10^{19}$  & $2.8\times 10^{21}$  \\
$^{188}_{76}$Os & $0^+$   & 13.24(27)    & 2143.2(9)  & $^{184}_{74}$W & $2^+$   & 111.2 keV & $1.3\times 10^{28}$  & $2.9\times 10^{29}$  \\
                &         &              &            &                & $4^+$   & 364.1 keV & $8.9\times 10^{33}$  & $1.9\times 10^{36}$  \\
$^{189}_{76}$Os & $3/2^-$ & 16.15(23)    & 1976.1(9)  & $^{185}_{74}$W & $3/2^-$ & g.s.      & $3.1\times 10^{29}$  & $2.4\times 10^{34}$  \\
$^{190}_{76}$Os & $0^+$   & 26.26(20)    & 1375.8(12) & $^{186}_{74}$W & $2^+$   & 122.6 keV & $1.6\times 10^{51}$  & $1.1\times 10^{54}$  \\
                &         &              &            &                & $4^+$   & 396.5 keV & $1.6\times 10^{65}$  & $5.8\times 10^{69}$  \\
$^{192}_{76}$Os & $0^+$   & 40.78(32)    & 361(4)     & $^{188}_{74}$W & $0^+$   & g.s.      & $1.4\times 10^{140}$ & $1.7\times 10^{153}$ \\
\noalign{\smallskip}
\hline
\noalign{\smallskip}
\multicolumn{7}{l}{$^a$ Recently the abundance of $^{184}$Os was measured more precisely in a sample} \\
\multicolumn{7}{l}{of ultra-pure osmium at the Curtin University as 0.0158(11)\% \cite{bel18}.} \\
\end{tabular}
\end{center}     
\end{table*}

The process of $\alpha$ decay can be accompanied by the emission of $\gamma$ quanta when the decay goes to excited
levels of a daughter nucleus (see Table \ref{tab:os})
or/and the daughter in the ground state is also unstable 
(as in case of $^{189}$Os and $^{192}$Os here). 
Thus, a possible strategy to detect such rare $\alpha$ decays is to 
look for the $\gamma$ rays following the $\alpha$ decay. 
An ultra-pure Os sample with natural isotopic composition is currently under
measurements at LNGS underground laboratory
with a low-background Broad Energy germanium (BEGe) detector \cite{bel18}.
This scheme was already used to search for double beta processes
in $^{184}$Os and $^{192}$Os nuclides \cite{bel13}.

\subsection{$^{180}$W and other W nuclides}
\label{w180}

The alpha decay is energetically allowed for all the five natural
isotopes of tungsten (see Table \ref{tab:alpha} where characteristics
of the isotopes are given). The $\alpha$ activity of tungsten with
half-life $T_{1/2}=2.2\times \delta \times 10^{17}$ yr (where
$\delta$ is the isotopic abundance of isotope) was postulated in an
early experiment by using the nuclear emulsion technique
\cite{Porschen:1953,Porschen:1956}. The authors considered two
possible explanations of the detected tracks with an energy of
$\sim 3$ MeV: 1) the activity should be attributed to a rare, and
as yet undiscovered, tungsten isotope (the assumption, however,
was called into question in \cite{Porschen:1956}); 2) the activity
is due to the alpha decay of $^{180}$W. Because the comparatively high
$Q_{\alpha}$ value for $^{180}$W, the result of Refs.
\cite{Porschen:1953,Porschen:1956} could be attributed to $\alpha$
decay of $^{180}$W with a half-life $2.6\times10^{14}$ yr
\cite{Danevich:2003a}. However, the observation in Refs.
\cite{Porschen:1953,Porschen:1956} was not supported by the work
\cite{Beard:1960a}, in which a cadmium tungstate (CdWO$_4$)
crystal scintillator (mass of 20.9 g) was simultaneously used as $\alpha$
source and as detector of the $\alpha$ decay events. After
193 h of measurements the limit: $T_{1/2}>1.0\times10^{15}$ yr, was
established. A similar restriction ($T_{1/2}>9.0\times10^{14}$ yr)
was also obtained in the experiment over 66.7 h with a ionization
counter (1200 cm$^2$ area) and a thin (83 mg/cm$^2$) sample of
W (79 mg total mass) enriched in $^{180}$W at 6.95\%
\cite{Macfarlane:1961}. These bounds were improved in the
measurements with two scintillators: CdWO$_4$ (mass of 452 g,
running time of 433 h), and $^{116}$CdWO$_4$ enriched in
$^{116}$Cd to 83\% (91.5 g, 951 h) \cite{Georgadze:1995a}.

\begin{table*}[!ht]
\caption{Limits and half-lives relatively to $\alpha$ decay of
natural tungsten isotopes.}
\label{tab:W-a}
\begin{center}
\resizebox{\textwidth}{!}{
\begin{tabular}{l|l|l|l|l}
 \hline
 $\alpha$ decay                     & Level of              & $T_{1/2}$ (yr)                    & Experimental technique                            & Reference (year) \\
                                    & daughter nucleus      & ~                                 & ~                                                 & \\
 \hline
 $^{180}$W$\rightarrow$$^{176}$Hf   & g.s. $0^+$            & $=2.6\times10^{14}$$^a$           & Nuclear emulsion                                  & \cite{Porschen:1953,Porschen:1956} (1953,1956) \\
 ~                                  & ~                     & $>1.0\times10^{15}$               & CdWO$_4$ crystal scintillator                     & \cite{Beard:1960a} (1960) \\
 ~                                  & ~                     & $>9.0\times10^{14}$               & Ionization counter, enriched $^{180}$W (6.95\%)   & \cite{Macfarlane:1961} (1961) \\
 ~                                  & ~                     & $>7.4\times10^{16}$               & CdWO$_4$ crystal scintillators                    & \cite{Georgadze:1995a} (1995) \\
 ~                                  & ~                     & $=1.1^{+0.9}_{-0.5}\times10^{18}$ & $^{116}$CdWO$_4$ crystal scintillators            & \cite{Danevich:2003a} (2003) \\
 ~                                  & ~                     & $=(1.8\pm0.2)\times10^{18}$       & CaWO$_4$ LT scintillating bolometers              & \cite{Cozzini:2004a} (2004) \\
 ~                                  & ~                     & $=1.0^{+0.7}_{-0.3}\times10^{18}$ & CaWO$_4$ scintillator                             & \cite{Zdesenko:2005a} (2005) \\
 ~                                  & ~                     & $=1.3^{+0.6}_{-0.5}\times10^{18}$ & ZnWO$_4$ scintillator                             & \cite{Belli:2011a} (2011) \\
 ~                                  & ~                     & $=(1.59\pm0.05)\times10^{18}$     & CaWO$_4$ LT scintillating bolometers                 & \cite{Munster:2014a} (2014) \\
 \hline
 $^{182}$W$\rightarrow$$^{178}$Hf   & g.s. $0^+$            & $>7.7\times10^{21}$               & CaWO$_4$ LT scintillating bolometers                 & \cite{Cozzini:2004a} (2004)  \\
 ~                                  & ~                     & ~                                 & ~                                                 & \\
 ~                                  & ~                     & ~                                 & ~                                                 & \\
  \hline
 $^{183}$W$\rightarrow$$^{179}$Hf   & g.s. $9/2^+$          & $>4.1\times10^{21}$              & CaWO$_4$ LT scintillating bolometers                  & \cite{Cozzini:2004a} (2004)  \\
 ~                                  & 375.0 keV $1/2^-$     & $>6.7\times10^{20}$              & ZnWO$_4$ crystal scintillators                     & \cite{Belli:2011b} (2011) \\
 ~                                  & ~                     & ~                                 & ~                                                 & \\
 \hline
 $^{184}$W$\rightarrow$$^{180}$Hf   & g.s. $0^+$            & $>8.9\times10^{21}$              & CaWO$_4$ LT scintillating bolometers                  & \cite{Cozzini:2004a} (2004)  \\
 ~                                  & ~                     & ~                                 & ~                                                 & \\
 ~                                  & ~                     & ~                                 & ~                                                 & \\
 \hline
 $^{186}$W$\rightarrow$$^{182}$Hf   & g.s. $0^+$            & $>8.2\times10^{21}$              & CaWO$_4$ LT scintillating bolometers                  & \cite{Cozzini:2004a} (2004)  \\
 ~                                  & ~                     & ~                                 & ~                                                 & \\
 ~                                  & ~                     & ~                                 & ~                                                 & \\
 \hline
\multicolumn{5}{l}{$^{a}$~Calculated in Ref. \cite{Danevich:2003a} taking into account the $^{180}$W isotopic abundance 0.0012(1).} \\
\end{tabular}}
\end{center}
\end{table*}

\begin{figure}[!ht]
\begin{center}
\vspace{-0.4cm}
\resizebox{0.5\textwidth}{!}{\includegraphics{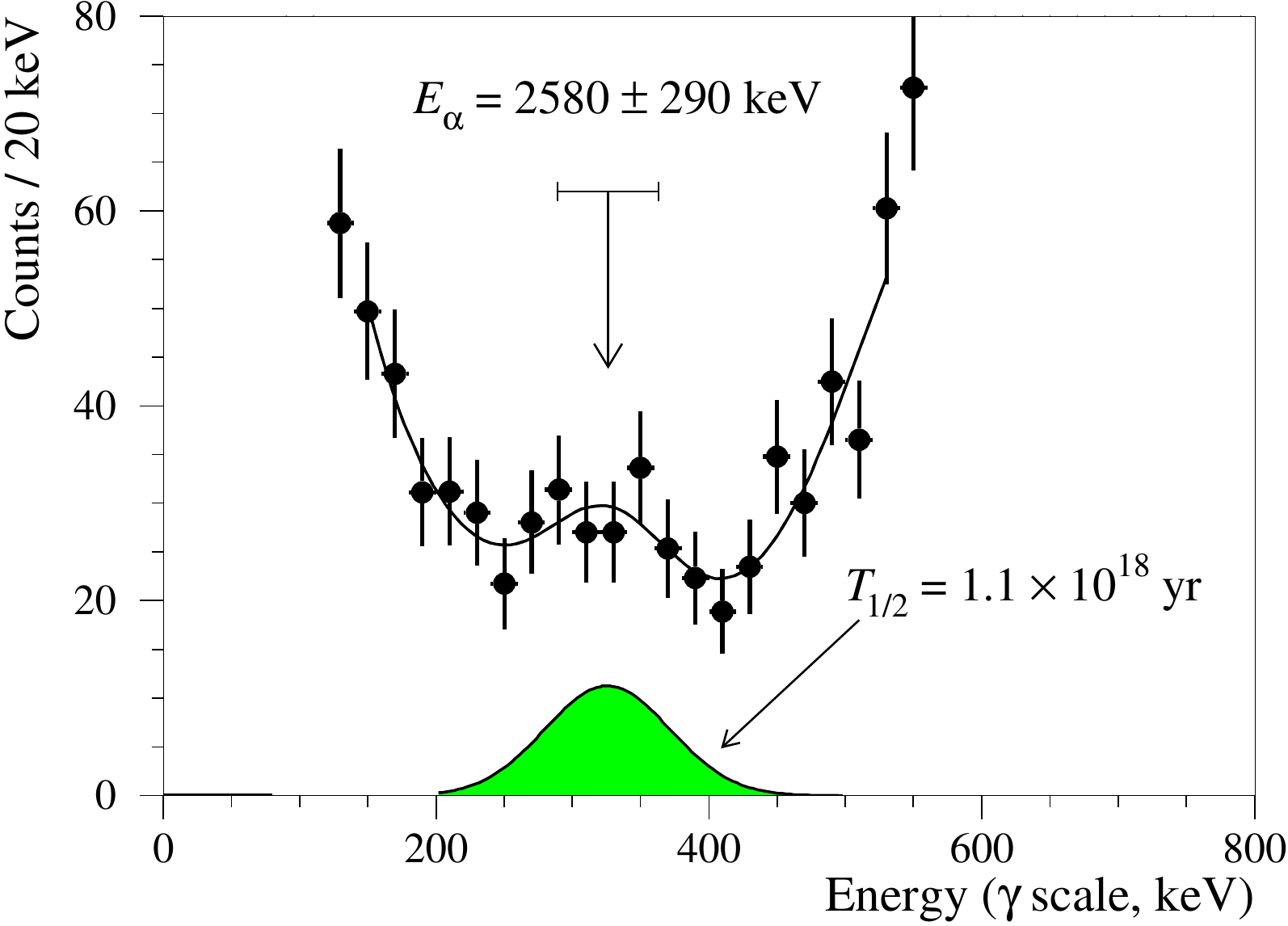}}
\end{center}
\vspace{-0.2cm}
\caption{(Color online) Part of the $\alpha$ spectrum measured
with the $^{116}$CdWO$_4$ detector over 2975 h together with the
fitting curve (solid line). The $\alpha$ peak of $^{180}$W with an
area of 64 counts corresponds to a half-life of
$T_{1/2}=1.1\times10^{18}$ yr. Figure taken from Ref. \cite{Danevich:2003a}.}
\label{fig:180w-a-scint}
\end{figure}

In 2003, the $\alpha$ decay of $^{180}$W was detected for the first
time \cite{Danevich:2003a} in the Solotvina $2\beta$ experiment
using $^{116}$CdWO$_4$ scintillator enriched in the isotope
$^{116}$Cd; the half-life \, $T_{1/2} = (1.1^{+0.9}_{-0.5}) \times 10^{18}$ yr was measured. 
Because of quenching of the light yield for $\alpha$ particles in scintillators \cite{Tretyak:2010}, the peak is shifted to lower energies 
in $\gamma$ energy scale. The discrimination of the events caused by $\alpha$ particles from those from $\beta / \gamma$'s (thanks to
the different shapes of their light signals) allowed to extract the $\alpha$ energy spectrum from all the accumulated events. The simultaneous 
coincidence of three conditions: (1) the peak is observed at the expected energy in $\gamma$ scale, (2) it is associated with 
$\alpha$ events, (3) the deduced half-life is in agreement with theoretical calculations, allowed to conclude that the peak 
is related with $^{180}$W $\alpha$ decay and not with other phenomena.
The $\alpha$ peak of $^{180}$W detected in the experiment is shown in
Fig. \ref{fig:180w-a-scint}.

\begin{figure}[htb]
\begin{center}
\vspace{-0.4cm}
\resizebox{0.48\textwidth}{!}{\includegraphics{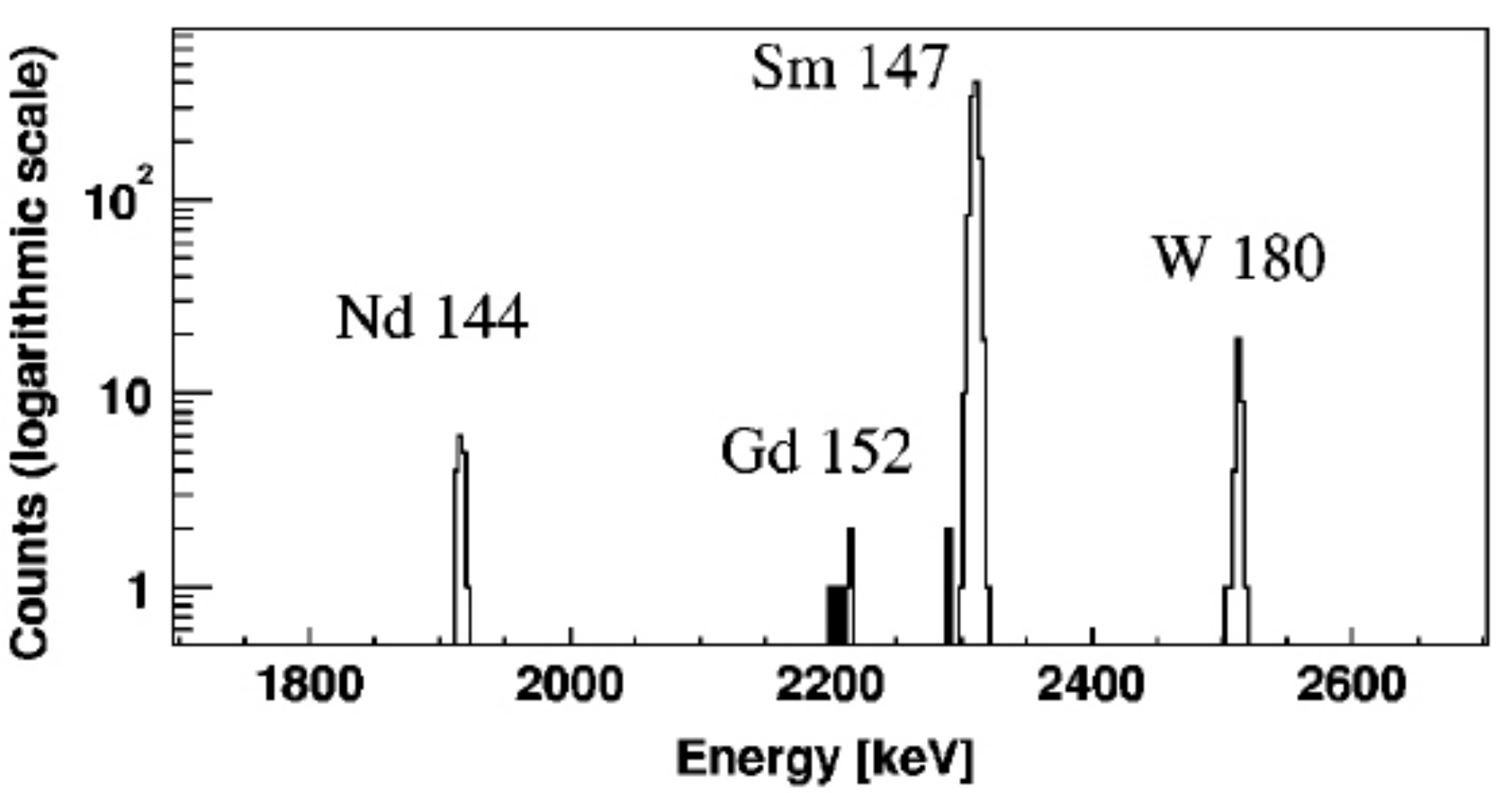}}
\end{center}
\vspace{-0.2cm}
\caption{Part of the $\alpha$ spectrum measured with the CaWO$_4$
low temperature scintillating bolometer in the experiment of Ref.
\cite{Cozzini:2004a}. The $\alpha$ peak of $^{180}$W is 
visible in the data thanks to the particle
discrimination and the energy resolution of the used technique.
Figure taken from Ref. \cite{Cozzini:2004a}.}
\label{fig:w180-a-scint-bolom}
\end{figure}

The observation of $^{180}$W decay was confirmed with low
temperature CaWO$_4$ scintillating bolometers 
($T_{1/2}=(1.8 \pm 0.2)\times10^{18}$ yr
\cite{Cozzini:2004a} and $T_{1/2}=(1.59\pm 0.05)\times10^{18}$ yr
\cite{Munster:2014a}). Fig. \ref{fig:w180-a-scint-bolom} shows 
some advantages of the cryogenic
scintillating bolometers technique (high energy
resolution and efficient particle discrimination). The decay was further
observed in the low background measurements with conventional scintillation detectors:
CaWO$_4$ ($T_{1/2}=(1.0^{+0.7}_{-0.3})\times10^{18}$ yr
\cite{Zdesenko:2005a}) and ZnWO$_4$
($T_{1/2}=(1.3^{+0.6}_{-0.5})\times10^{18}$ yr
\cite{Belli:2011a}).

The most stringent limits for the g.s. to g.s. $\alpha$ decays of
other W isotopes were set at level of
$T_{1/2}>10^{21}-10^{22}$ yr by using CaWO$_4$-based low
temperature scintillating bolometers \cite{Cozzini:2004a}, while
for the decay of $^{183}$W to the excited level of $^{179}$Hf
$(E_{exc}=375$ keV) the strongest limit $T_{1/2}>6.7\times10^{20}$
yr was obtained in Ref. \cite{Belli:2011b} by using a conventional
room temperature ZnWO$_4$ scintillation detector.

The results on $\alpha$ decay of natural tungsten isotopes 
are summarized in Table \ref{tab:W-a}, while
the theoretical predictions are reported in Table \ref{tab:alpha}.

The $\alpha$ decay of $^{180}$W to the first $2^+$ 88.3 keV excited
level of $^{176}$Hf is not observed yet. The detection of the decay
could be realized by specific high resolution scintillators or by scintillation bolometric technique with
tungstate crystal scintillators. The discrimination of
$\alpha~+~$recoils (decay to the ground state) and
$\alpha~+$ recoils + $\gamma$(conversion electrons) (decay to the
excited level) events would be a challenge for such a kind of
experiments because of the small energy of the excited level, and the 
typically rather poor energy resolution in the light channel of
scintillating bolometers. In fact, possible detectors for such experiments
are CaWO$_4$, ZnWO$_4$, CdWO$_4$; in particular, ZnWO$_4$ and
CdWO$_4$ scintillators have the advantage of very high level of radio-purity
\cite{Danevich:2018a}; but they have only modest energy resolution for 
alpha particles because of the anisotropy of their scintillation response \cite{Danevich:2003a,Danevich:2005a}
which is a disadvantage for an accurate particle discrimination.
The CaWO$_4$ scintillator, despite typically with higher radioactive
contamination \cite{Danevich:2018a}, might provide the 
best energy resolution of all tungstate scintillators for 
$\alpha$ particles \cite{Zdesenko:2005a}.

\subsection{$^{178m2}$Hf $\alpha$ decay}
\label{hf178m2}

The nuclear isomer $^{178m2}$Hf is an excited state of $^{178}$Hf with unique
combination of high excitation energy ($E_{exc}$ $=$ $2446.09(8)$~keV) and 
long half-life ($T_{1/2} = 31(1)$~yr) \cite{Ach09}. 
The latter is related with the high spin, $J^\pi = 16^+$, and high 
projection of the total angular momentum on the symmetry axis, $K=16$ of the isomer
(for the ground state $J^\pi = 0^+$, $K=0$).
The de-excitation of the $^{178m2}$Hf usually occurs through isomeric transitions
to $^{178}$Hf states with lower energies, lower spins and $K=8, K=0$ values. 
However, its decay is energetically possible also through $\alpha$ decay to
$^{174}$Yb (difference in energies between $^{178m2}$Hf and the ground state of
$^{174}$Yb is: $Q_\alpha = 4530$~keV), 
$\beta^-$ decay to $^{178}$Ta ($Q_{\beta^-} = 606$~keV),
electron capture to $^{178}$Lu ($Q_{\varepsilon} = 349$~keV) and
spontaneous fission $Q_{SF} \simeq 100$~MeV) \cite{Wang:2017}.
In 1980, only limits for these possible decays were established \cite{Kli80}:
$T_{1/2}(\alpha) > 6\times10^8$~yr (branching ratio: $b.r. < 5\times10^{-8}$),
$T_{1/2}(\beta^-) > 1\times10^4$~yr ($b.r. < 3\times10^{-3}$),
$T_{1/2}(\varepsilon) > 3\times10^3$~yr ($b.r. < 1\times10^{-2}$),
$T_{1/2}$(SF)$ > 1\times10^9$~yr ($b.r. < 3\times10^{-8}$).

In the work of 2007 \cite{Karamian:2007}, a thin source with a number 
of $^{178m2}$Hf nuclei equal 
to $2.8\times10^{13}$ was inserted between two solid-state track detectors CR-39.  
In the measurement of about 1 year duration, 307 excess $\alpha$ counts (in comparison with
background measurements) were observed and the half-life was deduced as:
$T_{1/2}(\alpha) = (2.5 \pm 0.5)\times10^{10}$~yr. 
Despite the largest energy release, the decay goes not directly to the ground state of $^{174}$Yb
because of the suppression due to the large spin variation ($16^+$ $\to$ $0^+$), but to few excited
$^{174}$Yb levels; as the theoretical estimations show,
the most probable transition is to the $6^+$ excited level at $E_{exc} = 526$~keV 
\cite{Karamian:2007}.

It is interesting to note that the high excitation energy and the long half-life
of $^{178m2}$Hf lead to the hope of creating a $\gamma$ ray laser (graser) and even to
the creation of non-conventional nuclear weapon (the so-called ``isomeric bomb'' 
with 1.33 GJ stored in 1 g of pure $^{178m2}$Hf, which is equivalent to 315 kg of trinitrotoluene) 
\cite{Sch04,Tka05}. It was supposed that the energy release can be triggered 
by external flux of particles as $\gamma$ rays and neutrons.

In 1999, Collins et al. reported evidence for accelerated decay of $^{178m2}$Hf irradiated
by X rays with an end point at $\simeq 90$~keV \cite{Col99}.
In this and subsequent experiments, the Collins' group observed an increase of the intensities
of different $\gamma$ peaks emitted in the $^{178m2}$Hf decay by $\simeq 2-6\%$;
the effect was seen for the 213, 217, 426, 495 keV $\gamma$ peaks, and also new peaks 
at 130 and 2457 keV of unknown origin were detected. 
However, (1) in different measurements different peaks were enhanced, (2)
other groups did not observe the effect, (3) the experimental results of 
Collins et al. contradicted the theoretical estimations by orders of magnitude
(see reviews \cite{Tka05,Car04,Har08} and one of the last works \cite{Kir15}
and Refs. therein).

It should be noted, however, that the theoretical analyses concerned the isomeric
transition in the decay of $^{178m2}$Hf, while it would be useful to analyse also
the possible enhancement of both the other two decays:
$^{178m2}$Hf $\beta^-$ decay and electron capture. 
In one of their experiments Collins et al. observed (at $6.5\sigma$ level)
a new $\gamma$ peak at 2457.20(22)~keV. A $^{178m2}$Hf sample was irradiated
at the SPring-8 synchrotron radiation source by monochromatic X rays with energy  
range $9.555 - 9.575$~keV \cite{Col05}. The sum of the energies of $^{178m2}$Hf and X rays:
2446.09(8) + 9.575 = 2455.67(8)~keV, is lower than the energy of the new line by
1.54(23)~keV (that was outside of calibration uncertainty)
and its appearance cannot be explained by X ray capture. 
It is interesting to note that $^{174}$Yb, daughter nuclide in the $^{178m2}$Hf $\alpha$ decay,
has an excited level at 2457(3)~keV with spin ($14^+$) \cite{Bro99} with the central value
coinciding with the energy of the new line. This coincidence is very curious, however, 
even if supposing that the 2457~keV level of $^{174}$Yb was populated in enhanced
$^{178m2}$Hf $\alpha$ decay, it is impossible to suggest a mechanism of direct
deexcitation of the 2457~keV level to the ground state of $^{174}$Yb.

\subsection{$^{151}$Eu $\alpha$ decay}
\label{eu151}

Both $^{151}$Eu (natural abundance 47.81(6)\% \cite{Meija:2016}) and $^{153}$Eu (52.19(6)\%)
have positive energy release to $\alpha$ decay, $Q_\alpha$: 1964.5(11) keV and 272.1(20) keV, respectively \cite{Wang:2017};
thus, they are potentially $\alpha$ radioactive.
The low $Q_\alpha$ of the $^{153}$Eu $\alpha$ decay implies a very long expected half-life far from the current experimental 
sensitivity, while the estimated half-life 
for the $^{151}$Eu has been explorable in recent years. It was measured for the first time
in 2007 \cite{bel07} with
a CaF$_2$(Eu) scintillator ($T_{1/2} = (5^{+11}_{-3}) \times 10^{18}$ yr)
and recently measured more precisely with a Li$_6$Eu(BO$_3$)$_3$ low-temperature
scintillating bolometer as $T_{1/2} = (4.6 \pm 1.2) \times 10^{18}$ yr \cite{cas14}.

\begin{figure*}[!ht]
\resizebox{1.0\textwidth}{!}{ 
  \includegraphics{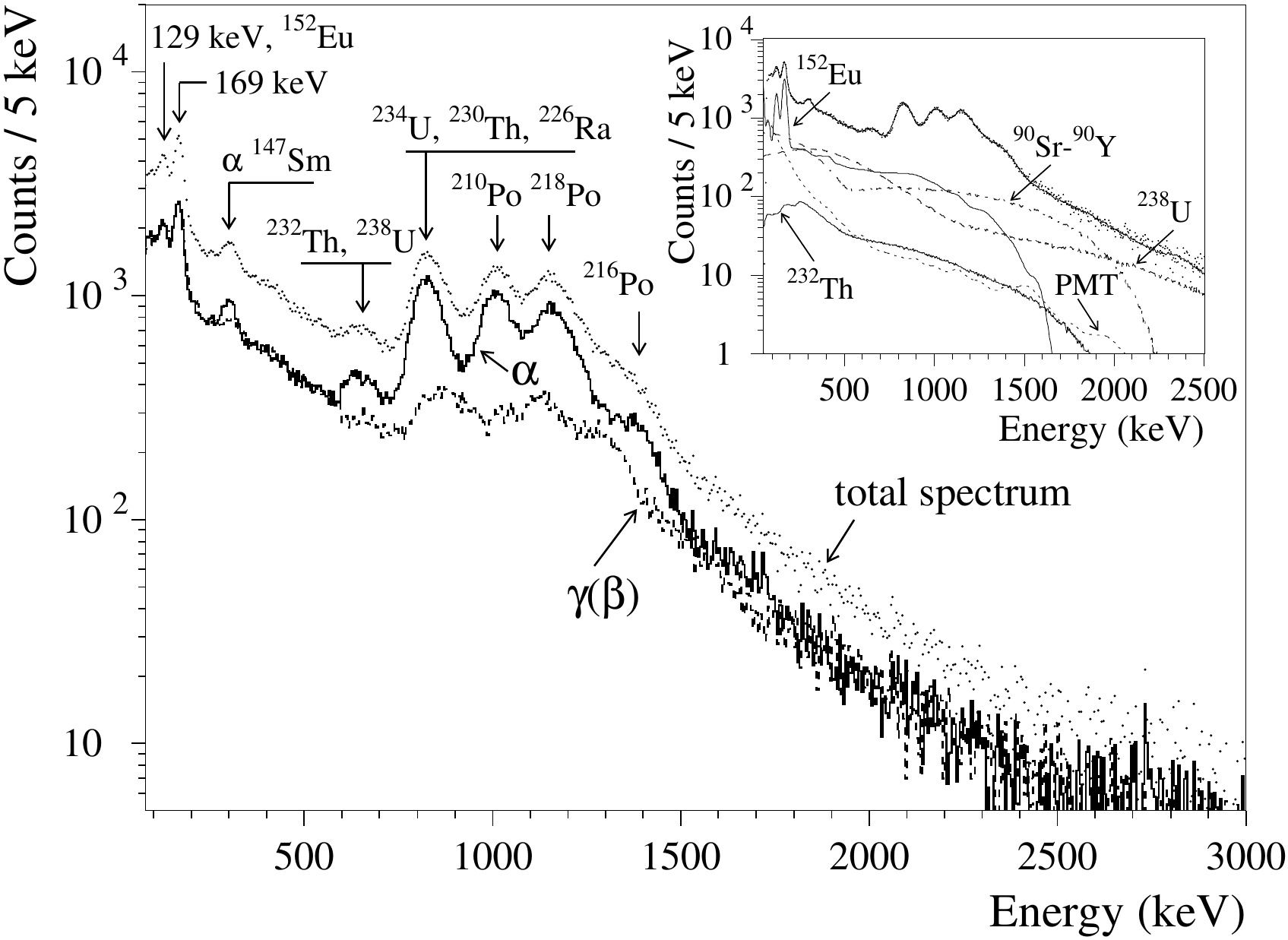}
  \includegraphics{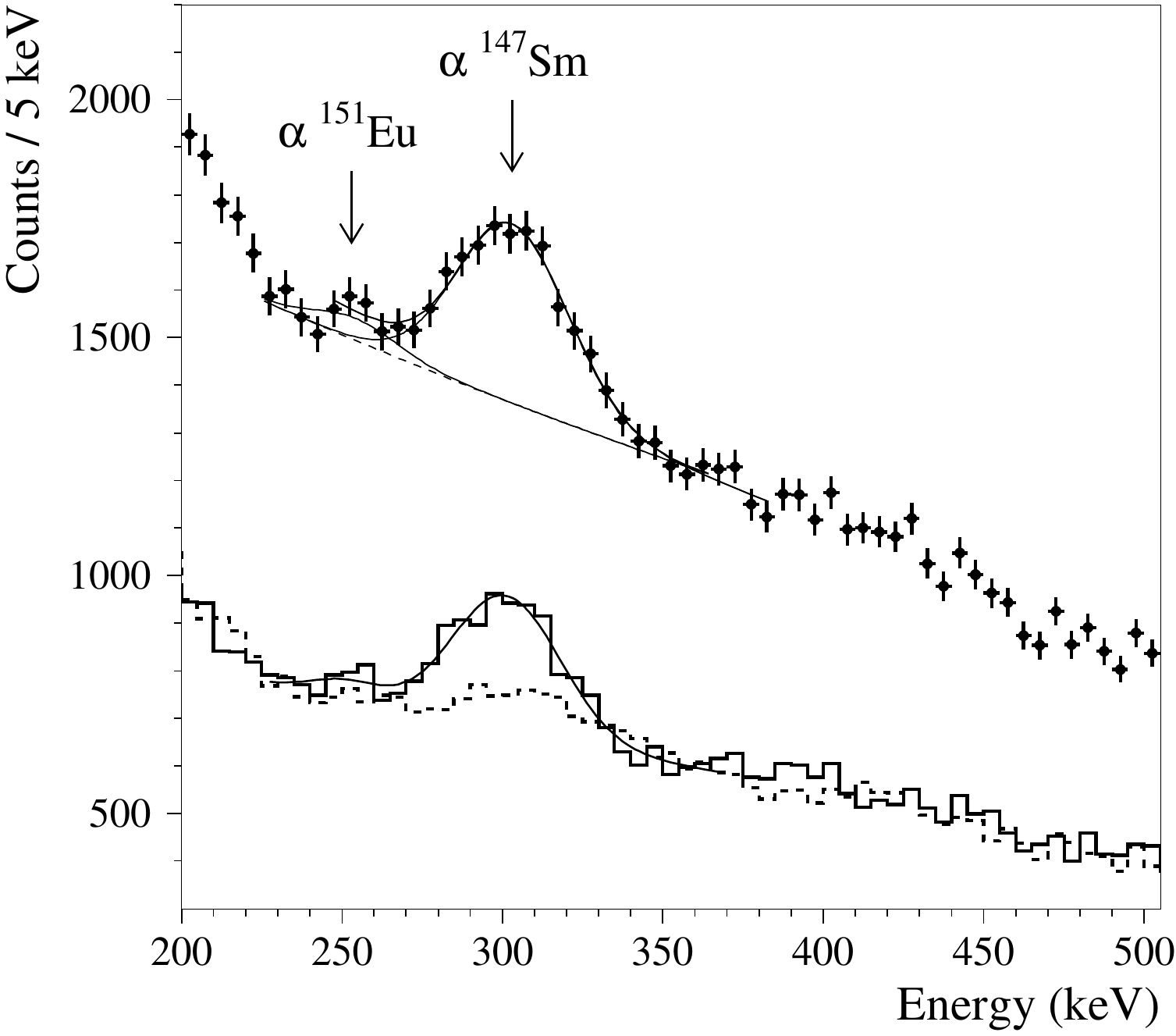}}
\caption{{\it Left:} energy distribution measured with the CaF$_{2}$(Eu) scintillator for 7426 h in the low background DAMA/R\&D set-up (points).
The energy distributions, obtained by applying the pulse-shape discrimination (PSD) technique, are shown by dashed line for $\gamma (\beta)$ 
component and solid line for $\alpha$ component.
(Inset) The fit of the total spectrum by simulated models in $90-2200$ keV energy interval is shown by solid line.
The most important components of $\gamma(\beta)$ background are also shown.
{\it Right:} low energy part of the energy distribution (points) and its $\gamma (\beta)$ (dashed line) and $\alpha$ (solid line) components obtained by applying the PSD.
The peculiarity on the left of the $^{147}$Sm peak can be attributed to the $\alpha$ decay of $^{151}$Eu with the half-life $T_{1/2}=5 \times 10^{18}$~yr
\cite{bel07}. Figure taken from Ref. \cite{bel07}.}
\label{figeu}
\end{figure*}

The first evidence \cite{bel07} was obtained with a low background \, europium-doped \, calcium fluoride, \, CaF$_2$(Eu), scintillator 
(370 g mass) optically coupled 
through 10 cm long TETRASIL-B light guide to low background PMT;
both the scintillation crystal and the light guide were wrapped by PTFE tape.
This detector was installed deep underground at LNGS in the low-background DAMA/R\&D set-up, where
an event-by-event data acquisition system recorded the amplitude and the arrival time of the events; 
in particular, the signals from the PMT were also recorded --  over a time window of 3125 ns --
by a 160~MSa/s Transient Digitizer.
The energy scale and the energy resolution of the CaF$_2$(Eu) detector for $\gamma$ quanta were measured with $^{22}$Na, 
$^{133}$Ba, $^{137}$Cs, $^{228}$Th and $^{241}$Am sources, while
the response to $\alpha$ particles was studied from 1 MeV to 5.25 MeV
by using a collimated $^{241}$Am $\alpha$ source and a set of absorbers.
The dependence of the $\alpha/\beta$ ratio (i.e. ratio of energy of $\alpha$ particle measured by the scintillator in $\gamma$ scale to its 
real energy) versus the energy of $\alpha$ particles was derived; in particular, $\alpha/\beta = 0.128(19)$
at the energy of $^{151}$Eu $\alpha$ particles was found, i.e. the expected energy of $^{151}$Eu $\alpha$ peak in $\gamma$ scale was 
245(36) keV. 

The precise concentration of Eu in the CaF$_2$(Eu) crystal was determined by Inductively Coupled Plasma - Mass Spectrometry analysis (ICP-MS,
Agilent Technologies mo\-del 7500a) to be: ($0.4\pm 0.1)\%$; the uncertainty also takes into account the 
uncertainty in the sample preparation procedure.

The data of the low background measurements with the CaF$_2$(Eu) detector were analyzed by several ap\-pro\-aches (time-amplitude analysis, pulse shape 
discrimination, removing double pulses, etc.); the presence of radioactive isotopes in the detector was studied as well. In particular,
to discriminate the events from $\alpha$ decays inside the crystal from the $\gamma (\beta)$ background, the optimal filter method was applied and the 
energy dependence of the shape indicators was measured.
The discrimination was rather low, however this procedure allowed to test the nature of the events in the energy distribution of Fig. \ref{figeu}{\it 
--left}. In the low energy part of the energy distribution measured in the low background set-up for 7426 h 
(shown in Fig. \ref{figeu}{\it --right}), a peculiarity at the energy near 250 keV is present 
which gives an indication on the existence of the process; furthermore,
the pulse-shape discrimination analysis has allowed to clarify the nature of events on the $\approx$250 keV peak. In fact,
as it can be seen on the bottom curve of Fig.~\ref{figeu}{\it --right}, this effect is present in the $\alpha$ component of the spectrum (solid line), 
while absent in the $\gamma(\beta)$ component (dashed line).
In this way, the half-life of $^{151}$Eu relatively to the $\alpha$ decay to the ground state of $^{147}$Pm was evaluated: 
$T_{1/2}^{\alpha}(g.s. \to g.s.)=(5_{-3}^{+11}) \times10^{18}$~yr.
Theoretical half-lives for $^{151}$Eu $\alpha$ decay calculated in different model frameworks are in the range of (0.3--3.6)$\times10^{18}$ yr;
in particular, the $^{151}$Eu half-life measured value was well in agreement with the calculations of Refs.~\cite{Poe83,Poe85}.

As mentioned above, several years later in Ref. \cite{cas14}, the $\alpha$ decay of $^{151}$Eu to the ground state of $^{147}$Pm was remeasured by using 
a 6.15 g Li$_6$Eu(BO$_3)_3$ crystal operated as a scintillating
bolometer. Germanium Neutron Transmutation Doped thermistors 
were used as temperature sensors coupled to the crystal and to the light detector (a high purity Ge wafer
44.5 mm diameter and 300 $\mu$m thickness);
the crystal was surrounded by a reflecting foil to maximize the light collection efficiency.

\begin{figure}[!ht]
\begin{center}
\resizebox{0.48\textwidth}{!}{ 
  \includegraphics{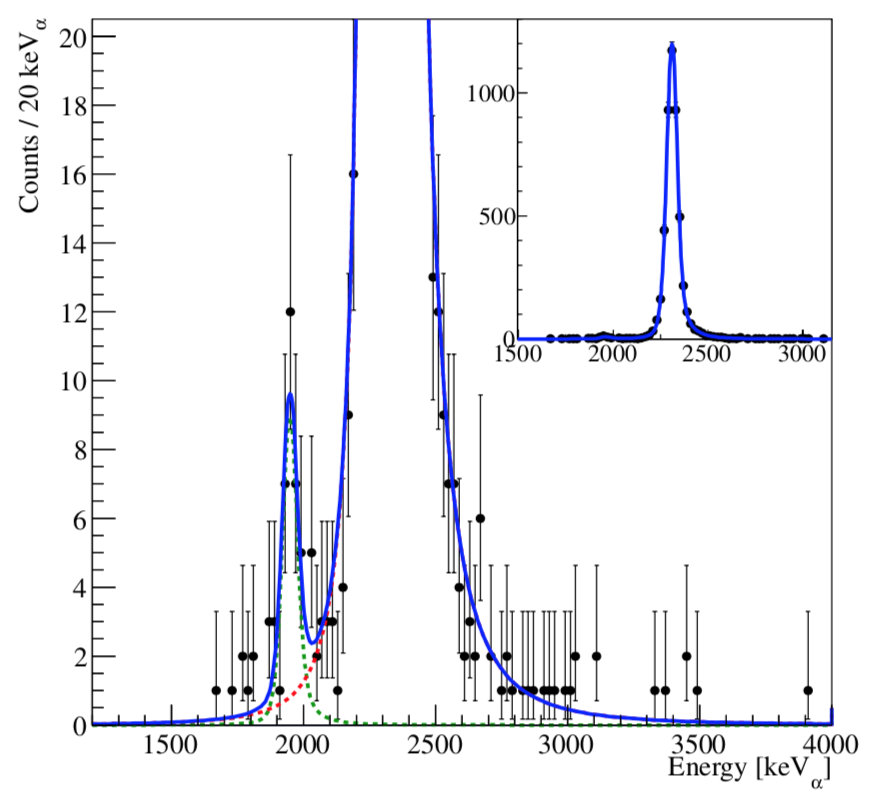}
}
\end{center}
\vspace{-0.2cm}
\caption{Zoom of the $\alpha$ energy spectrum around the region of interest. Two peaks are
visible: $^{147}$Sm at 2311 keV and one at 1949 keV, interpreted as $^{151}$Eu. In the inset 
the full y-scale plot is shown. The dashed lines represent the detector response function, while
the solid line is the sum of the response functions of the two peaks \cite{cas14}.
Figure taken from Ref. \cite{cas14}.}
\label{figeu_bol}
\end{figure}

The light vs bolometer signals were used for each event to significantly reduce the $\gamma / \beta$ 
events in the energy distribution; in Fig. \ref{figeu_bol} the best-fit $\alpha$ energy spectrum is shown from 1.2 MeV
to 4 MeV; the best-fit Q-value of the $^{151}$Eu
$\alpha$ transition is $[1948.9 \pm 6.9 (stat.) \pm 5.1 (syst.)]$ keV with 
an energy resolution (full width at half maximum, FWHM): (65$\pm$7) keV. The best-fit number of $^{151}$Eu events
corrected for the event selection efficiency was derived to be:
$(37.6 \pm 7.5)$ counts giving: 
$T_{1/2}^{\alpha}(g.s. \to g.s.)= [4.62 \pm 0.95(stat.) \pm 0.68(syst.)]) \times 10^{18}$ yr, which is in agreement with the first observation \cite{bel07} and predictions of nuclear theory using
the Coulomb and proximity potential model.

\subsubsection{$^{151}$Eu $\alpha$ decay to the first excited level of the $^{147}$Pm and limits on the $^{153}$Eu $\alpha$ decay }
\label{eu151other}

In addition to the decay of $^{151}$Eu to the $^{147}$Pm ground 
state, also the first excited level of $^{147}$Pm (5/2$^+$, $E_{exc}$=91
keV) can be populated\footnote{Higher $^{147}$Pm levels could be
populated too, but the probabilities for such $\alpha$ decays are much  
lower because of the exponential dependence 
on the energy release.}; its corresponding $Q_\alpha$ value is 1.873 MeV. 

Still in Ref. \cite{bel07} a study on such a process was performed. In particular, 
taking into account the $\alpha/\beta$ ratio, the energy of
$\alpha$ particles, which is 1.823 MeV, 
corresponds in the used CaF$_2$(Eu) detector to 234 keV
in the $\gamma$ scale. Then, in a subsequent
deexcitation process of the excited $^{147}$Pm level, some
particles will be emitted: either single gamma quantum ($E_\gamma=91$
keV) with a probability of 33\%, or conversion electron plus 
cascade
of X rays and Auger electrons with a probability of 67\% (coefficient
of conversion for this transition is equal to 2.06 
\cite{Derm92}).
Since all these low-energy particles are effectively absorbed
inside the crystal, the full energy observed by the CaF$_2$(Eu)
detector was equal to ($325 \pm 33$) keV. Such a peak, which 
should be on the right slope of the $^{147}$Sm peak,
is not evident in the experimental data (see Fig.~\ref{figeu}{\it --right}).
Fitting the background spectrum with a model consisting of two peaks 
($^{147}$Sm and the peak searched for) plus an exponential function, the
obtained upper limit on the $\alpha$ decay of $^{151}$Eu to the 
first excited level of $^{147}$Pm was determined as:
$T_{1/2}^{\alpha}(g.s. \to 5/2^+) \geq 6\times 10^{17}$~yr at 
68\% C.L.
This limit was one order of magnitude higher than the one reported in
Ref. \cite{Bell07} where 91 keV $\gamma$ rays
were searched for from a sample of Li$_6$Eu(BO$_3)_3$ crystal
measured with a ultra-low
background HPGe $\gamma$ spectrometer deep underground.

Some years later, in Ref. \cite{dan12}, the $\alpha$ decay of $^{151}$Eu to the first excited level of 
$^{147}$Pm was 
searched for at the HADES underground laboratory (about 500 m w.e.). A sample of high-purity europium oxide 
of natural isotopic composition and 303 g 
mass was measured over 2232.8 h with a high-energy-resolution ultra-low 
background HPGe detector (40 cm$^{3}$) with sub-micron dead-layer. The half-life limit
reached in this way 
was $T_{1/2}^{\alpha}(g.s. \to 5/2^+) \geq 3.7\times 10^{18}$~yr at 68\% C.L., approaching the theoretical predictions.

\begin{figure}[ht]
\begin{center}
\resizebox{0.48\textwidth}{!}{ 
  \includegraphics{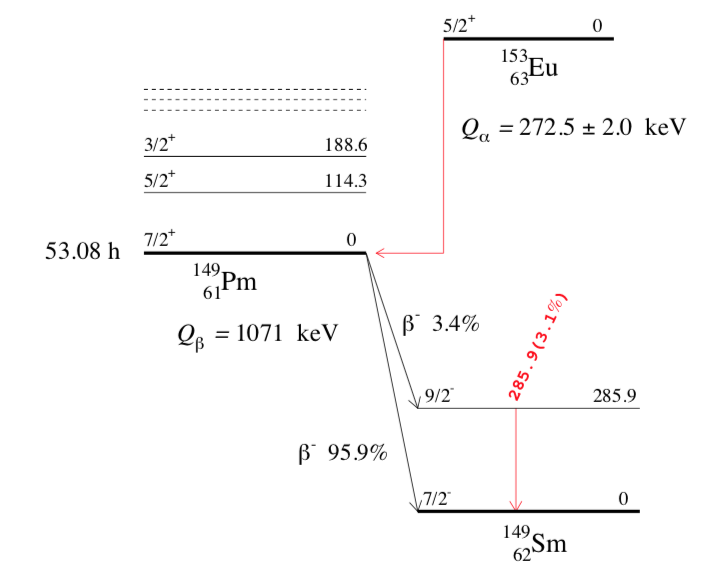}
}
\end{center}
\vspace{-0.2cm}
\caption{Simplified scheme of the $\alpha$ decay of $^{153}$Eu. The $\gamma$ ray produced by the transition following the
$\beta$ decay of $^{149}$Pm is highlighted. Figure taken from Ref. \cite{dan12}.}
\label{fig:153Eu_a}
\end{figure}

\begin{figure}[ht]
\begin{center}
\resizebox{0.48\textwidth}{!}{ 
  \includegraphics{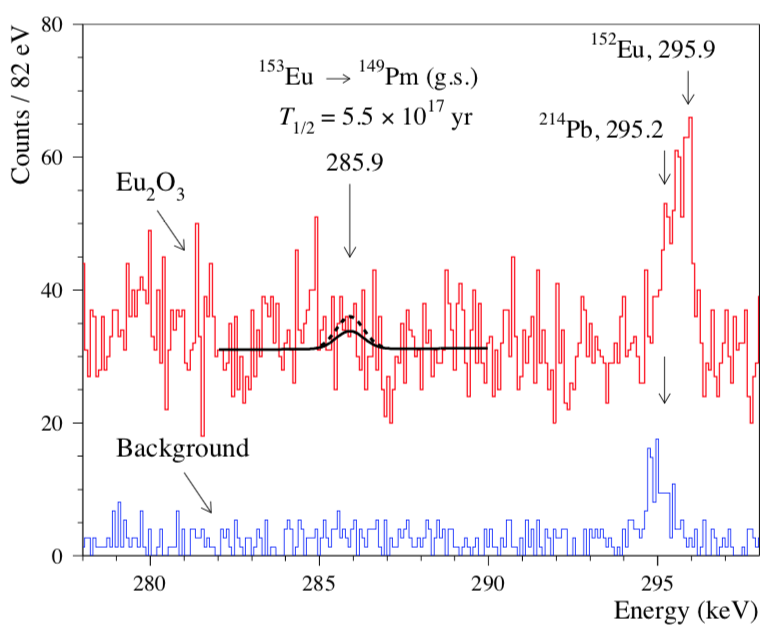}
}
\end{center}
\vspace{-0.2cm}
\caption{The energy spectrum measured in the experiment of Ref. \cite{dan12}. The fit of the spectrum in the energy interval 
282-290 keV is shown by solid line. The expected $\gamma$ peak at 285.9 keV from $\beta$ decay of $^{149}$Pm, see Fig. \ref{fig:153Eu_a},
is shown by the dashed line. The area of the peak corresponds to the half-life of $^{153}$Eu $T_{1/2} = 5.5 \times 10^{17}$ yr 
excluded at 68\% C.L. The background spectrum is normalized to the time of measurements with the sample. Figure taken from Ref. \cite{dan12}.}
\label{fig:153Eu_b}
\end{figure}

For completeness, we also remind that in the latter work from the same data a new half-life limit for decay of $^{153}$Eu was also determined
by studying the $\gamma$ radiations associated to the $\beta$ decay of the daughter nucleus, $^{149}$Pm, as shown in Fig. \ref{fig:153Eu_a}.
The energy spectrum measured in the experiment over 2232.8 h with the Eu$_2$O$_3$ sample in the energy interval 
278--298 keV is shown in Fig. \ref{fig:153Eu_b}; there is no evidence for the peak and 
new half-life limit for the decay of $^{153}$Eu was set to $\geq 5.5 \times 10^{17}$~yr at 68\% C.L.

\subsection{The $\alpha$ decay of samarium isotopes}
\label{samarium}

Samarium has seven naturally occurring isotopes (see Table \ref{tab:alpha}).
The natural Sm contains three radioactive isotopes: $^{147}$Sm (abundance 15.00\%), $^{148}$Sm (abundance 11.25\%) and $^{149}$Sm (abundance 13.82\%), which have long half-lives of 1.06(1) $\times$ 10$^{11}$ yr, 
7(3) $\times$ 10$^{15}$ yr, and $>$ 2 $\times$ 10$^{15}$  yr, respectively.
Potentially, other two isotopes can decay by emitting $\alpha$ particles: $^{150}$Sm and $^{152}$Sm.

For completeness let us mention that
the $\alpha$ decay from various excited states of samarium isotopes were investigated by thermal 
neutrons and resonance neutrons \cite{Kvi66,Oak67,Kvi70,Pop72}. While the $\alpha$ decay from the ground states have long life-times,
the decay from highly excited states is fast. The use of resonance neutrons allows to excite a number of discrete levels, but it is complicated 
because of the very small cross sections involved and the large $\gamma$-ray yield in competitive reactions, in fact the 
ratio $\sigma(n,\alpha)/\sigma(n,\gamma)$ is in the range of 10$^{-5}$--10$^{-9}$. The measurements regard the $\alpha$ widths, and the values measured
correspond to life-times of few ns ($\Gamma\times\tau$ = h/2$\pi$).

\subsubsection{$^{147}$Sm $\alpha$ decay}
\label{sm147}

To precisely know the half-life of {$^{147}$Sm is important because this isotope -- due to its long half-life -- is considered as a cosmochronometer; 
the dating method is based on the relative abundance of 
$^{143}$Nd to $^{144}$Nd \cite{Fau09}, with the first isotope deriving from the $\alpha$ decay of $^{147}$Sm. 
The $^{147}$Sm isotope is therefore suitable for dating ancient ores and meteorites.

Since the pioneering studies in 1933 by Hevesy et al. \cite{Hev33} and in 1934 by Curie and Joliot \cite{Cur34}, many  measurements of $\alpha$
radioactivity of  $^{147}$Sm have been performed (more than 30 measurements are available). 
Recently, a last high precision investigation of the $\alpha$ decay of the
$^{147}$Sm isotope ($J^{\pi}$= 7/2$^-$) 
into the ground and first excited states of $^{143}$Nd was performed by using $\alpha$ spectrometry and $\gamma$ spectroscopy \cite{wil17}. 
To measure the $\alpha$ activity a low-background twin Frisch-grid ionization chamber (TF-GIC) was used \cite{Har16}.
When no source is present the obtained background in the energy region (1 -- 9) MeV was of $(10.9 \pm 0.6)$ counts per day;
the main contribution was from the daughters of $^{222}$Rn ($^{218}$Po, $^{214}$Po and $^{210}$Po). There were also small contributions from $^{220}$Rn, 
coming from the  $^{232}$Th decay chain ($^{220}$Rn, $^{216}$Po and $^{212}$Po).
The background activity fell to ($8.3 \pm 0.7$) counts per day in the same energy region after the first 12 days of operation due to the radon daughters decay.

The measured $^{147}$Sm peak shows the expected maximum at 2.28 MeV. All events above 2.5 MeV were from Th and U contaminations, the majority of which are above 5 MeV. The energy resolution of the peak was $\sigma_E$ = 16.76(88) keV.

The combined half-life value of $^{147}$Sm on the four 
measured samples is
$T_{1/2}$ = (1.0787 $\pm$ 0.0095(stat.) $\pm$ 0.0244(syst.)) $\times$ 10$^{11}$ yr.
The quoted systematic uncertainty is the combination of the uncertainty of the detector efficiency and the uncertainty on the isotopic abundance. The systematic uncertainty is dominated by the precision on the activity of the calibration source which is 2.2\%. 
In Fig.~\ref{Sm1} the half-life values  \cite{Nic09} plotted as a function of year are shown as reported in Ref. \cite{wil17}.

\begin{figure}[ht]
\begin{center}
\resizebox{0.48\textwidth}{!}{ 
  \includegraphics{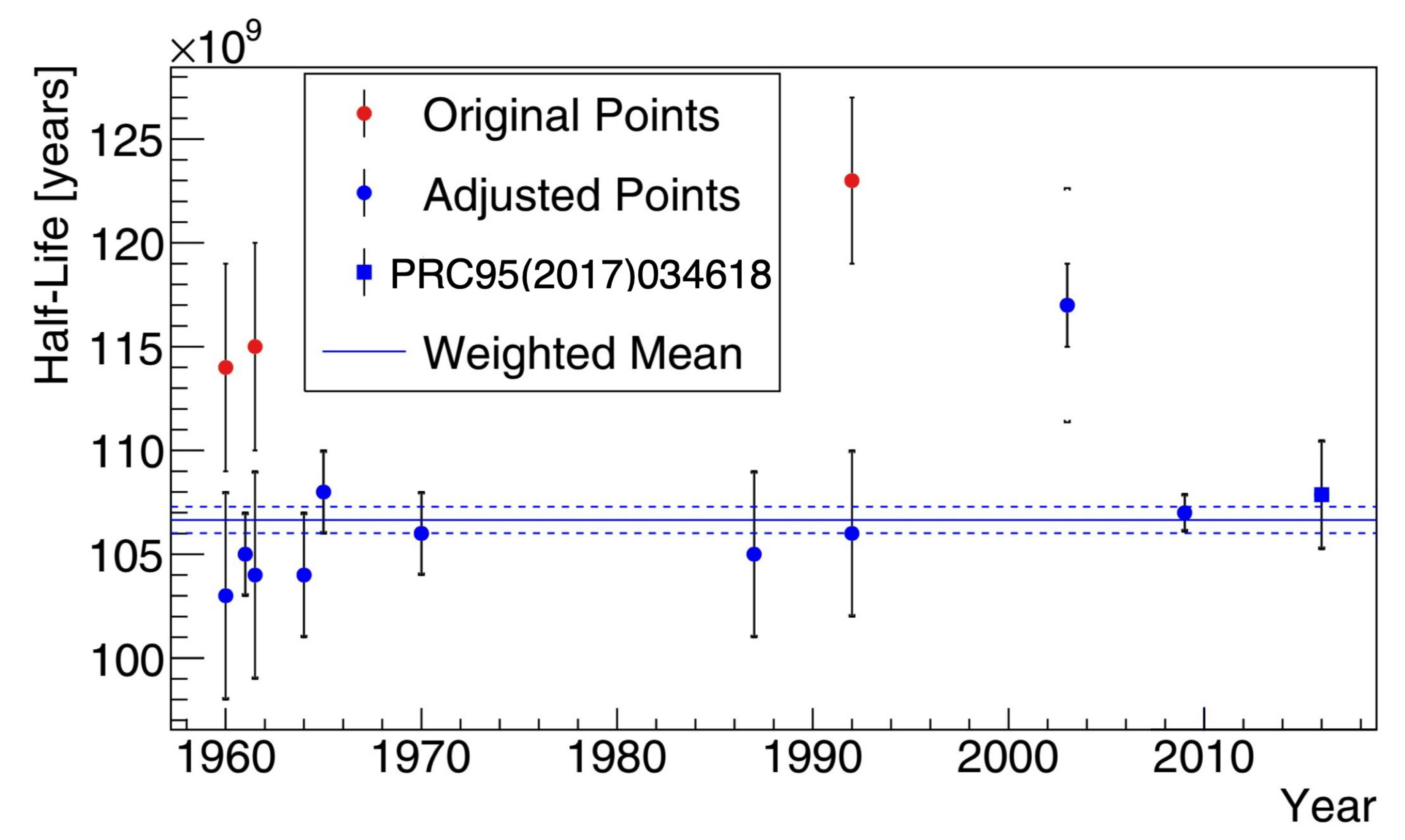}
}
\end{center}
\caption{The half-life values of $^{147}$Sm \cite{Nic09} plotted as a function of year are shown as reported in Ref. \cite{wil17}. The red circular points are the original published results. The blue circular points are the adjusted half-life values, by the method described 
in Ref. \cite{Nic09}. The weighted mean (uncertainties in squared braces) is calculated in Ref. \cite{wil17} using the method described in Ref. \cite{Raj92} with the adjusted half-life values.}
\label{Sm1}
\end{figure}

For a historical report on half-life measurements of $^{147}$Sm see also Ref. \cite{Nic09}.

An updated critical review and data analysis of the whole set of half-life-values obtained over time is available in Ref. \cite{Tav18}. 
In that paper the recommended value for the $\alpha$-decay half-life of $^{147}$Sm isotope is (1.063 $\pm$ 0.005) $\times$ 10$^{11}$ yr.
In particular the authors applied also a one-parameter, semi-empirical model for $\alpha$  emission from nuclei, developed in the framework of the quantum mechanical tunneling mechanism through a Coulomb-plus-centrifugal-plus-overlapping potential barrier, and evaluated a value of (1.082 $\pm$ 0.030) $\times$ 10$^{11}$ yr.

There is a finite probability of $\alpha$ decay into the excited state of $^{143}$Nd; this leads to a de-excitation through the emission of $\gamma$ radiation. The excited state transitions have been studied \, using low background $\gamma$
 spe\-ctro\-scopy \cite{wil17} and data were published only relatively to the first excited state $E_1 = 742.05(4)$ keV \cite{Bro12}
 (because of the lower $Q_\alpha$ values and potential total angular momentum changes). The samples of Sm$_2$O$_3$ were measured in a ultra-low 
background HPGe \cite{Koh09}. 
The measured spectrum shows few prominent peaks. These peaks are due to contamination from natural radioactive sources like Th and U decay chains as well as $^{40}$K. The peak at 2614.5 keV corresponds to the decay of $^{208}$Tl.
 The positron annihilation line from muon interactions is visible at 511 keV. The presence of natural
radioactivity makes a Compton continuum in the region of interest of
the expected signal.
The $\gamma$ spectrum measured in Ref. \cite{wil17} is shown in Fig.~\ref{Sm2}.

\begin{figure}[ht]
\begin{center}
\resizebox{0.48\textwidth}{!}{ 
  \includegraphics{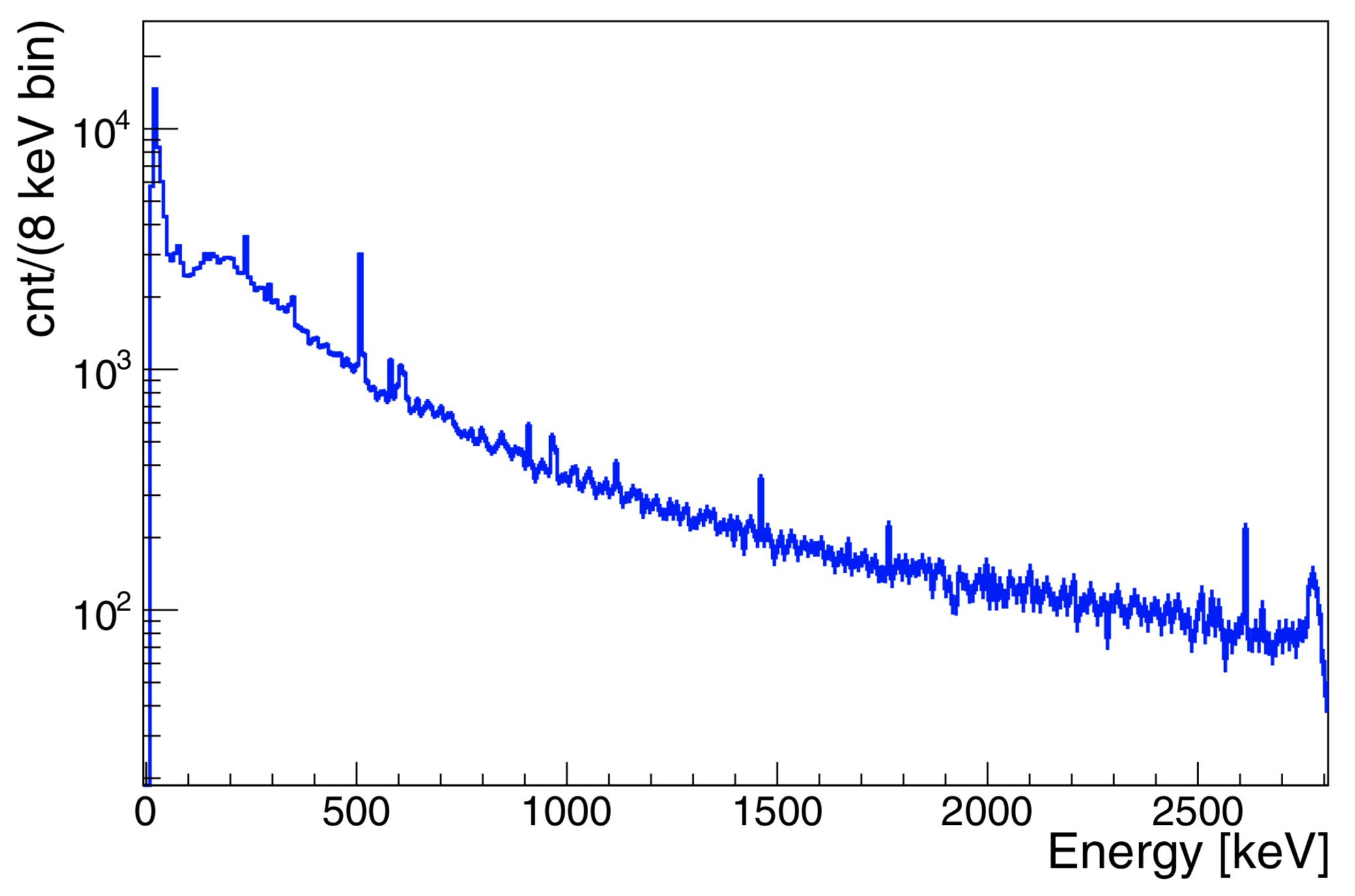}
}
\end{center}
\caption{The $\gamma$ spectrum from 50.74 g of Sm$_2$O$_3$, with a combined runtime of 62.91 d \cite{wil17}. Figure taken from Ref. \cite{wil17}.}
\label{Sm2}
\end{figure}

The region of interest (ROI) around 742 keV is shown in Fig.~\ref{Sm3}.

\begin{figure}[ht]
\begin{center}
\resizebox{0.48\textwidth}{!}{ 
  \includegraphics{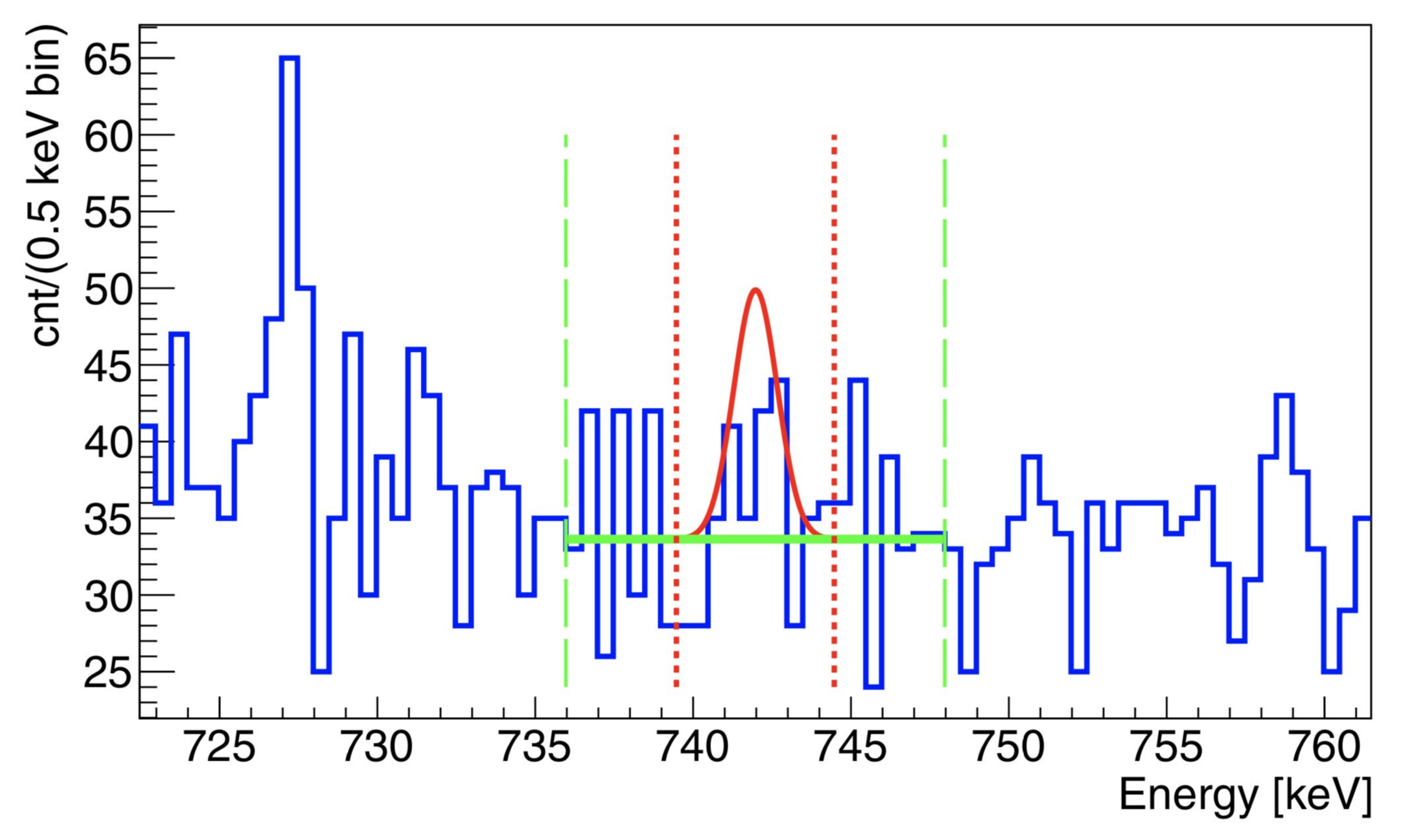}
}
\end{center}
\caption{The ROI (between the two red vertical lines) in the $\gamma$ spectrum for a potential $\alpha$ decay of $^{147}$Sm 
into the first excited state of $^{143}$Nd at 741.98(4) keV \cite{wil17}.  The line at 727.2 keV is most 
likely from $^{212}$Bi decay from the thorium decay chain. The efficiency at 742 keV was estimated to be 5.5\%. Figure taken from Ref. \cite{wil17}.}
\label{Sm3}
\end{figure}

The fit gave a background estimate in the ROI of 336(11) counts, while the measured number in the ROI was 352 counts.
The analysis resulted in a lower half-life limit of $T_{1/2} >  3.1 \times 10^{18}$ yr (at 90\% C.L.).
The theoretical prediction for this decay mode, taken from the Viola-Seaborg relation between the $Q_\alpha$ value and half-life
\cite{sah16}, gives an expected value for the half-life of the order of 10$^{24}$ yr. This is considered a rough estimate 
as the change in total angular momentum is not taken into account, but would modify the expected half-life even longer. 
The small $Q_\alpha$ value is also expected to have an important impact on the half-life. To reach the predicted half-life 
region lower background and larger exposure (mass, time) will be needed.

\subsubsection{$^{148}$Sm $\alpha$ decay}
\label{sm148}

Some measurements to study the alpha decay of $^{148}$Sm were performed in the past 
by two groups, and the half-lives were measured with quite large uncertainties: ($8\pm2)\times10^{15}$ yr \cite{kor68} and ($7\pm3)\times10^{15}$ yr \cite{gup70} due to
a rather small signal-to-background ratio.

Recently, the half-life of the $\alpha$ decay of $^{148}$Sm was measured at LNGS with improved accuracy by using a ZnWO$_4$ scintillating bolometer doped with
Sm, isotopically enriched in $^{148}$Sm to 95.54\% \cite{cas16}.
A ZnWO$_4$:$^{148}$Sm cylindrical crystal with a mass of 22.014 g was used as low temperature scintillating bolometer.
Defining the light yield as the ratio of the energy measured in the light detector 
 to the energy measured in the ZnWO$_4$:$^{148}$Sm crystal, it results that alphas have a low light yield and can be discriminated from the beta/gamma events.
The measured activity of (745.4 $\pm$ 0.5) mBq/kg for $^{147}$Sm was deduced from the peak at 2310.5 keV. The smaller peak just below 2 MeV was identified as related to the $^{148}$Sm alpha decay and was found consistent with the expected decay energy.
In order to evaluate the $^{148}$Sm $Q-$value and its half-life, a simultaneous unbinned and extended likelihood fit to the alpha energy spectrum was
done by using the response function.

The best fit value for the energy transition of the $^{148}$Sm alpha
decay was (1987.3 $\pm$ 0.5) keV, in agreement with the recommended value at
(1986.8$ \pm$ 0.4) keV \cite{Wang:2017}. The computed
 energy resolution for the $^{148}$Sm peak was (4.7 $\pm$ 0.5) keV (FWHM). Finally,
taking into account the ratio of the atomic concentration of $^{148}$Sm/$^{147}$Sm and the number of events in the two peaks, the obtained half-life was $T_{1/2} = (6.4^{+1.2}_{-1.3}) \times 10^{15}$ yr \cite{cas16}. 

The quoted uncertainties include both the statistical and the systematic contributions derived from the fit. 
The measured half-life is in agreement with the previously reported values  \cite{kor68,gup70}. Furthermore, it is in agreement with evaluation based on the cluster model \cite{Buc91,Buc92} and the semi-empirical formulae in Ref. \cite{Roy00}, giving 5.6$\times$10$^{15}$ yr and 8.0 $\times$10$^{15}$ yr, respectively.

The method has the potential to improve the sensitivity by three orders 
of magnitude, if the isotope mass is increased to few grams (it is less than 2 mg in this measurement).

\subsubsection{The $\alpha$ decay of other Sm isotopes}

The $^{149}$Sm is predicted to decay, but no decays have ever been observed. It is considered to undergo $\alpha$ decay to  $^{145}$Nd with a half-life 
over 2$\times 10^{15}$ yr, as reported in Ref. \cite{gup70}.

Recently a study on the extremely long-lived $\alpha$-decaying nuclei within a generalized density-dependent cluster model involving the experimental nuclear charge radii was performed \cite{qia14}
and calculations within the UMADAC model \cite{kin11,den09} and the analytic formulas given by Royer \cite{Roy00} are available. These three calculations give for the half-life of $^{149}$Sm:
5.8 $\times 10^{18}$ yr, 3.1$\times 10^{19}$ yr, 6.5 $\times 10^{18}$ yr, respectively (see also Table \ref{tab:alpha}).

\subsection{$^{144}$Nd and $^{146}$Nd $\alpha$ decay}
\label{nd146}

Neodymium has seven isotopes present in nature; five of them can
in principle be unstable undergoing an $\alpha$ decay. Some isotopes ($^{146}$Nd, $^{148}$Nd, and $^{150}$Nd)  
can undergo $2\beta^-$ decay; in particular, the $2\nu2\beta^-$ decay of $^{150}$Nd 
to the ground state and to the first excited level of $^{150}$Sm was measured in several direct experiments
in the ranges: 
$T_{1/2} (g.s. \rightarrow g.s.)          = (0.7 - 1.9) \times 10^{19}$ yr, and
$T_{1/2} (g.s. \rightarrow 0^+,~740.5$ keV$) = (0.47 - 1.4) \times 10^{20}$ yr, respectively \cite{Barabash:2018}.
Only a limit of $T_{1/2} > 3 \times 10^{18}$ yr is reported for the $2\beta^-$ decay of $^{148}$Nd \cite{bellotti:1982} (g.s. to g.s. transition).

The isotopes $^{143}$Nd, $^{145}$Nd, and $^{148}$Nd have either positive low $Q_\alpha$ values:
$(520.7 \pm 7.4)$ keV, $(1576.0 \pm 2.1)$ keV, and $(599.0 \pm 3.6)$ keV, respectively,
or low natural abundances: 12.173(26)\%, 8.293(12)\%, and 5.756(21)\%, respectively \cite{Meija:2016}; thus, they are classified as ``stable'' nuclei.
Some attempts to measure their $\alpha$ decay by a gridded ionization chamber have been done for the $^{145}$Nd isotope,
resulting in a limit on the half-life $ > 6 \times 10^{16}$ yr \cite{nd144_04}.

The half-life of the $\alpha$ decay of the $^{144}$Nd isotope to $^{140}$Ce has been adopted in Ref. \cite{Sonz:2001} to be:
$(2.29 \pm 0.16) \times 10^{15}$ yr; the $Q_\alpha$ value is $(1903.2 \pm 1.6)$ keV \cite{Wang:2017}.
This half-life is the unweighted average of the measured half-lives: 
$(2.65 \pm 0.37) \times 10^{15}$ yr \cite{nd144_01},
$(2.4  \pm 0.3)  \times 10^{15}$ yr \cite{Macfarlane:1961},
$2.2 \times 10^{15}$ yr \cite{Porschen:1956}, and
$1.9 \times 10^{15}$ yr \cite{nd144_04}. Three of them have been obtained in the decade 1950-1960
using either a large cylindrical ionization counter accommodating samples \cite{Macfarlane:1961},
or nuclear emulsion technique \cite{Porschen:1956}, or gridded ionization chambers \cite{nd144_04}.
A relatively more recent measurement was done in 1987 by using a cylindrical proportional counter
with a cathode source area of 1550 cm$^2$ \cite{nd144_01}.
The counting volume was enclosed by a brass cylinder, 10 cm in diameter and 63 cm long. 
The gold-plated tungsten anode wire had 50 $\mu$m diameter. The source material was deposited
on the inside of a thin stainless steel sheet with an area of approximately 1550 cm$^2$.
The used gas was a mixture of 90\% of Ar and 10\% of CH$_4$.  
The pressure was regulated at a 
few percent above atmospheric pressure. 
The spectrum obtained with a source of isotopically enriched Nd$_2$O$_3$ in $^{144}$Nd at 97.5\%
is shown in Fig. \ref{fg:144nd}.
The $^{144}$Nd $\alpha$ peak at $E_\alpha = 1.8$ MeV is clearly visible, together with a continuous 
background extending to about 8.8 MeV. This background 
is due to $^{232}$Th and its daughters, with its
last member $^{212}$Po contributing to the energy spectrum between 6.8 and 8.8 MeV. 
The obtained half-life is: $T_{1/2} = (2.65 \pm 0.37) \times 10^{15}$ yr \cite{nd144_01}.

\begin{figure}[ht]
\begin{center}
\resizebox{0.5\textwidth}{!}{ 
  \includegraphics{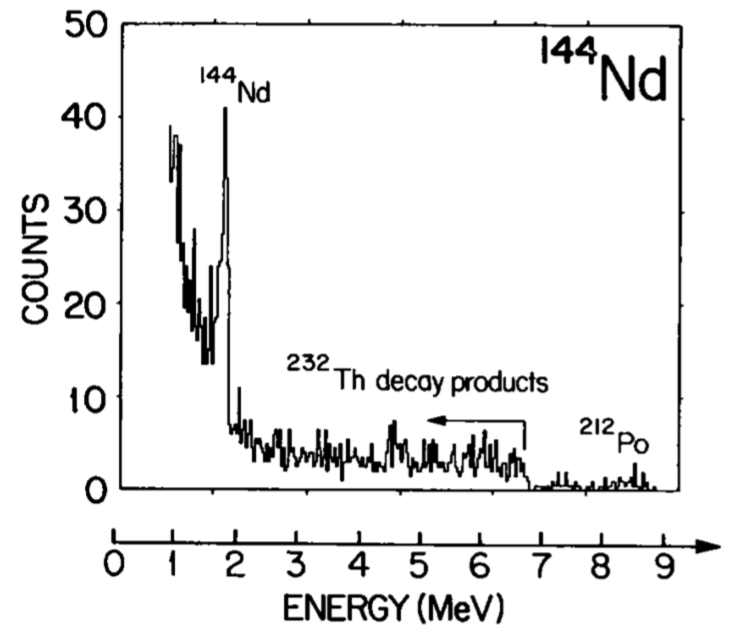}
}
\end{center}
\caption{Energy spectrum measured for 100 hours of data taking by a cylindrical proportional counter
with a cathode source containing $^{144}$Nd enriched at 97.5\% \cite{nd144_01}. The background above
2 MeV is due to the $^{232}$Th and its daughters. Figure taken from Ref. \cite{nd144_01}.}
\label{fg:144nd}
\end{figure}

The nuclide $^{146}$Nd is considered to be stable, however, according to the atomic mass 
evaluation \cite{Wang:2017} it has a positive $Q_\alpha$ value: $(1182.4 \pm 2.2)$ keV with respect to $\alpha$ decay. 
The decay to the ground state is expected to be accompanied by $\alpha$ decay into the first excited $2^+_1$ state 
of $^{142}$Ce. This state would de-excite via the emission of 641.282(9) keV $\gamma$ quanta.
The first search ever for its $\alpha$ decay into the first excited state 
of $^{142}$Ce by the detection of the 641.282 keV $\gamma$ quanta
was performed by using a low background HPGe detector and a sample of neodymium,
enriched in $^{146}$Nd at 97.20(1)\% \cite{Stengl:2015}. 
The measurements were done at the Felsenkeller Underground Lab of VKTA Rossendorf \cite{Stengl:2015}.

The measured spectrum of the HPGe with the enriched Nd sample 
around the energy region of interest after 144 hours of data taking is reported in Fig. \ref{fg:146nd}.
The most prominent line in this region is the 609.3 keV line from $^{214}$Bi from the natural 
decay chain of $^{238}$U, while no significant signal is present at the energy of the expected transition. 
Hence, the region has been fitted with a flat background and a Gaussian peak
and the lower limit on the half-life of this decay mode has been determined to be 
$T_{1/2} (0^+ \rightarrow 2^+_1) > 1.6 \times 10^{18}$ yr at 90\% CL \cite{Stengl:2015}.

\begin{figure}[ht]
\begin{center}
\resizebox{0.5\textwidth}{!}{ 
  \includegraphics{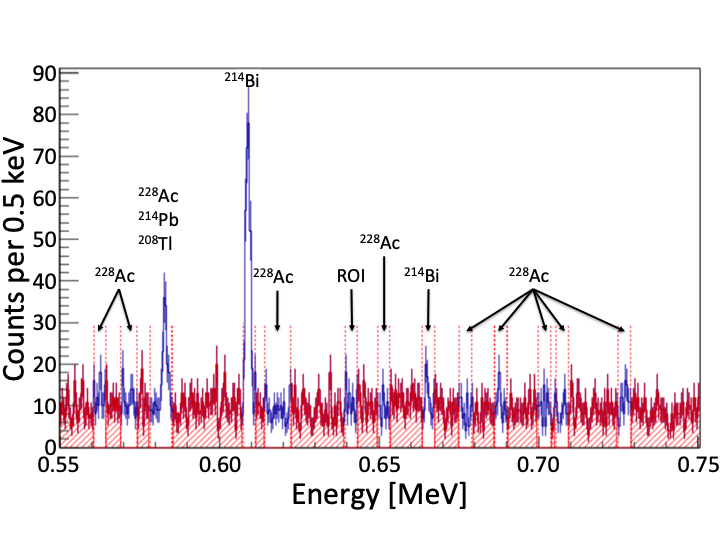}
}
\end{center}
\caption{Energy spectrum, measured by HPGe with the Nd sample enriched in $^{146}$Nd at 97.20(1)\%
around the expected gamma energy of the first excited state transition. No peak is seen in the ROI. 
The identified peaks (blue) have been subtracted before fitting the background. Figure taken from Ref. \cite{Stengl:2015}.}
\label{fg:146nd}
\end{figure}

The theoretical estimates of the half-life for the $\alpha$ decay of $^{146}$Nd to the ground state of $^{142}$Ce 
are in the region of about $10^{31 - 36}$ years (see also Table \ref{tab:alpha}), and even longer for the transition to the first excited state \cite{Stengl:2015},
thus far from the experimental sensitivities.

\subsection{The $\alpha$ decay of gadolinium and hafnium isotopes}
\label{gadolinium}

All six naturally abundant isotopes of Hf and the three isotopes of Gd are unstable (at least, potentially) with respect to $\alpha$ decay (see Table \ref{tab:alpha}). 
It is very peculiar that no measurements of the half-lives of the
$\alpha$ decay of gadolinium and hafnium isotopes are reported after 1961 (in their ground states; see Section \ref{hf178m2} on $^{178m2}$Hf).
In particular, the $\alpha$ decay of two nuclides, $^{152}$Gd and $^{174}$Hf, have been firstly
observed in 1959 by using nuclear emulsions \cite{Riezler:1959} and 
remeasured with higher accuracy in 1961 \cite{Macfarlane:1961}.
In the latter measurement a cylindrical ionization counter was used, and the samples -- enriched 
in the isotopes of interest -- were deposited uniformly over a 
substrate of metallic sheets which fitted concentrically into the counter. The total active area was 1200 cm$^2$
and the typical thickness of the samples was between 10 and 100 $\mu$g/cm$^2$.

The energy spectrum collected after 109.3 hours using the enriched gadolinium oxide sample
is reported in Fig. \ref{fig:gd_hf}. The peak of the $\alpha$ particles due to the $^{152}$Gd decay is well 
evident and corresponds to the half-life of 
$T_{1/2} = (1.08 \pm 0.08) \times 10^{14}$ yr \cite{Macfarlane:1961}.
The authors verified, by repeating the measurement adding samarium oxide
to the sample, that there was no contamination in the peak due to $\alpha$ particles 
from samarium decay \cite{Macfarlane:1961}.
Fig. \ref{fig:gd_hf} also shows the energy spectrum collected in 236 hours with a sample of 100 $\mu$g/cm$^2$
containing the enriched hafnium oxide sample and a weak reference source of $^{210}$Po.
The measured half-life of $^{174}$Hf was $T_{1/2} = (2.0 \pm 0.4) \times 10^{15}$ yr \cite{Macfarlane:1961}.

\begin{figure}[ht]
\begin{center}
\resizebox{0.42\textwidth}{!}{ 
  \includegraphics{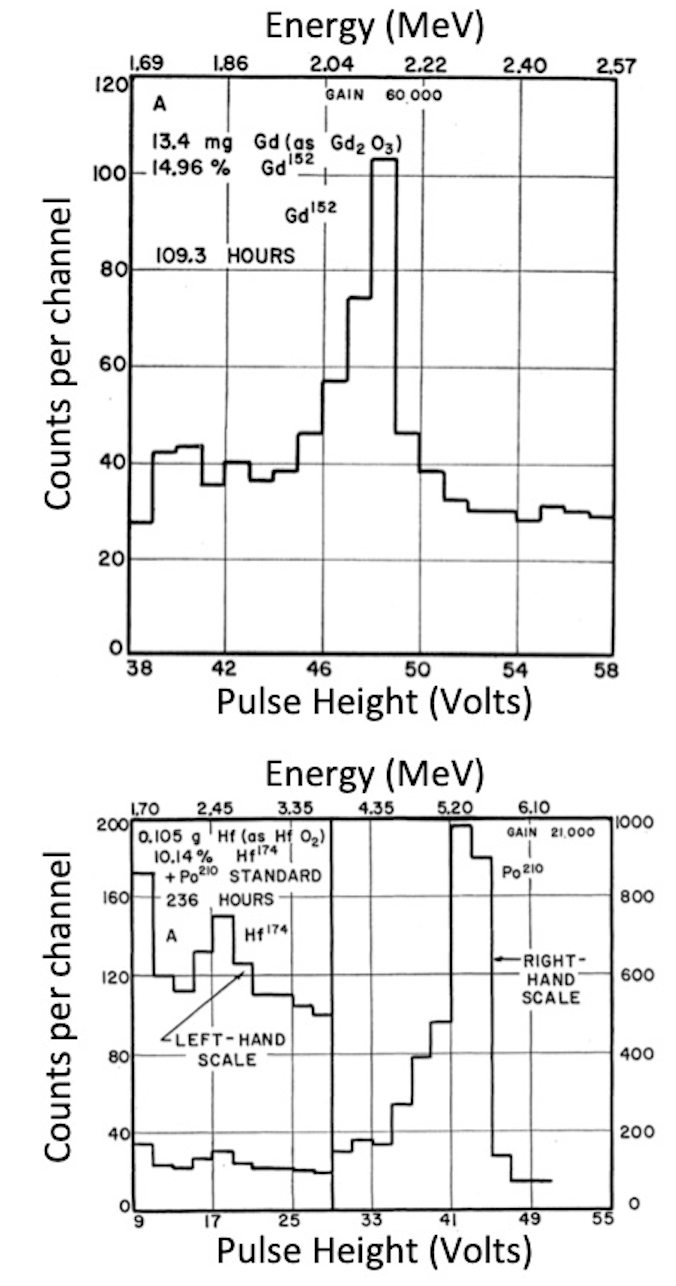}
}
\end{center}
\caption{Energy spectra collected with the enriched samples of $^{152}$Gd (Top) and $^{174}$Hf (Bottom) 
by the large cylindrical ionization counter of  
Ref. \cite{Macfarlane:1961}. The peaks due to the $\alpha$ particles from the decay of $^{152}$Gd, $^{174}$Hf and $^{210}$Po, 
used as reference, are evident. Figures taken from Ref. \cite{Macfarlane:1961}.}
\label{fig:gd_hf}
\end{figure}

The half-lives measured in this experiment are still the reference ones, as reported in
the table values \cite{aud17}.
It is our opinion that a more precise determination of them is recommended by exploiting the present technologies 
and facilities.
No measurements of the $\alpha$ decay of the other naturally occurring isotopes of gadolinium and hafnium
have been performed until now at our knowledge.

\section{Search for very long living $\alpha$ decays}
\label{long_alpha}

The $\alpha$ decay of some isotopes shown in Table \ref{tab:alpha} has neither been
observed, nor any attempts to get lower limit on the half-life have been performed, to our knowledge.
The theoretical estimations of their half-life is far from any reasonable experimental sensitivity
(see Table \ref{tab:alpha}).
These isotopes are: $^{165}$Ho, four isotopes of dysprosium ($^{158}$Dy, $^{160}$Dy, $^{161}$Dy, $^{162}$Dy), 
two of neodymium ($^{143}$Nd, $^{148}$Nd),
two of gadolinium ($^{154}$Gd, $^{155}$Gd), 
two of thallium ($^{203}$Tl, $^{205}$Tl), 
five of erbium ($^{164}$Er, $^{166}$Er, $^{167}$Er, $^{168}$Er, $^{170}$Er),
$^{152}$Sm,
$^{197}$Au,
five isotopes of mercury ($^{198}$Hg, $^{199}$Hg, $^{200}$Hg, $^{201}$Hg, $^{202}$Hg), the two isotopes of lutetium ($^{175}$Lu, $^{176}$Lu),
all of them because of 
their low $Q_\alpha$ values (see Table \ref{tab:alpha}).

Moreover, no attempt for half-life measurements has been done for $^{150}$Sm, $^{169}$Tm, the two isotopes 
of rhenium ($^{185}$Re, $^{187}$Re),
of tantalum ($^{180m}$Ta, $^{181}$Ta),
of iridium ($^{191}$Ir, $^{193}$Ir), and $^{196}$Hg, either.
In particular, $^{187}$Re is undergoing $\beta$ decay.

In the following, we review the cases of those isotopes for which measurements on their decay were performed,
setting lower limits on $T_{1/2}$.

\subsection{$^{142}$Ce $\alpha$ decay}
\label{ce142}

The cerium isotopes $^{136}$Ce, $^{138}$Ce, and $^{142}$Ce, with natural abundances 0.185(2)\%, 0.251(2)\%, and 11.114(51)\%, respectively, \cite{Meija:2016}
are good candidates for $2\beta$ decay; competitive experimental results are reported in Ref. \cite{ber97,dan01,bel03,bel09,bel14,bel17}. 
Moreover, the $^{142}$Ce isotope can decay to the ground state of $^{138}$Ba with emission of an 
$\alpha$ particle. The corresponding energy release is equal to $(1303.5 \pm 2.5)$ keV \cite{Wang:2017}, thus 
the expected energy of the $\alpha$ particle is $E_\alpha = 1267$ keV. The decay has not yet been observed and only 
experimental limits have been set. Firstly a limit was set in the sixties: $T_{1/2}^\alpha \ge 5 \times 10^{16}$ yr \cite{Macfarlane:1961}
by using a large cylindrical ionization counter, accommodating samples up to 1200 cm$^2$ in area.

A more stringent limit has been achieved by using a more sensitive set-up \cite{bel03} at LNGS.
In this experiment the used detector was a CeF$_3$ crystal scintillator ($2 \times 2 \times 2 $ cm$^3$, mass of 49.3 g) \cite{bel03},
that served as source and detector simultaneously. The crystal was directly coupled to two low radioactive (to enhance the sensitivity to such rare events)
PMTs, working in coincidence.
The detector was surrounded by Cu bricks and sealed in a low radioactive Cu box continuously flushed with high purity nitrogen gas 
(stored in bottles deeply underground for a long time to allow the decaying of short/medium half-lives cosmogenic isotopes) 
to avoid presence of residual environmental radon. 
The Cu box was surrounded by a passive shield made of 10 cm of high purity Cu, 15 cm of low radioactive lead, 
1.5 mm of cadmium and $4-10$ cm polyethylene (paraffin) to reduce the external background. The whole shield was closed inside a Plexiglas box, 
also continuously flushed by high purity nitrogen gas.
The performances of this CeF$_3$ crystal scintillator were investigated \cite{bel03}. In particular, 
the relative light output for $\alpha$ particles as compared with that for $\beta$ particles (or $\gamma$ rays), 
named $\alpha / \beta$ ratio, was measured at various $\alpha$ energies either by $\alpha$'s, collimated and attenuated by mylar foils, from a $^{241}$Am source
or by peaks due to internal contamination of the CeF$_3$ crystal itself. The dependence can be written as \cite{bel03}:
$$ \alpha / \beta = 0.084+1.09\times 10^{-5}E_\alpha, $$
where $E_\alpha$ is in keV.
Finally, the possibility of a pulse-shape discrimination between $\alpha$ particles and $\gamma$ quanta, and
the radioactive contamination of the crystal were also studied \cite{bel03}. 

\begin{figure}[ht]
\begin{center}
\resizebox{0.35\textwidth}{!}{ 
  \includegraphics{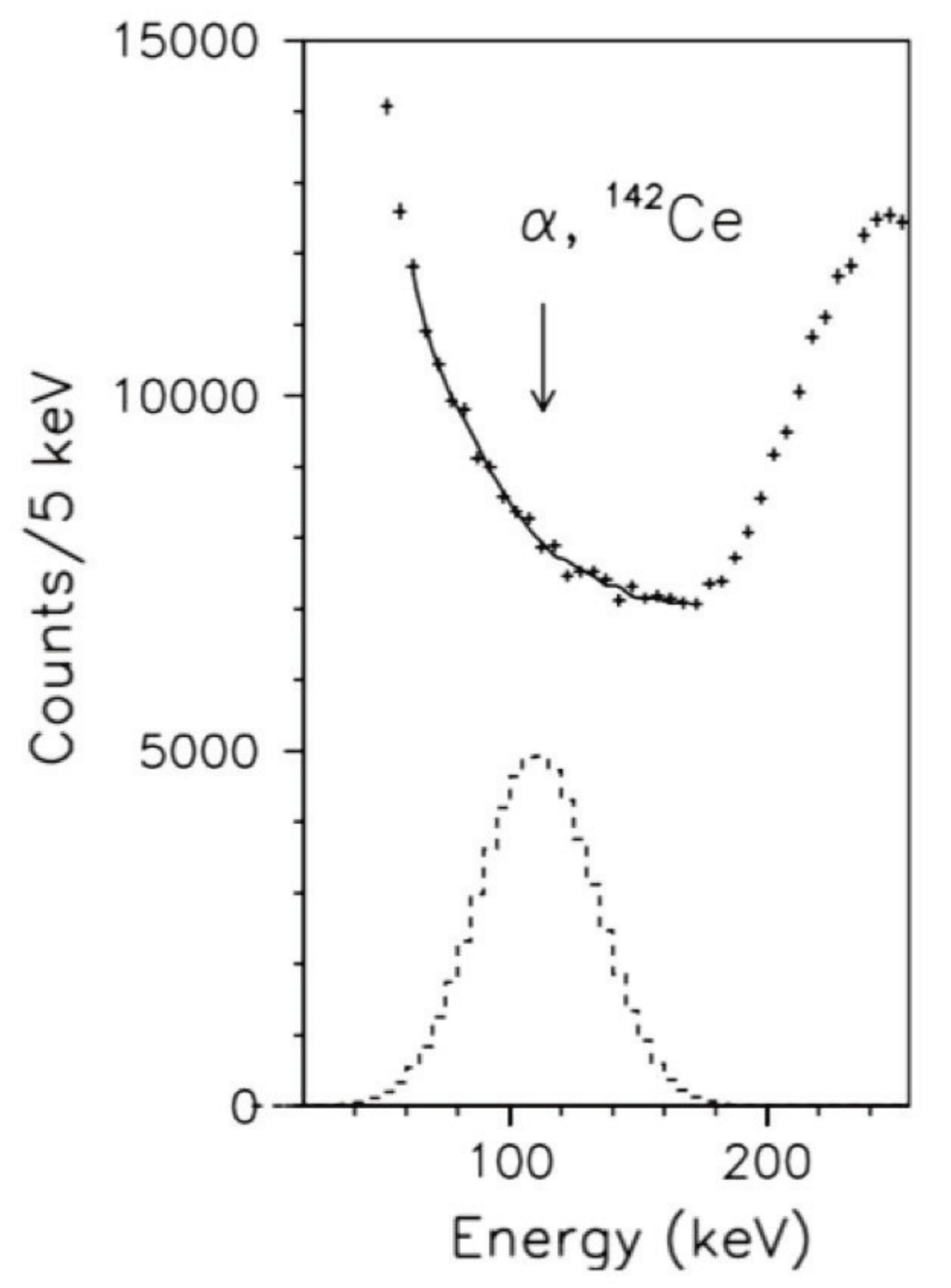}
}
\end{center}
\vspace{-0.4cm}
\caption{Low energy part of the background spectra measured for 2142 h in the low background set-up with the CeF$_3$ 
scintillator after the rejection of the PMT noise and the double pulses \cite{bel03}. The spectrum contains at least 99\% of $^{142}$Ce 
$\alpha$ peak. The fit of the spectrum is shown by solid line, the simulated $\alpha$ peak of $^{142}$Ce corresponding to
half-life $T_{1/2} = 5 \times 10^{16}$ yr is shown by dashed line as reference. Figure taken from Ref. \cite{bel03}.}
\label{fg:142ce}
\end{figure}

Considering the natural abundance, \, the number of $^{142}$Ce nuclei in the CeF$_3$ crystal was $1.67 \times 10^{22}$. 
The search for the $^{142}$Ce $\alpha$ decay was performed by using the energy spectrum obtained by discarding the 
double pulses, that identify events due to known background from radioactive isotopes, such as those due to the following chains:
$^{212}$Bi $\to$ $^{212}$Po $\to$ $^{208}$Pb, and 
$^{220}$Rn $\to$ $^{216}$Po $\to$ $^{212}$Pb. Taking into account the $\alpha/\beta$ ratio,
the $\alpha$ peak of $^{142}$Ce can be expected \, at the energy $(127 \pm 7)$ keV \, (with FWHM $=53$ keV), 
where the experimental $\alpha$ spectrum contains at least 99\% of $\alpha$ events. A fitting procedure
was applied to search for the $^{142}$Ce $\alpha$ decay. The best fit in the (60--165) keV energy interval gives a peak's area equal to 
$(-790 \pm 1010)$ counts and leads to $\lim S = 972$ counts at 90\% C.L. Thus, the lower bound on the half-life of the $^{142}$Ce $\alpha$ decay is \cite{bel03}:

\begin{equation}
T_{1/2}^\alpha \ge 2.9 \times 10^{18} \;\; \textrm{yr} \;\;\; \textrm{at 90\% C.L.}
\end{equation}

A part of the experimental distribution together with the fitting curve is presented in Fig. \ref{fg:142ce}. 
The simulated $\alpha$ peak of $^{142}$Ce with $T_{1/2} = 5 \times 10^{16}$ yr 
(that is also the half-life limit of Ref. \cite{Macfarlane:1961}) is shown as reference.
The present experimental limit 
is still much lower than the existing theoretical estimates, which are in the range of 
$\approx 10^{27}$ yr \cite{alb88,Poe83,Buc91}, as shown in Table \ref{tab:alpha}.

\subsection{The $\alpha$ decay of dysprosium isotopes}
\label{dysprosium}

Five of the seven natural dysprosium isotopes are potentially
unstable undergoing $\alpha$ decay. Furthermore, the
transitions to the excited levels of daughter nuclei are
allowed for all the isotopes. Characteristics of the possible alpha decays are
presented in Table \ref{tab:alpha}.
In Ref. \cite{Dy} a search for $\alpha$ decay of the dysprosium
isotopes to the excited levels of the daughter gadolinium isotopes
was realized by analysing the low background energy spectrum
measured by a low background HPGe $\gamma$ detector 
with a purified 322 g Dy$_2$O$_3$ sample for 2512 h. 
\begin{figure}[ht]
\begin{center}
\resizebox{0.48\textwidth}{!}{ 
  \includegraphics{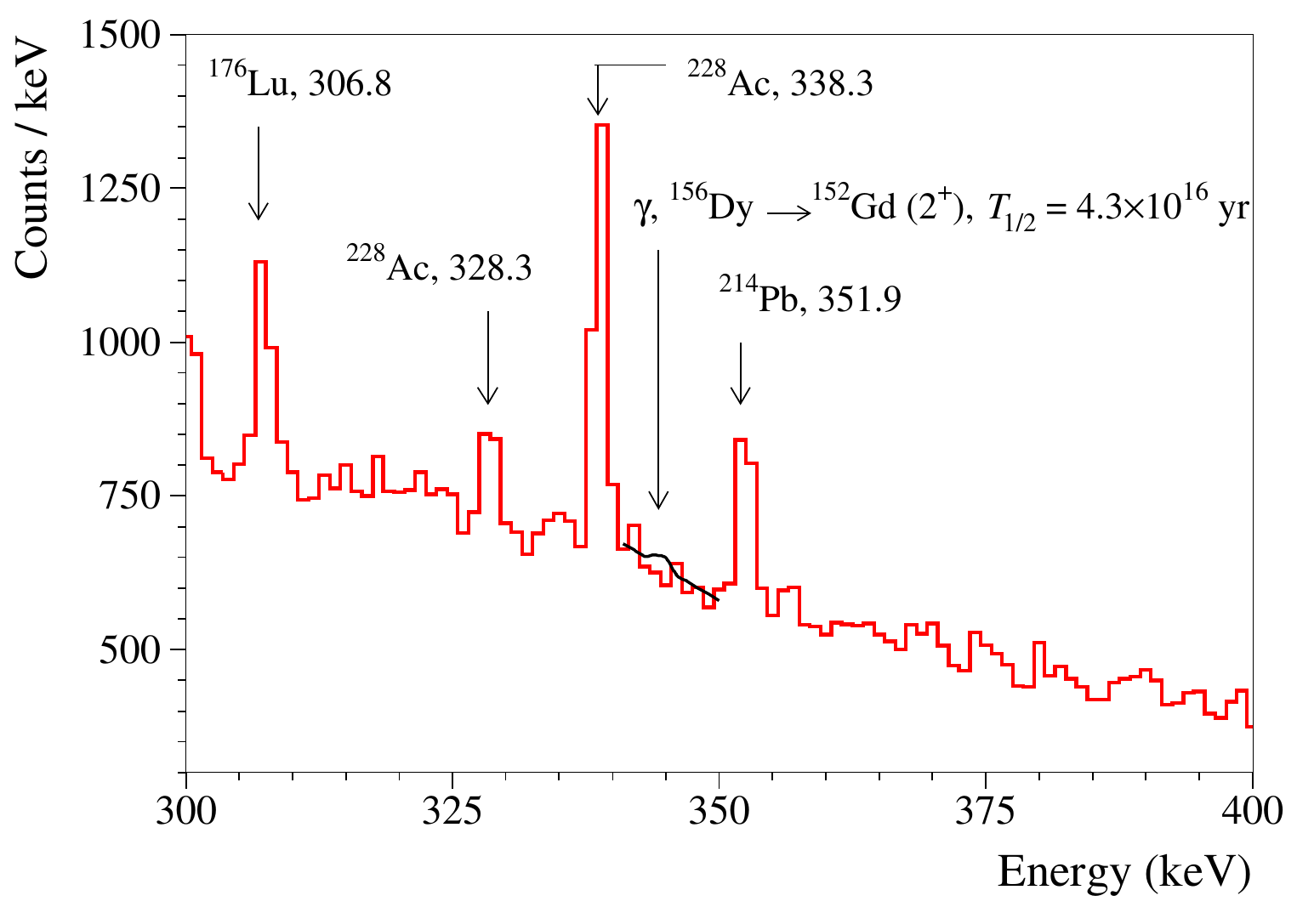}
}
\end{center}
\caption{(Color online) A part of the energy spectrum measured by the HPGe
spectrometer with a Dy$_2$O$_3$ sample for 2512 h. An
expected $\gamma$ peak from $\alpha$ decay of $^{156}$Dy to the
first excited level of $^{152}$Gd corresponding to the half-life
$T_{1/2} = 3.8\times10^{16}$ yr (excluded at 90\% C.L.) is shown
by the solid line \cite{Dy}. Figure taken from Ref. \cite{Dy}.}
\label{figdy}
\end{figure}
No peculiarities, that
could be interpreted as $\gamma$ peaks from $\alpha$ decays of the
dysprosium isotopes to the lowest excited levels of the daughter
nuclei, were observed in the energy spectrum measured with  
HPGe spectrometer; 
thus, only limits were set on these decays. 
As an example in Fig. \ref{figdy} is shown a part of the energy spectrum accumulated by the HPGe  
spectrometer.
 
The results for the various decay modes are summarized
in Tables \ref{tab:alpha} and \ref{tbdy} for the g.s. to g.s. transitions and for 
the transitions to the lowest excited levels of the daughter nuclei, respectively.
In these Tables are given also the theoretical estimates that were calculated by using the
approach used in \cite{Poe83}.

\begin{table*}[!ht]
\caption{Characteristics of possible alpha decays of dysprosium
isotopes to the lowest excited levels of the daughter nuclei \cite{Dy}.}
\begin{center}
\begin{tabular}{l|l|l|l|l|l}
 \hline
  Alpha decay                         & Excited level         & Detection     & $\lim S$  & Experimental              & Theoretical    \\
  $Q_{\alpha}$ (keV) \cite{Wang:2017} & of daughter           & efficiency    & ~         & limit on                  & estimation of  \\
  $\delta$ (\%) \cite{Meija:2016}     & nuclei (keV),         & of the $\gamma$ & ~       & $T_{1/2}$ (yr)            & $T_{1/2}$ (yr) \\
   ~                                  & yield of $\gamma$     & quanta        & ~         & ~                         &  \cite{Poe83} \\
   ~                                  & per $\alpha$ decay    &  ~            & ~         & ~                         & ~ \\
  \hline
  $^{156}$Dy$\rightarrow$$^{152}$Gd   & $2^+$, 344.3          & 2.1\%         & 31        & $\geq3.8\times10^{16}$    & $2.3\times10^{34}$ \\
  1753.0(3)                           & 0.962 \cite{Art96}    & ~             & ~         & ~                         & ~ \\
  0.056(3)                            & ~                     & ~             & ~         & ~                         & ~ \\
    \hline
  $^{158}$Dy$\rightarrow$$^{154}$Gd   & $2^+$, 123.1          & 0.9\%         & 31        & $\geq1.3\times10^{16}$    & $1.4\times10^{68}$ \\
  873.7(24)                           & 0.455 \cite{Rei09}    & ~             & ~         & ~                         & ~ \\
  0.095(3)                            & ~                     & ~             & ~         & ~                         & ~ \\
    \hline
  $^{160}$Dy$\rightarrow$$^{156}$Gd   & $2^+$, 89.0           & 0.4\%         & 230       & $\geq8.5\times10^{15}$    & $5.4\times10^{126}$ \\
  437.3(11)                           & 0.203 \cite{Rei03}    & ~             & ~         & ~                         & ~ \\
  2.329(18)                           & ~                     & ~             & ~         & ~                         & ~ \\
    \hline
  $^{161}$Dy$\rightarrow$$^{157}$Gd   & $5/2^-$, 54.5         & 0.2\%         & 83        & $\geq3.5\times10^{16}$    & $1.3\times10^{147}$ \\
  342.8(11)                           & 0.075 \cite{Hel04a}   & ~             & ~         & ~                         & ~ \\
  18.889(42)                          & ~                     & ~             & ~         & ~                         & ~ \\
    \hline
  $^{162}$Dy$\rightarrow$$^{158}$Gd   & $2^+$, 79.5           & 0.3\%         & 113       & $\geq1.0\times10^{17}$    & $1.7\times10^{1414}$ \\
  83.2(11)                            & 0.142 \cite{Hel04}   & ~             & ~         & ~                         &  ~ \\
  25.475(36)                          & ~                     & ~             & ~         & ~                         & ~ \\
    \hline

\end{tabular}
\end{center}
\label{tbdy}
\end{table*}

As shown in the Table \ref{tbdy}, the theoretical estimates disfavour
realistic observations of
alpha decays of all naturally abundant dysprosium isotopes to the
excited levels of daughter nuclei. At the same time the detection of the
$^{156}$Dy decay to the ground state of $^{152}$Gd (supposing one
will apply crystal scintillators from dysprosium enriched in
$^{156}$Dy) looks possible. In fact, a half-life
$6.8\times10^{24}$ yr was estimated by using the approach of Ref. \cite{Poe83}.
The strong advantage of scintillation methods is an almost 100\%
detection efficiency to alpha particles and pulse-shape
discrimination capability \cite{Danevich:2003a,bel07}. 

As far as we know, there was only one attempt many years ago to measure the half-life of the $\alpha$ decay of the 
$^{156}$Dy to the ground state of $^{152}$Gd. It was done by using Ilford C2 emulsion plates in contact with the
dysprosium sample \cite{Riezler:1958}. The absence of $\alpha$ tracks gave a rough estimate of the half-life 
lower limit: $T_{1/2} > 10^{18}$ yr \cite{Riezler:1958}. 
To our knowledge no attempt was made to investigate the $\alpha$ decay of the other Dy isotopes --
also disfavoured by the lower $Q_\alpha$ values with respect to $^{156}$Dy (see Table \ref{tab:alpha}) -- to the ground 
state of the daughter nuclei.

In conclusion, the development of
enrichment methods for dysprosium isotopes and of radiopure crystal
scintillators containing Dy is mandatory to observe $\alpha$
activity of natural dysprosium.

\subsection{The $\alpha$ decay of erbium isotopes}
\label{erbium}

Erbium has six naturally occurring isotopes (see Table \ref{tab:alpha}).
They are stable, but potentially all of them can decay emitting $\alpha$ particles \cite{aud17}, 
in particular:
$^{162}$Er is considered to undergo $\alpha$ decay to $^{158}$Dy (with a half-life larger than 1.40 $\times 10^{14}$ yr).

The only experimental data are available from Ref. \cite{Porschen:1956}, where the $\alpha$ activity
was studied for some elements.
To detect the $\alpha$ particles, Ilford plates were used (with the element impregnated in emulsions or mixed in a gel form). After the development, 
the plates were screened with a  Leitz-Ortholux microscope. 
Only traces fully contained in the plates were considered in order to evaluate the energy.
All the experimental parameters were considered (effects on the preparation of the emulsions, the fixing bath, the post treating bath, humidity etc.). The mass of the element impregnated in the emulsions was around 5-20 mg in 800 mg of emulsion, in order to have a negligible variation of the emulsion density. In case of gel 
the density was determined by measuring it.
Also some blank measurements were performed without impregnating the emulsion with the element under study. Most of the traces are due to radioactive contaminants.
Some radioactive impurities (Sm, Th, U, Po, etc.) were present in traces in the examined samples, so the 
measurements were performed in two energy intervals (1.5--2.5) MeV and (2.5--3.7) MeV, where the first interval is more affected by the Sm impurity (it is present at level of $10^{-4} - 10^{-5}$).
Twenty-three elements were examined and for 19 of them no alpha activity was found. For these elements only a lower
limit for life-time was set. 

In Table 4 of Ref. \cite{Porschen:1956} a lower limit value around 10$^{17}$ is quoted for the Er element, value that is at level of sensitivity of the applied experimental approach. 

Considering that from the theory-phenomenology of the $\alpha$ decay (Gamow, Royer, Viola-Seaborg etc.), the smal\-ler the value of the energy released in the decay 
the larger the half-life of the decay becomes. The shortest half-life (in principle the first for which one could hope to reach a suitable experimental sensitivity)
is that of the $^{162}$Er isotope (see Table \ref{tab:alpha}). Considering its natural abundance (0.139(5)\%),
the limit of Ref. \cite{Porschen:1956} corresponds to $T_{1/2} > $ 1.4 $\times$ 10$^{14}$ yr, which is in fact the value quoted in Ref. \cite{aud17}.

In literature no other experimental determination is at hand, and certainly with the updated advanced experimental techniques presently available this limit could be strongly improved. 
However, the modeling estimates seem to set the half-life of these decays to very high values, as reported in Table \ref{tab:alpha}
and e.g. in Ref. \cite{qia14}, 
where the quoted calculated values are in the range ($10^{28} - 10^{30}$) for $^{162}$Er and
even ($10^{39} - 10^{41}$) for $^{164}$Er.

\subsection{The $\alpha$ decay of ytterbium isotopes}
\label{ytterbium}

Ytterbium has seven naturally occurring isotopes (see Table \ref{tab:alpha}).

They are stable, but potentially all of them can decay emitting $\alpha$ particles \cite{aud17}, 
in particular:
$^{168}$Yb is considered to undergo $\alpha$ decay to $^{164}$Er (with a half-life over $1.3 \times 10^{14}$ yr).

The only experimental data are available from Ref. \cite{Porschen:1956} obtained with the same experimental approach already described in the Er case. 

In case of Yb, in Table 4 of Ref. \cite{Porschen:1956} a lower limit around 10$^{17}$ is quoted, a value that is at the level of sensitivity of the applied experimental approach. 

Therefore, considering that for Yb isotopes (see Table \ref{tab:alpha}) the highest $Q_\alpha$ value is that of  the $^{168}$Yb isotope (natural abundance 0.123(3)\%),
the limit of Ref. \cite{Porschen:1956} corresponds to a $T_{1/2} > $ 1.3 $\times$ 10$^{14}$ yr, which is in fact the value quoted in Ref. \cite{aud17}.
Also in case of Yb isotopes the theoretical estimates seem to set the life-time of these decays to higher values. The quoted calculated values (see Table \ref{tab:alpha}
and e.g. Ref. \cite{qia14}) 
are in the range ($10^{23} - 10^{25}$) for $^{168}$Yb.

\subsection{Search for $\alpha$ decay of lead isotopes}
\label{lead}

After the discovery of the $\alpha$ decay of $^{209}$Bi in 2003 \cite{Marcillac:2003a}, the
lead is considered to be the heaviest stable element.
As shown in Table \ref{tab:alpha}, lead has four naturally occurring isotopes
with $Q_\alpha$ larger than zero and, therefore, potentially unstable undergoing $\alpha$ decay.
Given the theoretical predictions for the half-lives, there is no possibility to
observe their $\alpha$ decays with the present sensitivities.
Moreover, let us stress the important role of $^{204}$Pb in lead geochronology as a standard reference \cite{Dick:05}.

\begin{figure}[ht]
\begin{center}
\resizebox{0.45\textwidth}{!}{ 
  \includegraphics{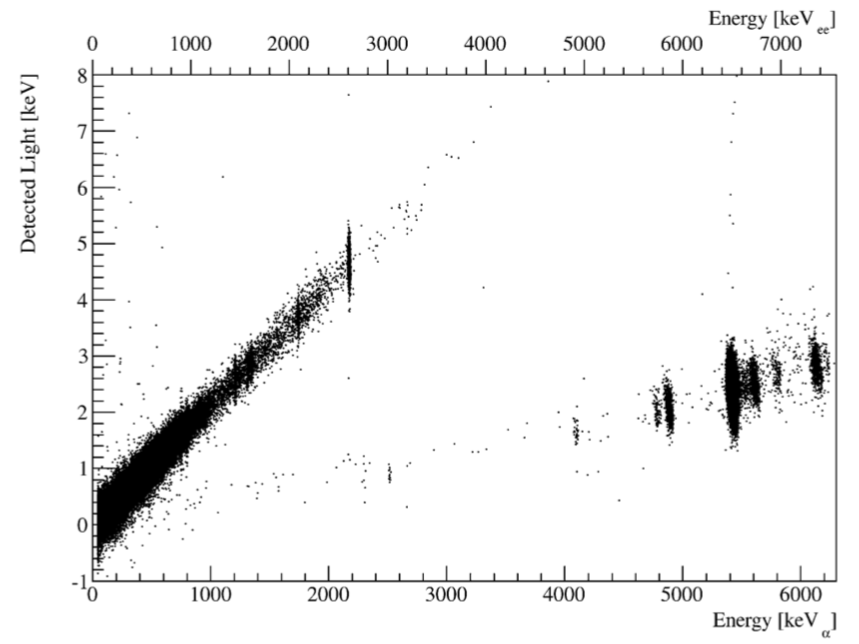}
}
\end{center}
\caption{Light-heat scatter plot measured over 586 h. The two horizontal axes are calibrated for $\alpha$ and $\beta/\gamma$, respectively.
Figure taken from Ref. \cite{bee13}.}
\label{fig:pb1}
\end{figure}

\begin{figure}[ht]
\begin{center}
\vspace{-0.6cm}
\resizebox{0.45\textwidth}{!}{ 
  \includegraphics{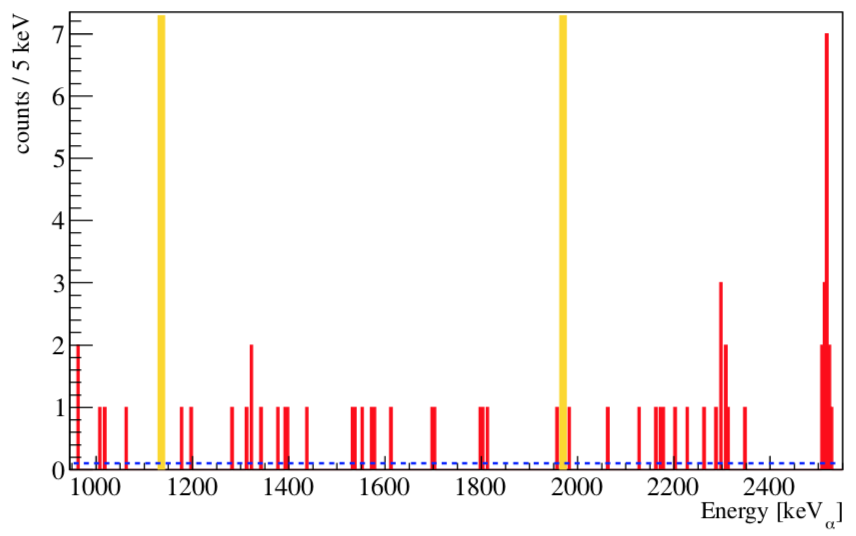}
}
\end{center}
\caption{Energy spectrum of $\alpha$ particles in the PbWO$_4$ experiment. The two bands are the ROI for $^{204}$Pb and $^{206}$Pb decays.
Figure taken from Ref. \cite{bee13}.}
\label{fig:pb2}
\end{figure}

A possible evidence of the $\alpha$ decay of $^{204}$Pb was reported in the fifties \cite{Riezler:1958}, by using Ilford C2 emulsion plates in contact with the
enriched (at 27.0\%) sample of $^{204}$Pb. However, such measurement was reconsidered on the basis of more accurate mass determination
of the parent and daughter nuclei, leading to a half-life limit for the $\alpha$ decay of $^{204}$Pb to $T_{1/2} > 1.4 \times 10^{17}$ yr.

In 2013 new limits on the $\alpha$ decay of naturally occurring
Pb isotopes were obtained in Ref. \cite{bee13},
by using a 454.1 g PbWO$_4$ crystal as a bolometer at LNGS underground laboratory. 
The crystal was grown using low background ancient Roman lead.
The scintillation light of the bolometer was detected by a Ge absorber working as a bolometer.
The scatter plot of the events in the plane light vs heat is reported in Fig. \ref{fig:pb1}; the horizontal axes are calibrated
for $\alpha$ and $\beta/\gamma$, respectively.
The separation shown in Fig. \ref{fig:pb1} allows the rejection of the background and the selection of the $\alpha$ particles,
whose energy spectrum is reported in Fig. \ref{fig:pb2}.
New upper limits on the half-lives of the four natural lead isotopes have been estimated 
in the range of $T_{1/2} > 1.4 \times 10^{20} - 2.6 \times 10^{21}$ yr at 90\% C.L.
with PbWO$_4$ scintillating bolometer \cite{bee13} (see also Table \ref{tab:alpha}).

\vspace{0.6cm}
\section{The rare $\beta$ decays}
\label{intro_beta}

Another hot field of interest in nuclear physics is the investigation of nuclear beta decays (weak interaction
processes inside the nuclei), and in particular the study of rare and highly forbidden beta decays.
Beta radiation was observed long time ago \cite{rut99}; however, our knowledge of this phenomenon still can be and should be improved. 
Most of the beta decay transitions, which are observed in nature, are of allowed type, or in few cases transitions with low level of forbiddenness. More recently rare beta decays of long-living nuclides are under investigation. 
They are beta decay 
($\beta^-$ or $\beta^+/\varepsilon$) transitions suppressed by low $Q_\beta$ value and/or high level of forbiddenness due to the large change in spin and parity; thus they correspond to
long half-lives.

The $\beta$ decays is well approximated by a point-like interaction vertex, since the massive $W^{\pm}$ boson has 
$\approx 80$ GeV/c$^2$ mass  while the decay energy ($\approx 1$ MeV) is small.
In addition to the general information, which can be derived by the experiments as shown later in this paper, 
two arguments are intriguing topics: 
\begin{itemize}
\item 1) the very accurate determination of the shape of the beta spectrum at the end-point, 
      in order to get information on the neutrino mass with direct and model-independent approach;
\item 2) the investigation of the role of the axial-vector coupling, $g_A$.
\end{itemize}
For the first case the most suitable 
nuclei/transitions 
are those with as low as possible $Q-$value and allowed, but 1st (2nd, ...) forbidden could be manageable as well;
at present the exploited candidates are: $^3$H ($Q_\beta = 18.594(8)$ keV) and $^{187}$Re ($Q_\beta = 2.469(4)$ keV).
As regards the axial-vector coupling, the value $g_A = 1.27$ extracted from the 
partially conserved vector current hypothesis 
holds for bare nucleons only; on the other hand, in the $0\nu 2\beta$ decay rate $\lambda \propto (g_A)^4$,
thus, a possible -- even small -- quenching of $g_A$ can play a strong impact on the sensitivities of the present and future
$0\nu 2\beta$ experiments. This quenching from the free-nucleon value can arise from nuclear medium effects 
and/or from nuclear many-body effects.
Moreover, the effective value of $g_A$ can also depend on the energy scale of the process and it can be different 
for $\beta$ and $2\beta$ decays (for exhaustive reviews see Refs. \cite{suh19,Suhonen:2017b}).
A good probe to study the effective value of $g_A$ can be the $\beta$ decay of the
$^{113}$Cd isotope (4th forbidden non-unique, $Q_\beta = (323.83\pm 0.27)$ keV \cite{Wang:2017}). 

Some rare $\beta$ decays ($T_{1/2} > 10^{10}$ yr) are poorly investigated;
for some of them the $\beta$ decay is not yet observed, as in the case of $^{50}$V, $^{123}$Te, and $^{180m}$Ta. The interest in 
$\beta$ decays increased in the recent past 
because sometimes they cause significant backgrounds in searches for and investigations of rare effects, 
like solar neutrinos (e.g. $^{14}$C in Borexino \cite{aa02}), double beta ($2\beta$) decay (e.g. $^{39}$Ar, $^{42}$Ar/$^{42}$K 
in GERDA \cite{aa03}, $^{113}$Cd in $^{106}$Cd double beta decay experiments) 
or dark matter experiments, especially based on Ar (e.g. $^{39}$Ar, $^{42}$Ar/$^{42}$K in DarkSide \cite{aa04}). 
While in the above-mentioned cases $\beta$ decays constitute one of the main features in the measured spectra, other single $\beta$ 
decays create not so large, but noticeable, backgrounds in many experiments: $^{40}$K, $^{90}$Sr/$^{90}$Y, $^{137}$Cs, $^{214}$Bi 
and others. Quite often their energy spectra have not-allowed shape and are classified as forbidden (unique or non-unique),
and sometimes they are not well studied. For example, $^{214}$Bi is one of the main backgrounds 
in $2\beta$ experiments due to its high energy release, $Q_\beta$ = 3270 keV; in 19.1\% of the cases it decays to the ground state 
of $^{214}$Po with change in spin and parity $1^{-} \rightarrow 0^+$, $\Delta J^{\Delta \pi} = 1^{-}$, classified as 1-forbidden 
non-unique (1 FNU); however, to our knowledge, theoretical calculations of its shape 
appeared only recently \cite{kosten:2018}, after Ref. \cite{tretyak:2017} which drew attention to this problem.
In addition, it was not well measured experimentally. Its shape in graphical form can be found only in old papers (see \cite{aa05,aa06,aa07}), 
from which it could be concluded that it is not far from the allowed. It is clear that the good knowledge of the shapes of single 
$\beta$ decays is very important for a proper fitting of experimental spectra and correct estimation of weak effects possibly 
present there. 

After a summary of the classification of single $\beta$ decays and shapes of their energy spectra, we review recent achievements 
in studies and searches for rare $\beta$ processes.

A collection of partial half-lives (or limits) on $\beta$ decays of all the naturally occurring nuclides with
$Q_\beta > 0$ is reported in Table \ref{tab:beta_nucl}.

\onecolumn
\begin{landscape}
\begin{table}
\caption{Collection of partial half-lives (or limits) on rare $\beta$ decays. The branching fractions for the considered decay channel 
(and, if present, for other decay modes in brackets) are reported. Only nuclides with natural abundance greater than zero (i.e.
naturally present in nature) and with $Q_\beta > 0$ are considered. The nuclide with the lowest experimental total $T_{1/2}$ reported in this table 
is $^{40}$K with the measured total half-life of $(1.248 \pm 0.003) \times 10^{9}$  yr. The limits, when not specified, are at 90\% C.L.}
\label{tab:beta_nucl}
\begin{center}
\resizebox{\textwidth}{!}{
\begin{tabular}{r|c|r|r|cr|cr}
\hline\noalign{\smallskip}
   Nuclide transition               & Branching                             & Natural                     &  $Q_\beta$ (keV)         & \multicolumn{4}{c}{Partial $T_{1/2}$ (yr)}                            \\
                                    & ratio, $\eta$ \cite{aud17}            & abundance \cite{Meija:2016} &  \cite{Wang:2017}        & \multicolumn{2}{c|}{experimental} & \multicolumn{2}{c}{theoretical}   \\
\hline
\multicolumn{7}{l}{$\beta^-$ decay:}     \\
$^{ 40}$K  $\rightarrow ^{ 40}$Ca   & 89.28(13)\%        &   0.0117(1)\%            &   1310.89(6)   &   $(1.248 \pm 0.003) \times 10^{9}/\eta$         & \cite{aud17}           &   $2.08\times10^{8}$               & \cite{Suh07}  \\
$^{ 48}$Ca $\rightarrow ^{ 48}$Sc   & $<20\%$ $^a$       &   0.187(21)\%            &   279(5)       &   $> 2.5 \times 10^{20}$ $^b$                    & \cite{Bak02}           &   $1.1^{+0.8}_{-0.6}\times10^{21}$ & \cite{Aun99}  \\
                                    &                    &                          &                &                                                  &                        &   $(2.6-7.0) \times 10^{20}$       & \cite{Haa14}  \\
                                    &                    &                          &                &                                                  &                        &   $4.9\times10^{20}$               & \cite{Kos17}  \\
$^{ 50}$V  $\rightarrow ^{ 50}$Cr   & $<1.4 \%$          &   0.250(10)\%            &   1038.06(30)  &   $>1.9 \times 10^{19}$                          & \cite{Lau18}           &   $9.0\times10^{17}$               & \cite{Nis79}  \\
                                    &                    &                          &                &                                                  &                        &   $(2.0-2.3)\times 10^{19}$        & \cite{Haa14b} \\
                                    &                    &                          &                &                                                  &                        &   $1.5\times10^{19}$               & \cite{Kos17}  \\
$^{ 87}$Rb $\rightarrow ^{ 87}$Sr   & 100\%              &   27.83(2)\%             &   282.275(6)   &   $(4.97 \pm 0.03) \times 10^{10}$               & \cite{aud17}           &   $5.1\times10^{8}$                & \cite{Sas69}  \\
                                    &                    &                          &                &                                                  &                        &   $4.0\times10^{10}$               & \cite{Kop72}  \\
                                    &                    &                          &                &                                                  &                        &   $(0.6-1.2)\times10^{10}$         & \cite{Szy76}  \\
                                    &                    &                          &                &                                                  &                        &   $3.6\times10^{9}$                & \cite{Kos17a} \\
                                    &                    &                          &                &                                                  &                        &   $4.6\times10^{9}$                & \cite{Kos17}  \\
$^{ 96}$Zr $\rightarrow ^{ 96}$Nb   & $<38\%$ $^c$       &   2.80(2)\%              &   163.97(10)   &   $> 3.8 \times 10^{19}$ $^d$                    & \cite{aa19}            &   $2.4\times10^{20}$               & \cite{Hei07}  \\
                                    &                    &                          &                &                                                  &                        &   $1.3\times10^{20}$               & \cite{Kos17}  \\
$^{113}$Cd $\rightarrow ^{113}$In   & 100\%              &   12.227(7)\%            &   323.83(27)   &   $(8.04 \pm 0.05) \times 10^{15}$               & \cite{Belli:2007a}     &   $1.05\times10^{16}$              & \cite{Mus07}  \\
                                    &                    &                          &                &                                                  &                        &   $1.4\times10^{17}$               & \cite{Kos17a} \\
$^{115}$In $\rightarrow ^{115}$Sn   & 100\%              &   95.719(52)\%           &   497.489(10)  &   $(4.41 \pm 0.25) \times 10^{14}$               & \cite{Pfe78,Pfe79}     &   $3.57\times10^{14}$              & \cite{Mus07}  \\
                                    &                    &                          &                &                                                  &                        &   $3.0\times10^{15}$               & \cite{Kos17a} \\
         $\rightarrow ^{115}$Sn$^*$ & $10^{-4}\%$        &                          &                &   $(4.2  \pm 0.5) \times 10^{20}$ $^e$           & \cite{Wie09,And11}     &   -- & \\
$^{138}$La $\rightarrow ^{138}$Ce   & 34.4(5)\%          &   0.08881(71)\%          &   1052(4)      &   $(1.02 \pm 0.01) \times 10^{11}/\eta$          & \cite{aud17}           &   $9(3)\times10^{15}$              & \cite{Kos17c} \\
$^{176}$Lu $\rightarrow ^{176}$Hf   & $100\%$ $^f$       &   2.599(13)\%            &   1194.1(9)    &   $(3.684 \pm 0.018) \times 10^{10} $            & \cite{aud17,Hult:2014} &   -- & \\
$^{180m}$Ta $\rightarrow ^{180}$W   & -- $^f$            &   0.01201(32)\%          &   778.6(20)    &   $> 1.7 \times 10^{17}$                         & \cite{chan:2018}       &   $5.4\times10^{23}$               & \cite{Eji17}  \\
$^{187}$Re $\rightarrow ^{187}$Os   & $100\%$                    &   62.60(5)\%     &   2.467(2)     &   $(4.33 \pm 0.07) \times 10^{10} $              & \cite{aud17,Basu:2009} &   -- & \\
                                    & $(\alpha<0.0001\%)$        &                  &                &                                                  &                        &      & \\
\hline\noalign{\smallskip}
\multicolumn{7}{l}{$\beta^+/\varepsilon$ decay:}  \\
$^{ 40}$K  $\rightarrow ^{ 40}$Ar   & $\eta_\varepsilon= 10.72(13)\%$, &   0.0117(1)\%    &   1504.40(6)   &   $(1.248 \pm 0.003) \times 10^{9}/\eta$        & \cite{aud17}           &   -- & \\
                                    & $\eta_{\beta^+}= 0.00100(13)\%$  &                  &                &                                                 &                        &   -- & \\
$^{ 50}$V  $\rightarrow ^{ 50}$Ti   & $\eta_\varepsilon>98.6 \%$,      &   0.250(10)\%    &   2207.6(4)    &   $(2.67^{+0.16}_{-0.18}) \times 10^{17}$       & \cite{Lau18}           &   $1.5\times10^{16}$          & \cite{Nis79}  \\
                                    &                                  &                  &                &                                                 &                        &   $(3.6-5.1)\times10^{17}$    & \cite{Haa14b} \\
                                    & $\eta_{\beta^+}=?$               &                  &                &   --                                            &                        &   -- & \\
$^{123}$Te $\rightarrow ^{123}$Sb   & $\eta_\varepsilon= 100\%$        &   0.89(3)\%      &   51.91(7)     &   $> 9.2 \times 10^{16}$                        & \cite{Ale03}           &   $(4.2-7.2)\times10^{19}$    & \cite{Civ01}  \\
$^{138}$La $\rightarrow ^{138}$Ba   & $\eta_\varepsilon= 65.6(5)\%$,   &   0.08881(71)\%  &   1742(3)      &   $(1.02 \pm 0.01) \times 10^{11}/\eta$         & \cite{aud17}           &   $8(2)\times10^{15}$         & \cite{Kos17c} \\
                                    & $\eta_{\beta^+}=?$               &                  &                &                                                 &                        &   -- & \\
$^{176}$Lu $\rightarrow ^{176}$Yb   & -- $^f$                          &   2.599(13)\%    &   109.1(12)    &   --                                            &                        &   -- & \\
$^{180m}$Ta $\rightarrow ^{180}$Hf  & -- $^f$                          &   0.01201(32)\%  &   921.8(20)    &   $> 2.0 \times 10^{17}$                        & \cite{chan:2018}       &   $1.4\times10^{20}$          & \cite{Eji17}  \\
\noalign{\smallskip}\hline\noalign{\smallskip}
\noalign{\smallskip}\hline
\end{tabular}}
\end{center}     
The results in Refs. \cite{Kos17a,Kos17} are given for $g_A=g_V=1.0$. \\
$^a$ The $2\beta$ decay of $^{48}$Ca is observed with $T_{1/2}=(6.4^{+1.4}_{-1.1}) \times 10^{19}$ yr \cite{Arn16}. \\
$^b$ Decay to the 131 keV excited level of $^{48}$Sc (the most probable). \\
$^c$ The $2\beta$ decay of $^{96}$Zr is observed with $T_{1/2}=(2.3 \pm 0.2) \times 10^{19}$ yr \cite{aa20}. \\
$^d$ Decay to the 44 keV excited level of $^{96}$Nb  (the most probable).\\
$^e$ Decay to the first excited level ($E_{exc} = 497$~keV) of $^{115}$Sn. The average of the most precise measurements is reported. \\
$^f$ The $\alpha$ decay is also energetically possible. See Table \ref{tab:alpha}.
\end{table}
\end{landscape}
\twocolumn

\subsection{The shapes of $\beta$ spectra}
\label{shape_beta}

The beta decays are classified as allowed or forbidden with some level of forbiddenness depending on the change in spin $J$ and parity 
$\pi$ between parent and daughter nuclei, as reported in Table \ref{tab:beta}.

For unique decays, the rate of decay and the shape of the energy spectrum are defined by only one nuclear matrix element. 
The shape of the $\beta$ spectrum in general is described as: 


\begin{table*}[!ht]
\caption{Classification of $\beta$ decays on the basis of the change in spin $J$ and parity 
$\pi$ between the parent and the daughter nuclei.}
\label{tab:beta}  
\begin{center}     
\begin{tabular}{llll}
$\Delta J^{\Delta \pi} $            &  $\Delta \pi $         &                                 & forbiddenness   \\
\noalign{\smallskip}\hline
  $0^+, 1^+ $                       &                        &  -- allowed;                    &                 \\
  $0^-, 1^-, 2^+, 3^-, 4^+, ...$    & $(-1)^{\Delta J}  $    &  -- forbidden non-unique (FNU); &  $\Delta J$     \\
  $2^-, 3^+, 4^-, ...$              & $(-1)^{\Delta J - 1}$  &  -- forbidden unique (FU);      &  $\Delta J - 1$ \\
\noalign{\smallskip}\hline
\end{tabular}
\end{center}     
\end{table*}


\begin{figure*}
\resizebox{1.0\textwidth}{!}{ 
  \includegraphics{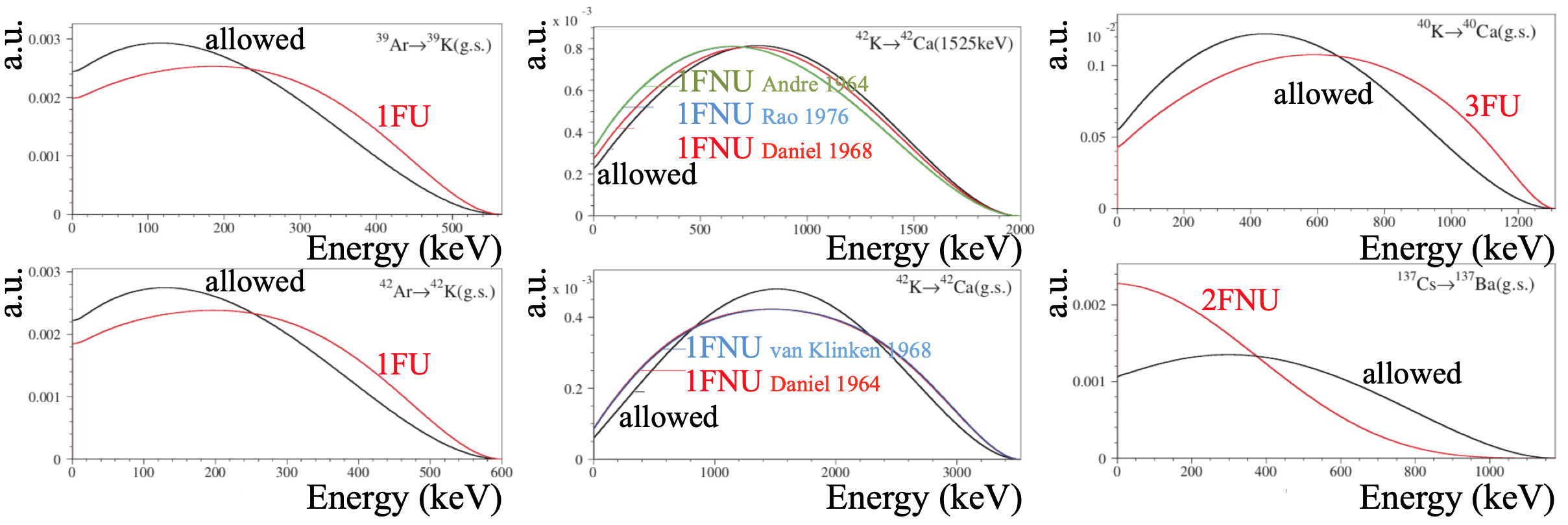}
}
\caption{Shapes of the $\beta$ spectra of $^{39,42}$Ar, $^{42}$K, $^{40}$K and $^{137}$Cs. The calculated allowed spectra (black curves) 
are reported for comparison with the expected spectra on the basis of the degree of forbiddenness. Figure taken from Ref. \cite{tretyak:2017}.}
\label{fg:shape}       
\end{figure*}

$$ \rho (E) = \rho_{allowed} (E) \times C(E),$$ 

\noindent where $$\rho_{allowed} (E) = F(Z,E) W P(Q_\beta - E)^2$$ is the distribution for 
the allowed spectrum; $W(P)$ is the total energy (momentum) of the $\beta$ particle; $F(Z, E)$ is the Fermi function:
$$ F(Z, E) = const \cdot P^{2\gamma - 2} \mathrm{exp}^{(\pi s)} \left|{\Gamma(\gamma+is)}\right|^2,$$ 

\noindent where: $\gamma = \sqrt{1 - (\alpha Z)^2}$; $s = \alpha Z W / P$; $\alpha = 1/137.036$ is the fine structure constant; 
$Z$ is the atomic number of the daughter nucleus
($Z > 0$ for $\beta^-$ and $Z < 0$ for $\beta^+$ decay); $\Gamma$ is the gamma function; $W$ is given in $m_ec^2$ units 
and $P$ in $m_ec$ units. 

Finally, $C(E)$ is an empirical correction factor depending on the energy $E$ of the $\beta$ particle and is often written as:
\begin{align}
  C(E) & =  C_1(E)                  & \textrm{for FNU decays} \nonumber \\
       & =  C_1(E) \cdot C_2(E)     & \textrm{for FU decays}  \nonumber
\end{align}

The function $C_1(E)$ is often parametrized in one of the two possible forms:
\begin{eqnarray}
  \textrm{either  } C_1(E) & = &  1 + a_1/W + a_2W + a_3W^2 + a_4W^3    \nonumber \\
  \textrm{or  }     C_1(E) & = &  1 + b_1P^2 + b_2Q^2,                  \nonumber
\end{eqnarray}
where $Q$ is the momentum of (anti)neutrino, in $m_ec$ units.

The correction factor $C_2(E)$ is parametrized differently for each FU degree: 
\begin{align}
C_2 & =  P^2 + c_1Q^2                                       & \textrm{for 1 FU} \nonumber \\
C_2 & =  P^4 + c_1P^2Q^2 + c_2Q^4                           & \textrm{for 2 FU} \nonumber \\
C_2 & =  P^6 + c_1P^4Q^2 + c_2P^2Q^4 + c_3Q^6               & \textrm{for 3 FU} \nonumber \\
C_2 & =  P^8 + c_1P^6Q^2 + c_2P^4Q^4 + c_3P^2Q^6 + c_4Q^8   & \textrm{for 4 FU} \nonumber 
\end{align}
Sometimes alternative forms are used: 
for 1 FU it is $C_2 = Q^2 + \lambda_2 P^2$, 
for 2 FU it is the analogous expression with $\lambda_2, \lambda_4$, 
and so on, where $\lambda_i$ are the Coulomb functions calculated in Ref. \cite{aa08}.

The coefficients $a_i, b_i, c_i$ above reported should be evaluated theoretically (they are mixture of products of phase space factors with different nuclear matrix elements) 
or extracted from experimental measurements. Compilations of the experimental 
$a_i, b_i, c_i$ can be found in Ref. \cite{aa09,aa10,aa11,aa12}.

As examples, the left panels of Fig. \ref{fg:shape} show the $\beta$ spectra for $^{39}$Ar and $^{42}$Ar in comparison with the allowed shapes. In both cases 
$\Delta J^{\Delta \pi} = 2^-$, so such decays are classified as 1 FU; the correction factor is given as $C(E) = Q^2 + \lambda_2P^2$.

The middle panels of Fig. \ref{fg:shape} show the $\beta$ spectra of $^{42}$K (daughter of long-living $^{42}$Ar), measured in few experiments. 
For the transition $^{42}$K $\to$ $^{42}$Ca (ground state, probability 81.90\%), which is also 1 FU, 
the correction factor is $C(E) = (Q^2 + \lambda_2P^2)(1 + aW)$. 
The transition $^{42}$K $\to$ $^{42}$Ca (excited level with $E_{exc} = 1525 $ keV, 17.641\%)
is classified as 1 FNU $(\Delta J^{\Delta \pi} = 0^-)$, 
and $C(E) = 1 + a_1/W + a_2W + a_3W^2$. The values of $a, a_i$ and the references to the original works can be found in Ref. \cite{aa11}.

The right panels show the spectra of $^{40}$K and $^{137}$Cs decays. For $^{40}$K $\to$ $^{40}$Ca (g.s., 89.28\%, $\Delta J^{\Delta \pi} = 4^-$, 3 FU);
the correction factor is $ C(E) = P^6 + c_1P^4Q^2 + c_2P^2Q^4 + c_3Q^6$ with $c_1 = c_2 = 1, c_3 =7$ as measured in Ref. \cite{aa13}.
For $^{137}$Cs $\to$ $^{137}$Ba (g.s., 5.3\%, $\Delta J^{\Delta \pi} = 2^+$, 2 FNU) the correction factor is
$ C(E) = 1 + a_1/W + a_2W + a_3W^2$ with $a_1 = 0, a_2 = -0.6060315, a_3 = 0.0921520$ as measured in Ref. \cite{aa14}.

This is a demonstration that the forbidden $\beta$ spectra can significantly deviate from the allowed shapes, 
and this has to be taken into account in the simulations of the response of the detectors.
It should be also noted that sometimes even decays that are classified as allowed have deviations from 
the allowed shape, as in the case of $^{14}$C, where in the theoretical description the first order terms
mutually compensate each other and only the second order terms determine the $T_{1/2}$ and the shape.

\section{Investigations of rare $\beta$ decays}
\label{inv_beta}

In the following various experimental results on rare beta decay will be presented considering that the careful determination of the shape, of the end-point behaviour of the beta spectrum as well as accurate $Q_\beta$ value determination are crucial.

Five naturally occurring isotopes, $^{40}$K, $^{87}$Rb, $^{138}$La, $^{176}$Lu, and $^{187}$Re,
$\beta$ unstable and with long half-lives (as reported in Table \ref{tab:beta_nucl}), are not considered in the following
since their decay parameters are well established (very often they create undesirable background in the investigations of rare effects such as $2\beta$ decay, dark matter, etc.). 
In fact $^{40}$K, despite the low natural abundance, produces 
a very distinctive $\gamma$ ray of 1460.851 keV following its decay; this allowed a precise determination of its half-life.
A similar situation is for $^{138}$La, whose signature of the decay is rather peculiar giving $\gamma$ rays of 1435.795 keV and
788.744 keV (see e.g. Ref. \cite{Bern:2005}), and for $^{176}$Lu, producing during its $\beta$ decay
$\gamma$ rays at 306.78, 201.83 and 88.34 keV.
On the contrary, $^{87}$Rb and $^{187}$Re are not accompanied by $\gamma$ radiation; however, they have large natural abundances
(see Table \ref{tab:beta_nucl}), that facilitated the study of their decay.

\subsection{The highly forbidden $\beta$ decay of $^{48}$Ca }
\label{ca48}

In the case of $^{48}$Ca, the double $\beta$ decay channel
competes with the ordinary single $\beta$ decay channels. In addition, 
the single $\beta$ decay of $^{48}$Ca can populate few excited levels of $^{48}$Sc
(see Fig.~\ref{fig:Ca1}).

\begin{figure}[ht]
\begin{center}
\resizebox{0.48\textwidth}{!}{ 
  \includegraphics{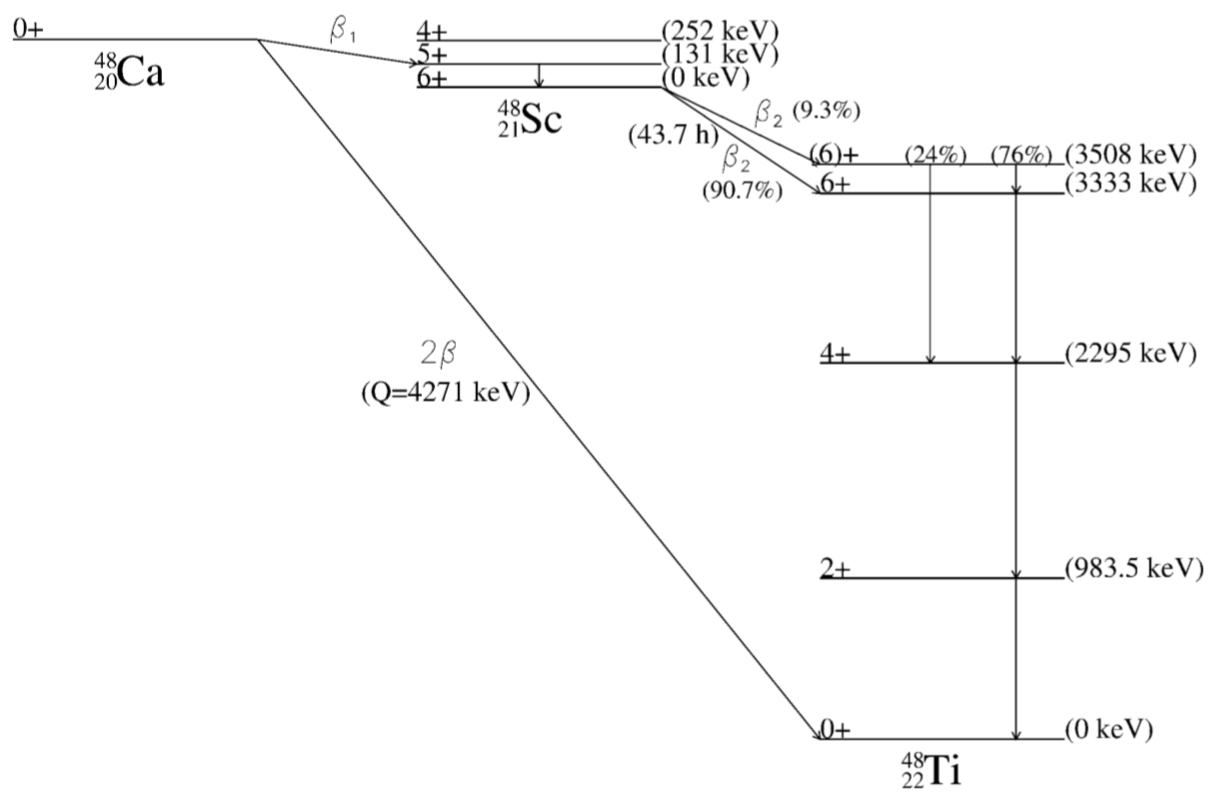}
}
\end{center}
\caption{The single (and double) $\beta$ decay of $^{48}$Ca. Figure taken from Ref. \cite{Ber02}.}
\label{fig:Ca1}
\end{figure}

The mechanism of the ordinary $\beta$ decay of $^{48}$Ca results from two 
features of the decay. Firstly, the large change in angular momentum 
($\Delta J =$ 4, 5, or 6) between the initial and final nuclear states, and secondly the energy release which is small. The experimentally measured $Q_\beta$ value for 
the 6$^{th}$-forbidden ground-state-to-ground-state transition is (279 $\pm$ 5) keV \cite{Wang:2017}.
With this value, the $Q_\beta$ values for the 4$^{th}$-forbidden branches are evaluated to be
 148 keV (5$^+$) and 27 keV (4$^+$) with an uncertainty of 5 keV each one. 
The probability of $\beta$ decay is determined by the change $\Delta J^{\Delta \pi}$ (an increase in $\Delta J$ by 1 unit leads to the increase of $T_{1/2}$ by $\sim 3$ orders of magnitude) 
and by the energy release ($T_{1/2} \sim 1/Q_\beta^5$). 
The transition to the 131 keV level is estimated as the most probable and, after some initial studies by Warburton \cite{War85} and by Aunola, Suhonen, and Siiskonen \cite{Aun99}, 
a value $T_{1/2} = (2.6 - 7.0) \times 10^{20}$ yr was estimated \cite{Haa14}. The process is not yet observed; however, it should be noted that the second order process -- two neutrino 2$\beta$ decay to $^{48}$Ti -- is faster in this case, as already underlined; it was already observed with $T_{1/2}$= 6.4 $\times$ 10$^{19}$ yr \cite{Arn16}.
The shape of $^{48}$Ca $\beta$ spectrum was calculated in Ref. \cite{Kos17}. 

The potentiality of the $\beta$--$\gamma$ coincidence technique was exploited to search for the ordinary highly forbidden $\beta$ decay (see e.g. Ref. \cite{Ber02}). 
In fact, e.g. after a $^{48}$Ca $\beta$ decay, a $\gamma$ quantum of 131 keV is expected from the excited nuclear level of $^{48}$Sc and, subsequently,  $\gamma$ quanta of 984, 1037 and 1312 keV are emitted when the daughter $^{48}$Sc decays to the excited nuclear levels of $^{48}$Ti.

In Ref. \cite{Ber02}, by using a 1.11 kg CaF$_2$(Eu) scintillator partially surrounded by low background 
NaI(Tl) detectors, the process was investigated by searching in the experimental data for the
$\beta$ decay of the daughter nuclei $^{48}$Sc to $^{48}$Ti with $T_{1/2}$ = 43.7 h. With large probability
(89.98\%) the decay goes to the excited 6$^+$ level of  $^{48}$Ti with excitation energy 3333 keV. 
In this case the emitted electron has energy up to 656 keV, and three $\gamma$ quanta are emitted 
as well with energies 1037, 1312 and 984 keV \cite{Burrow:2006}. In the remaining 10.02\% of the cases the 
daughter nucleus $^{48}$Sc decays to $^{48}$Ti  level with energy 3508 keV
 (electron energy up to 482 keV); the subsequent emitted $\gamma$ quanta are listed in Fig.~\ref{fig:Ca1}. Thus, the coincidence technique can be used to determine the half-life of the $\beta$ decay of $^{48}$Ca since the $^{48}$Ca--$^{48}$Sc chain is in equilibrium. With this procedure a limit of $T_{1/2}$ $>$ 2.4 $\times$ 10$^{18}$ yr (at 90\% C.L.) \cite{Ber02} was obtained for the
 highly forbidden  $\beta$ decay of $^{48}$Ca. 

However, the best experimental limit $T_{1/2} > 2.5 \times$ 10$^{20}$ yr was obtained with 
a 400 cm$^3$ HPGe detector and 24.558 g of enriched  $^{48}$CaF$_2$ powder, by looking for 
$\gamma$ rays following $\beta^-$ transitions of $^{48}$Ca \cite{Bak02}. The obtained 
limit is not far from the above mentioned theoretical estimations.

\subsection{Electron capture and $\beta$ decay of $^{50}$V }
\label{v50}

$^{50}$V can decay both through electron capture to $^{50}$Ti with energy release of
$Q(\varepsilon) = 2207.6(4)$~keV and through $\beta^-$ decay to $^{50}$Cr with
$Q(\beta^-) = 1038.06(30)$~keV \cite{Wang:2017} (see Fig. \ref{fig:50V}).
In both cases the change in spin for the g.s. to g.s. transitions
is $6^+$ $\to$ $0^+$, and transitions to the first $2^+$ excited levels of $^{50}$Ti (1553.8~keV) or
$^{50}$Cr (783.3~keV) are more probable (by orders of magnitude). 
With $\Delta J^{\Delta \pi} = 4^+$ these decays are classified as 4th-forbidden non-unique.
$^{50}$V is one of the only 3 nuclei where $\beta$ processes with 
$\Delta J^{\Delta \pi} = 4^+$ were measured (the other two are $^{113}$Cd and $^{115}$In). 

\begin{table*}
\caption{Summary of the investigations of $^{50}$V decay.}
\label{tab:50V}  
\begin{center}     
\begin{tabular}{llll}
\hline
\noalign{\smallskip}
Experimental technique                          & \multicolumn{2}{c}{$T_{1/2}$ (yr)}                                          & Year [Ref.] \\
~                                               & $\varepsilon$                         & $\beta^-$                           & ~ \\
\noalign{\smallskip}
\hline
\noalign{\smallskip}

Geiger counter (1200 cm$^2$ surface),           & $>3.0\times10^{15}$                   & $>3.0\times10^{14}$                 & 1955 \cite{Hei55} \\
Fe-V 12.7 kg (80\% V)                           & ~                                     & ~                                   & ~                 \\
\noalign{\smallskip}

Prop. counter, V$_2$O$_5$ $\simeq10$ mg/cm$^2$; & $(4.0\pm1.1)\times10^{14}$            & $>2.4\times10^{14}$                 & 1957 \cite{Glo57} \\
NaI 103 cm$^3$, V$_2$O$_5$ 110 g                & ~                                     & ~                                   & ~                 \\
\noalign{\smallskip}

Prop. counter, NaI 79 cm$^3$,                   & $(4.8\pm1.2)\times10^{14}$            & $-$                                 & 1958 \cite{Bau58} \\
V$_2$O$_5$ 500 g                                & ~                                     & ~                                   & ~                 \\
\noalign{\smallskip}

NaI 1173 cm$^3$, V$_2$O$_5$ 500 g               & $>8.0\times10^{15}$                   & $>1.2\times10^{16}$                 & 1961 \cite{McN61} \\
\noalign{\smallskip}

NaI 6260 cm$^3$, V 4998 g, 128 h                & $(8.9\pm1.6)\times10^{15}$            & $(1.8 \pm 0.6)\times10^{16}$        & 1962 \cite{Wat62} \\
\noalign{\smallskip}

NaI 824 cm$^3$, V 5000 g, $\simeq 10$ m w.e.    & $>9.0\times10^{16}$                   & $>6.9\times10^{16}$                 & 1966 \cite{Son66} \\
\noalign{\smallskip}

Ge(Li) 70 cm$^3$, V 4000 g, 1173 h              & $>8.8\times10^{17}$                   & $>7.0\times10^{17}$                 & 1977 \cite{Pap77} \\
\noalign{\smallskip}

HPGe 100 cm$^3$, V 4245 g, 3252 h               & $1.5^{+0.3}_{-0.7}\times10^{17}$      & $>4.3\times10^{17}$                 & 1984 \cite{Alb84} \\
\noalign{\smallskip}

HPGe 208 cm$^3$, V 100.6 g, 193 h,              & $1.2^{+0.8}_{-0.4}\times10^{17}$      & $>1.2\times10^{17}$                 & 1985 \cite{Sim85} \\
Windsor salt mine $\simeq 700$ m w.e.           & ~                                     & ~                                   & ~                 \\
\noalign{\smallskip}

3 HPGe 560 cm$^3$, V 337.5 g, 1109 h,           & $(2.05\pm0.49)\times10^{17}$          & $8.2^{+13.1}_{-3.1}\times10^{17}$   & 1989 \cite{Sim89} \\
Windsor salt mine $\simeq 700$ m w.e.           & ~                                     & ~                                   & ~                 \\
\noalign{\smallskip}

HPGe 357 cm$^3$, V 255.8 g, 2347 h,             & $(2.29\pm0.25)\times10^{17}$          & $>1.5\times10^{18}$                 & 2011 \cite{Dombrowski:2011} \\
Asse salt mine 1200 m w.e.                      & ~                                     & ~                                   & ~                 \\
\noalign{\smallskip}

HPGe $\simeq 400$ cm$^3$, V 815.5 g, 5774 h,    & $2.67^{+0.16}_{-0.18}\times10^{17}$   & $>1.9\times10^{19}$                 & 2019 \cite{Lau18} \\
LNGS 3600 m w.e.                                & ~                                     & ~                                   & ~                 \\
\noalign{\smallskip}

\hline
\noalign{\smallskip}
\end{tabular}
\end{center}     
\end{table*}

\begin{figure}[!ht]
\begin{center}
\resizebox{0.48\textwidth}{!}{\includegraphics{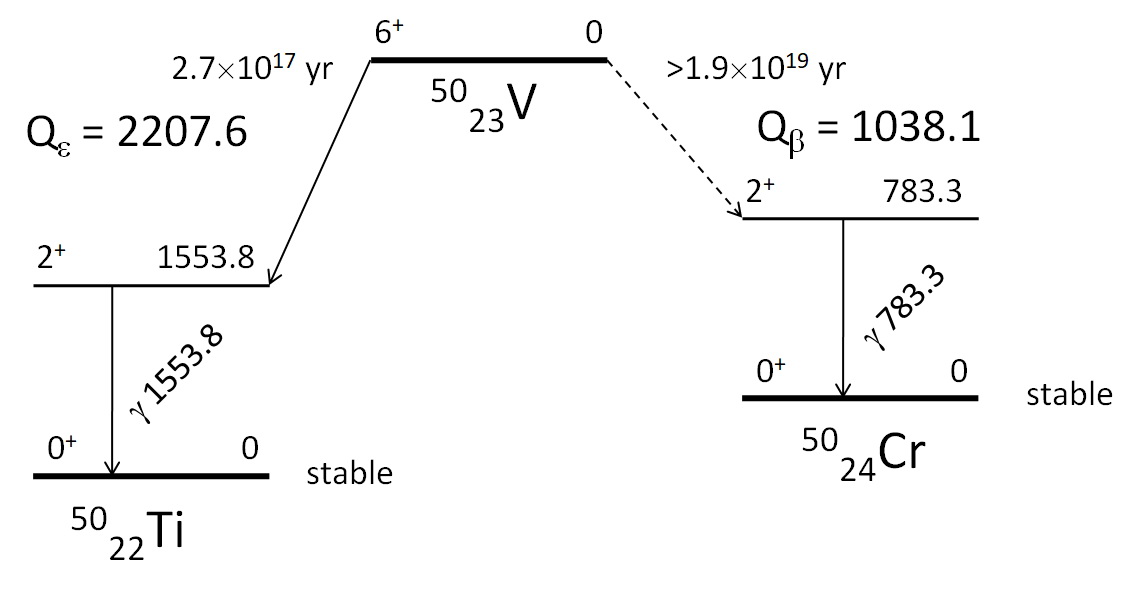}}
\end{center}
\caption{Scheme of the $^{50}$V decay. Energies are in keV. Data are from \cite{Ele11,Wang:2017,Lau18}.}
\label{fig:50V}
\end{figure}

The low natural abundance of $^{50}$V ($\delta = 0.250(10)\%$ \cite{Meija:2016}) 
and the large expected half-lives make its experimental investigations rather difficult. 
Different approaches were used to search for $^{50}$V decay.
On the first stages, Geiger counters and proportional counters with thin V-containing foils 
were applied to look for X rays emitted in result of electron capture 
and for electrons emitted in $\beta^-$ decay.
Later on, NaI, Ge(Li) and HPGe detectors and massive V sources were used to look for the 783 and 1554 keV 
$\gamma$ rays; these techniques were more sensitive than the previous ones. 

To our knowledge, the first search for the decay of $^{50}$V was performed in 1949 \cite{Lel49}; 
a thin-walled beta counter and also photographic plates were used; 
the activity was not observed ($T_{1/2}$ limits were not reported).
The summary of the subsequent experiments is given in Table~\ref{tab:50V}.
Early reports on positive observations of electron capture and $\beta^-$ at level of $T_{1/2} \sim 10^{14}-10^{16}$~yr 
\cite{Glo57,Bau58,Wat62} were not supported in later investigations. 
The detection of the 783~keV $\gamma$ peak expected in $^{50}$V $\beta^-$ decay by Simpson et al. in 1989 \cite{Sim89} 
with $T_{1/2}(\beta^-)=(8.2^{+13.1}_{-3.1})\times10^{17}$~yr was quite uncertain and not confirmed
in the subsequent investigations; the best known limit obtained in a recent experiment \cite{Lau18} is
$T_{1/2}(\beta^-) > 1.9\times10^{19}$~yr.

On the contrary, the 1554~keV $\gamma$ peak, expected in the electron capture branch of the $^{50}$V decay, is confidently observed 
in all experiments since 1984, with the recent, most accurate result of $T_{1/2}(\varepsilon)=(2.67^{+0.16}_{-0.18})\times10^{17}$~yr \cite{Lau18}.

The half-life of the $^{50}$V was calculated in nuclear shell model by Nishimura in 1979 \cite{Nis79} as:
$T_{1/2}(\varepsilon)=1.5\times10^{16}$~yr, 
$T_{1/2}(\beta^-)=9.04\times10^{17}$~yr,
which is in disagreement with the experimental values.
Recent calculations of Haaranen et al. \cite{Haa14b} for the electron capture branch:
$T_{1/2}(\varepsilon)=(5.13\pm0.07)[(3.63\pm0.05)]\times10^{17}$~yr given for $g_A=1.00[1.25]$
is in good agreement with the measured value \cite{Lau18}.
The predicted value for $\beta^-$ decay  
$T_{1/2}(\beta^-)=(2.34\pm0.02)[(2.00\pm0.02)]\times10^{19}$~yr \cite{Haa14b}
is not far from the best achieved limit $1.9\times10^{19}$~yr \cite{Lau18},
that gives hope to observe the effect in near future.

The shape of $^{50}$V $\beta^-$ spectrum was calculated in Ref. \cite{Kos17}.

Recently, an YVO$_4$ crystal (5~cm$^3$, $\oslash1.8\times2$~cm, mass of 22~g) was tested as  
cryogenic scintillating bolometer \cite{Pat18}; the light yield for beta/gamma particles
was measured to be 
$(59.4\pm0.1)$~keV/MeV and for alpha particles
$9-12$~keV/MeV (dependent on the energy). 
The full width at half of maximum (FWHM) for the 661 keV line of $^{137}$Cs was determined to be $(9\pm2)$~keV.
The sensitivity of the measurements of the YVO$_4$ scintillating bolometer with an additional detector 
registering the 783~keV $\gamma$ quantum was estimated at level of $T_{1/2} \sim 1\times10^{20}$~yr
which could allow the observation of $^{50}$V $\beta^-$ decay for the first time and the measurement of its $\beta^-$ shape.

\subsection{$^{96}$Zr $\beta$ decay}
\label{zr96}

The situation of $^{96}$Zr is analogous to the case of $^{48}$Ca
(the decay scheme is shown in Fig. \ref{Zr-96}): 
the single $\beta$ decay to $^{96}$Nb is possible,
but not yet observed ($T_{1/2} > 3.8 \times 10^{19}$ yr \cite{aa19} for the most probable transition to the 44 keV excited level), 
while $2\nu2\beta$ decay has been already measured with $T_{1/2} = (2.3 \pm 0.2) \times 10^{19}$ yr \cite{aa20}.
The shape of the $^{96}$Zr $\beta$ spectrum was calculated in Ref. \cite{Kos17}.

\begin{figure}[!ht]
\begin{center}
\resizebox{0.45\textwidth}{!}{ 
  \includegraphics{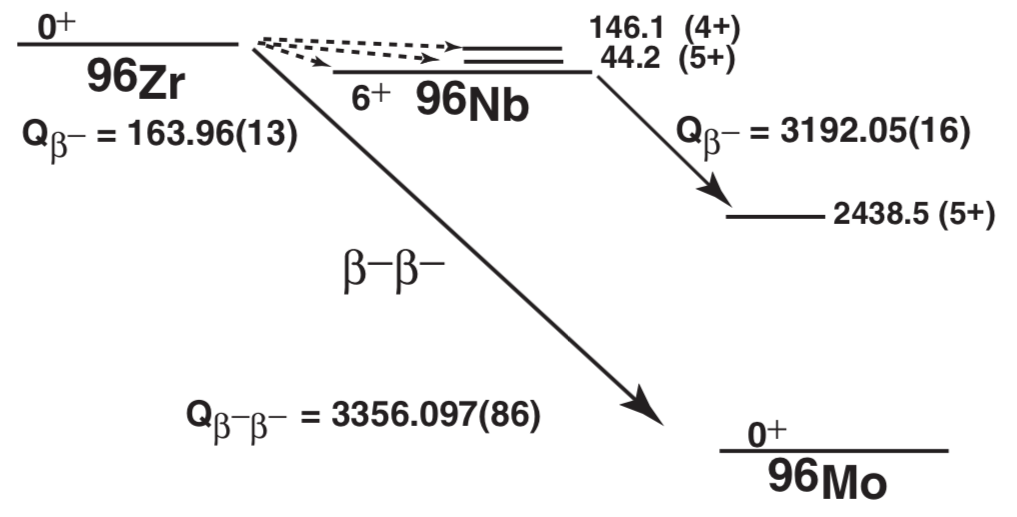}
}
\end{center}
\caption{Decay scheme for the A=96 triplet showing the energy position of $^{96}$Zr with respect to its neighbor $^{96}$Nb and $^{96}$Mo;
the energies are in keV. Figure taken from Ref. \cite{aa19_2}.}
\label{Zr-96}
\end{figure}

Recently single and double beta decay $Q$ values among the triplet $^{96}$Zr, $^{96}$Nb, and $^{96}$Mo have been determined in Ref. \cite{aa19_2}
by means of high-precision mass measurements using the JYFLTRAP mass spectrometer at the IGISOL facility of
the University of Jyv\"{a}skyl\"{a} \cite{aa19_3,aa19_4};
in particular, the energy of the decay has been measured as $Q_\beta(^{96}$Zr$) = 163.96(13)$ keV.
 
This single beta decay is of special importance being the main branch fourfold unique forbidden, which is an alternative decay path to the
$^{96}$Zr double beta decay; thus, its observation can provide one of the most direct tests of the neutrinoless
double beta decay nuclear-matrix-element calculations, as these
can be simultaneously performed for both decay paths with no further assumptions. In the latter reference the theoretical single beta
decay rate has been re-evaluated using a shell-model approach, which also indicates $^{96}$Zr single
beta decay life-time reachable in near future for an experimental verification; in fact, for the most probable transition to the 44 keV excited level the value:
$T_{1/2} = 11 \times 10^{19}$ yr, has been estimated.
The uniqueness of the decay also supports the relevance of a dedicated experiment which could allow an investigation into the origin of the 
quenching of the axial-vector coupling constant $g_A$.

\subsection{$^{113}$Cd $\beta$ decay}
\label{cd113}

The isotope $^{113}$Cd is one of only three nuclei (the two others are
$^{50}$V and $^{115}$In) with four-fold forbidden $\beta$ decays
that can be practically investigated. The high order of
forbiddenness, together with rather low energy of the decay:
$Q_{\beta}=323.83(27)$ keV \cite{Wang:2017}, leads to a large
half-life: in the range of $\sim10^{16}$ yr.

After a number of unsuccessful attempts since 1940 (see
\cite{Martell:1950,Watt:1962} and references in Ref.
\cite{Greth:1970}), the $\beta$ activity of $^{113}$Cd was
detected for the first time in 1970 \cite{Greth:1970} by using a
sample of cadmium oxide enriched in the isotope $^{113}$Cd
installed in a proportional counter. The experiment has measured
the $^{113}$Cd half-life as $T_{1/2} = (9.3\pm1.9) \times10^{15}$ yr.
The $\beta$ spectrum of $^{113}$Cd was observed for the first time
with the help of a small 0.27 cm$^3$ CdTe detector in Ref.
\cite{Mitchell:1988}, as reported in Fig. \ref{fig:cd113-b-cdte}.

\begin{figure}[htb]
\begin{center}
\resizebox{0.4\textwidth}{!}{\includegraphics{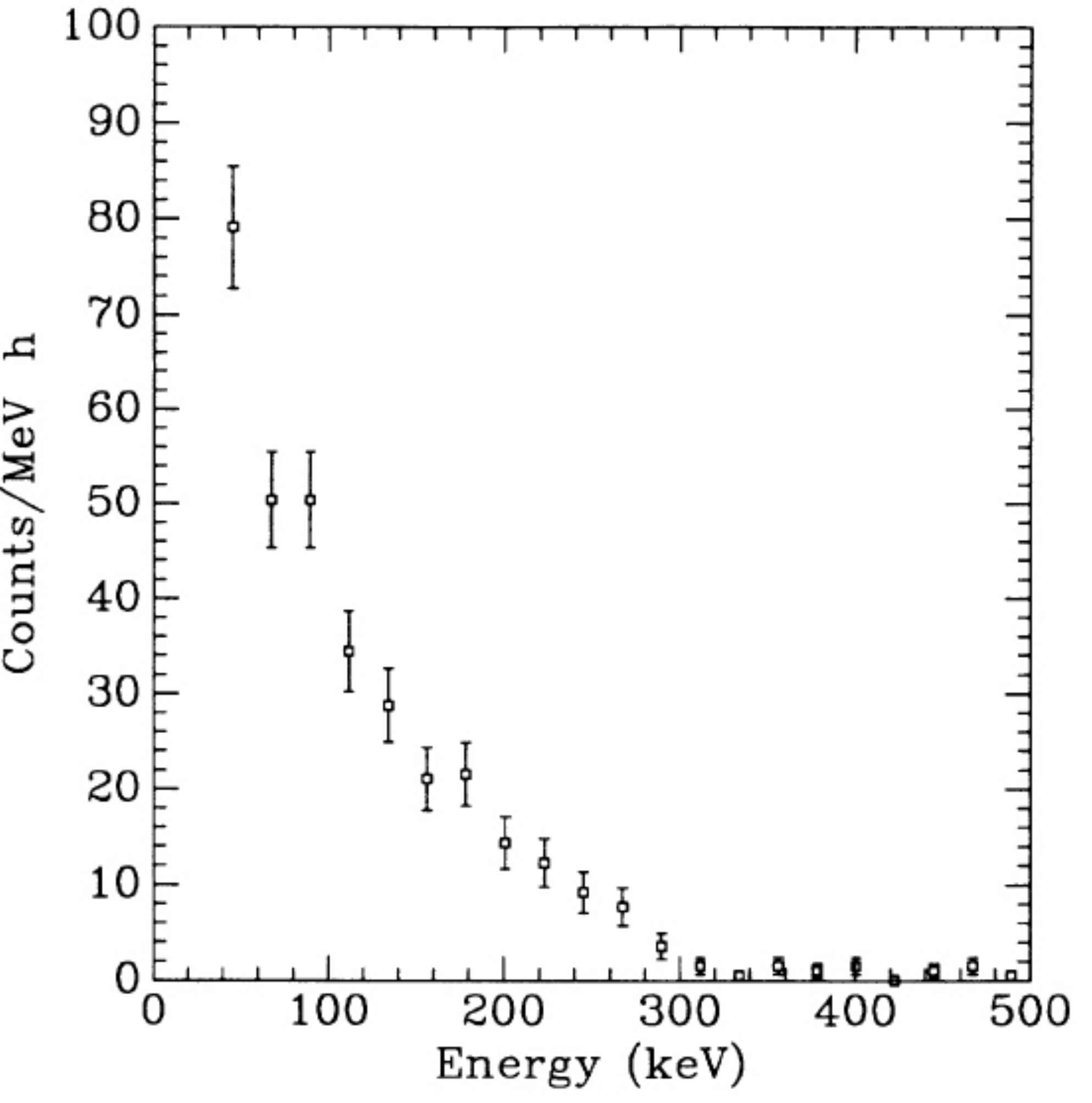}}
\end{center}
\caption{Low-energy spectrum from the CdTe detector 16 mm diameter
and 2 mm thick, with an active area of 1.33 cm$^2$, measured over
87.4 h live time. The increase in counts below 320 keV is ascribed
to the $\beta$ decay of $^{113}$Cd with half-life
$T_{1/2} \sim 10^{16}$ yr. Figure taken from Ref. \cite{Mitchell:1988}.}
\label{fig:cd113-b-cdte}
\end{figure}

Both the half-life and the spectral shape of $^{113}$Cd were
investigated by using CdWO$_4$ low temperature bolometer in Refs.
\cite{Alessandrello:1994a,Alessandrello:1994b}. Further
improvement of the accuracy was achieved
with a CdWO$_4$ crystal scintillator at the Solotvina underground
laboratory \cite{Danevich:1996a} and in the experiment
\cite{Goessling:2005} with a CdZnTe detector.

\begin{table*}[ht]
\caption{Half-lives (limits) relatively to $\beta$ decay of $^{113}$Cd. }
\begin{center}
\begin{tabular}{l|l|l}
 \hline
  $T_{1/2}$ (yr)                                                            & Experimental technique                                & Reference (year) \\
  \hline
 $>6\times10^{14}$                                                          & Cd sample in Geiger-M\"{u}ller counter                & \cite{Martell:1950} (1950) \\
  \hline
 $>1.3\times10^{15}$                                                        & CdO sample in proportional counter                    & \cite{Watt:1962} (1962) \\
  \hline
 $=(9.3\pm1.9)\times10^{15}$                                                & Enriched $^{113}$CdO sample in proportional counter   & \cite{Greth:1970} (1970) \\
  \hline
 $=(4-12)\times 10^{15}$                                                    & CdTe detector                                         & \cite{Mitchell:1988} (1988)\\
  \hline
 $=[9.3\pm0.5(\mathrm{stat})\pm1(\mathrm{syst})]\times 10^{15}$             & CdWO$_4$ LT bolometer                                 & \cite{Alessandrello:1994a,Alessandrello:1994b} (1994)\\
 \hline
 $=(7.7\pm0.3)\times 10^{15}$                                               & CdWO$_4$ scintillator                                 & \cite{Danevich:1996a} (1996)\\
 \hline
 $=[8.2\pm0.2(\mathrm{stat})^{+0.2}_{-1.0}(\mathrm{syst})]\times 10^{15}$   & CdZnTe detector                                       & \cite{Goessling:2005} (2005)\\
 \hline
$=(8.04\pm0.05)\times 10^{15}$                                              & CdWO$_4$ scintillator                                 & \cite{Belli:2007a} (2007)\\
 \hline

\end{tabular}
\end{center}
\label{tab:113cd}
\end{table*}

\begin{figure}[!ht]
\begin{center}
\resizebox{0.48\textwidth}{!}{\includegraphics{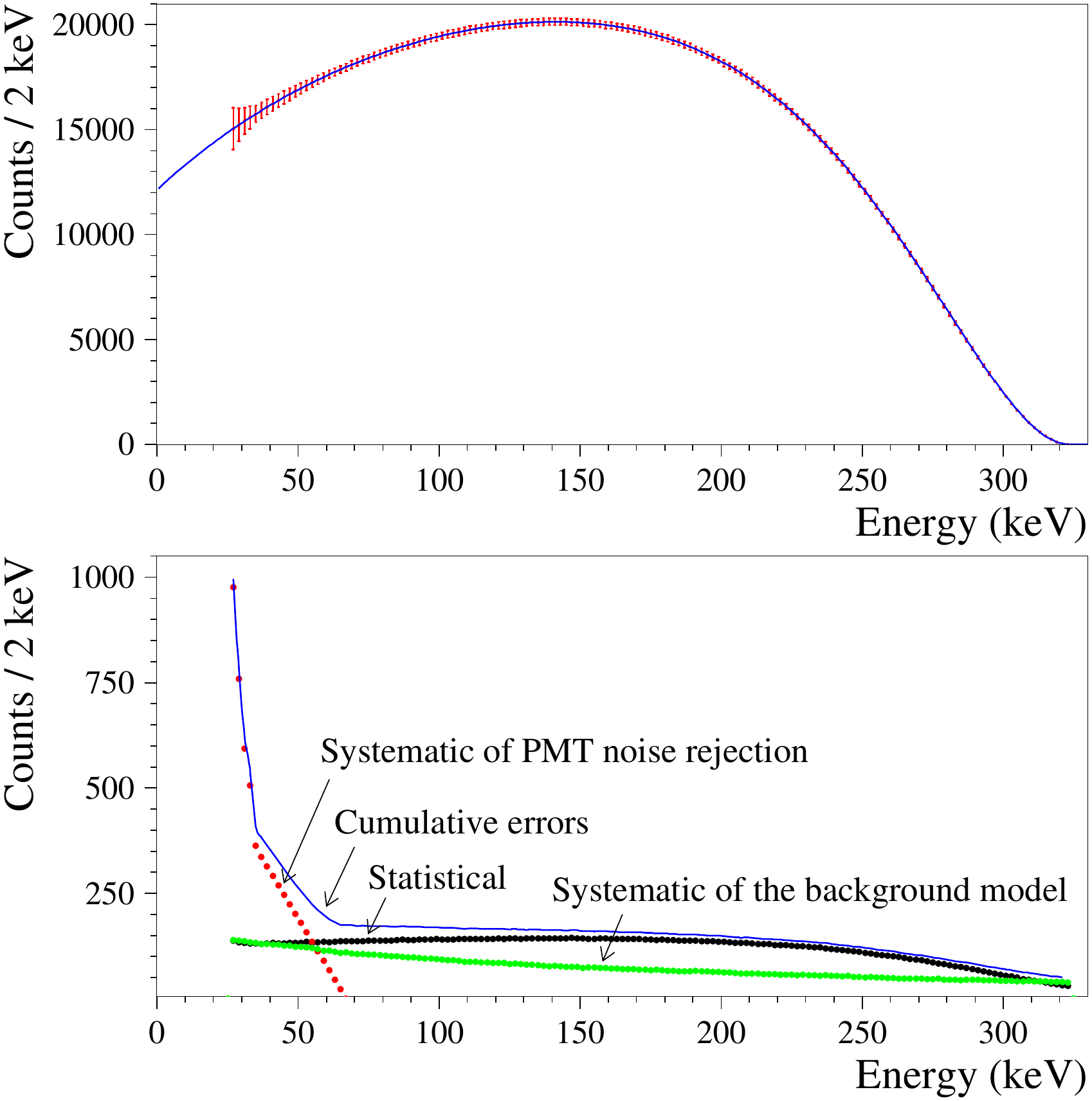}}
\end{center}
\caption{The analytical beta spectrum of $^{113}$Cd obtained after
de-convolution of the experimental $\beta$ spectrum
\cite{Belli:2007a}, with uncertainties (upper panel). Uncertainties of different
origin and the cumulative uncertainties of the spectrum are shown in the
upper panel (lower panel).}
\label{fig:cd113}
\end{figure}

The most accurate measurement of the $^{113}$Cd $\beta$ decay was
realized in the experiment \cite{Belli:2007a} at the LNGS underground
laboratory by using the same CdWO$_4$ crystal scintillator as 
in Ref. \cite{Danevich:1996a} (to minimize its cosmogenic activation,
the crystal has been stored underground at the Solotvina
laboratory after the experiment \cite{Danevich:1996a}, and then
was transported in a thick-wall iron container to the LNGS
laboratory). The shape of the spectrum was precisely measured, and
the half-life of $^{113}$Cd was determined as $T_{1/2} = (8.04 \pm
0.05) \times 10^{15}$ yr \cite{Belli:2007a}. The high accuracy of the experiment was
achieved thanks to the deeper experimental site, the better
discrimination capability between the photomultiplier noise and
the scintillation events, the larger exposure, and the lower level of
background. In addition, in the experiment \cite{Belli:2007a} the
energy resolution was slightly better than that in the Solotvina
experiment \cite{Danevich:1996a}. A summary of the experiments
for the $\beta$ decay of $^{113}$Cd is given in
Table \ref{tab:113cd}.

As noted in Refs. \cite{Haaranen:2016,Haaranen:2017,Kos17a}, the investigations of the spectral shape of non-unique 
forbidden $\beta$ decays have the potential to give an estimate of
the weak interaction coupling
constants $g_A$ and $g_V$ by comparing a theoretical shape with
the experimental spectrum. The approach provides an important
possibility to solve the $g_A$ problem of the neutrinoless double
beta decay
\cite{Barea:2013,DellOro:2014,Suhonen:2017a,Suhonen:2017b}.
However, the spectrum measured in \cite{Belli:2007a} and utilized
in the work \cite{Haaranen:2016} to estimate the $g_A$ constant,
was without systematic uncertainties which limited the capability of the
approach to find the most suitable nuclear models and to estimate
$g_A$. We have re-analyzed the data of the experiment
\cite{Belli:2007a} by adding the systematic uncertainties. Besides, we
have corrected the energy scale of the $\beta$ spectrum by taking
into account that the beta-decay energy measured in
\cite{Belli:2007a} ($Q_{\beta} = [344.9 \pm 0.2(\mathrm{stat}) \pm
21(\mathrm{syst})]$ keV) is substantially higher than the
current table value: $Q_{\beta}=323.83(27)$~keV \cite{Wang:2017}.
We ascribe this excess to the quenching of the scintillation
light efficiency of the CdWO$_4$ scintillator for gamma quanta
(used to calibrate the detector energy scale). The quenching is
a well known effect in scintillation detectors (see, e.g.,
\cite{Tretyak:2010,Belli:2014}). In case of $\beta$ particles
the effect is related to the non-proportionality in the
scintillation response \cite{Dorenbos:1995}, which is present also
in CdWO$_4$ crystal scintillators \cite{Bizzeti:2012}. Thus, we have
transformed the energy spectra measured in \cite{Belli:2007a} 
to obtain the beta-spectrum endpoint at the table value 
of $^{113}$Cd. The result is presented in Fig.
\ref{fig:cd113} together with the estimated systematic uncertainties. The
main sources of the systematic uncertainties are due to the PMT noise 
(the effect is essential at low energies), and
ambiguity of the background model subtracted from the experimental
data.

\subsection{$^{115}$In $\beta$ decay}
\label{in115}

\subsubsection{$^{115}$In $\beta$ decay to the ground state of $^{115}$Sn}
\label{in115_gs}

$^{115}$In is present in nature with a 
large abundance of $\delta  = 95.719(52)\%$ \cite{Meija:2016}. However, it is 
unstable in relation to beta decay to $^{115}$Sn, 
with $Q_\beta = (497.489 \pm 0.010)$~keV \cite{Wang:2017}. It has one of the 
longest half-lives of the observed single $\beta^-$ decays;
this is due to the large difference in spin between the parent ($9/2^+$) 
and the daughter nuclei ($1/2^+$ for the ground state of $^{115}$Sn).
With $\Delta J^{\Delta \pi}=4^+$, the decay is classified as 4-fold forbidden 
non-unique, similar to that of $^{113}$Cd. 

The $^{115}$In half-life was measured for the first time in 1950 as 
$T_{1/2}  = (6\pm2)\times10^{14}$~yr \cite{Martell:1950}
with the subsequent results as:
$\approx1\times10^{14}$~yr \cite{Coh51},
$(6.9\pm1.5)\times10^{14}$~yr \cite{Bea61},
$(5.1\pm0.4)\times10^{14}$~yr \cite{Watt:1962}.
Currently, the recommended value 
$T_{1/2}  = (4.41\pm0.25)\times10^{14}$~yr 
was obtained in 1978 \cite{Pfe78,Pfe79}; no results have been published since then.

While for $^{113}$Cd the shape of the $\beta$ spectrum was measured 
in a rather large number of works, on the contrary the shape of $^{115}$In 
was measured only in \cite{Pfe79}\footnote{A crude measurements of the 
$^{115}$In $\to$ $^{115}$Sn $\beta$ shape were done in \cite{Bea61}.}. 
This experiment, done with liquid scintillator (LS) loaded by In at 51.2 g/L, 
had a number of drawbacks: 
the measurements were performed at sea level (thus the background was quite high); 
the quenching of low energy electrons, which is rather strong for LS, was not taken into account; 
the energy resolution was not known exactly; 
and the energy threshold was quite high (around 50~keV). 
Recently, this decay was measured again in ultra-low background conditions 
with a scintillating LiInSe$_2$ bolometer
($(8 \times 15 \times 19)$ mm$^3$, 10.2 g) 
at the Modane underground laboratory \cite{Led18}.
The shape of the spectrum was obtained with $\simeq 20$~keV energy threshold;
a preliminary value of the half-life is 
$T_{1/2}  = (2.9\pm0.1(stat.))\times10^{14}$~yr. 
Theoretical calculations of the $^{115}$In $\beta$ shape can be found in Refs. \cite{Haaranen:2016,Haaranen:2017,Kos17a}.

\subsubsection{$^{115}$In $\rightarrow ^{115}$Sn$^*$ $\beta$ decay}
\label{in115-2}

Until 2005, the $^{115}$In $\to$ $^{115}$Sn decay was considered as 100\% 
g.s. to g.s. transition, but in 2005 also the decay to the first 
excited level of $^{115}$Sn ($E_{exc} = (497.334 \pm 0.022)$~keV \cite{Bla12}) 
was observed for the first time \cite{Cattadori:2005,Cat07}.
A sample of metallic indium (mass of 929 g) was measured for 2762 h with 4 HPGe detectors 
($\simeq 225$ cm$^3$ each) installed deep underground at LNGS. 
The branching ratio of the decay was estimated as $b.r. = (1.18 \pm 0.31)\times10^{-6}$
which corresponds to the partial half-life $T_{1/2} = (3.73 \pm 0.98)\times10^{20}$~yr 
\cite{Cattadori:2005}. 
The atomic mass difference $\Delta m_a$ between $^{115}$In and $^{115}$Sn 
(equal to the energy release $Q_\beta$ in $^{115}$In $\beta$ decay) at the time of 
measurements \cite{Cattadori:2005} was known with quite low accuracy: 
$Q_\beta = (499 \pm 4)$~keV \cite{Aud03}; 
however, in 2009 it was measured with extremely high accuracy of 10~eV: 
$Q_\beta = (497.489 \pm 0.010)$~keV \cite{Mou09}. 
Taking into account that the energy of the $^{115}$Sn first excited level was known as 
$E_{exc} = (497.334 \pm 0.022)$~keV \cite{Bla12}, the energy release in the decay 
$^{115}$In $\to$ $^{115}$Sn$^*$ was equal to 
$Q^*_\beta = (155 \pm 24)$~eV, that is the lowest known $Q_\beta$ value among the observed 
$\beta$ decays (the next one is $Q_\beta = (2.467 \pm 0.002$)~keV for $^{187}$Re \cite{Wang:2017}). 
The observation \cite{Cattadori:2005} was confirmed in measurements of In sample 
(2566 g) at the HADES underground laboratory with 3 HPGe detectors; 
slightly more precise values of half-life were obtained: 
$T_{1/2} = (4.1 \pm 0.6)\times10^{20}$~yr \cite{Wie09} and 
$T_{1/2} = (4.3 \pm 0.5)\times10^{20}$~yr \cite{And11}.

Very recently the energy of the excited $^{115}$Sn level was precisely remeasured 
in two experiments.
In the first one, a 30 mg Sn target enriched at 70\% in $^{114}$Sn
(the natural isotopic abundance of $^{114}$Sn is: $\delta(^{114}$Sn) = 0.66\% \cite{Meija:2016})
was irradiated by cold neutrons; the excited $^{115}$Sn level was populated in 
the $^{114}$Sn($n,\gamma$)$^{115}$Sn reaction. The result is: 
$E_{exc} = (497.316 \pm 0.007$)~keV, that leads to $Q^*_\beta = (173 \pm 12)$~eV \cite{Urb16}.

In the second one, a 230 mg Sn sample enriched in $^{115}$Sn to 50.7\% 
(natural isotopic abundance: $\delta(^{115}$Sn) = 0.34\%) was 
irradiated by a proton beam. The $^{115}$Sb \, radioactive \, isotope, \, created \, in 
the $^{115}$Sn(p,n)$^{115}$Sb reaction, decays with $T_{1/2} = 32$ min to $^{115}$Sn 
populating the 497~keV level with $\simeq 96$\% probability. The measurements
gave an even more precise value
$E_{exc} = (497.342 \pm 0.003)$~keV leading to $Q^*_\beta = (147 \pm 10)$~eV \cite{Zhe18}.
It is important to note that also the energy of the $^{115}$Cd 492~keV $\gamma$ quantum was measured in this experiment
as $E_\gamma = (492.350 \pm 0.003)$~keV, which is in excellent agreement with
the table value: $(492.351 \pm 0.004)$~keV \cite{Bla12}\footnote{~$^{115}$Cd was 
produced in the reaction $^{114}$Cd$(n,\gamma)^{115}$Cd with the Cd target enriched in 
$^{114}$Cd to 99\% (natural isotopic abundance: $\delta(^{114}$Cd) = 28.8\%).}.
This $^{115}$Cd $\gamma$ peak is at the distance of only 5~keV from the 497~keV peak of the $^{115}$Sn,
which increases the credibility of the obtained $Q^*_\beta$ value.

The decay scheme of $^{115}$In $\to$ $^{115}$Sn is shown in Fig. \ref{fig:115In}.

\begin{figure}[!ht]
\begin{center}
\resizebox{0.45\textwidth}{!}{\includegraphics{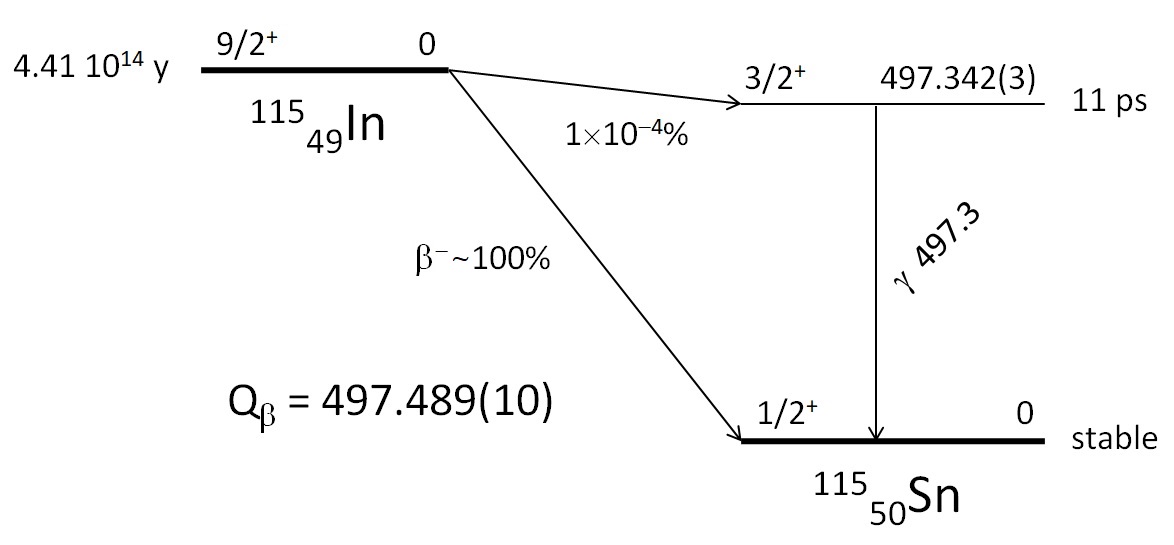}}
\end{center}
\caption{Scheme of the $^{115}$In decay. The energies are in keV. The data are from \cite{Mou09,Bla12,Zhe18}.}
\label{fig:115In}
\end{figure}

The half-life for the $^{115}$In $\to$ $^{115}$Sn$^*$ decay was calculated in \cite{Wie09,Mus10a}
as a function of the $Q_\beta^*$ value, however for $Q_\beta^* \simeq 150$~eV there is a
disagreement by $\simeq 1$ order of magnitude which could be related to atomic effects,
essential for such low energies and not accounted for in the calculations. 
The shape of the $^{115}$In $\to$ $^{115}$Sn$^*$ spectrum ($\Delta J^{\Delta \pi}=3^+$, 
classified as 2-fold forbidden unique) was calculated in \cite{Dvo11}. Deviations from
the theoretical shape could be used to limit (or to measure) the neutrino mass
\cite{Cattadori:2005,Cat07}, however, taking into account the high $T_{1/2}$ and the low 
$Q_\beta^*$, such measurements will be extremely difficult.

There are also other nuclides which are candidates for the lowest $Q_\beta^*$ value in
their decays to excited levels of daughter nuclei:
$^{77}$As, $^{79}$Kr, $^{109}$In, $^{111}$In, $^{115}$Cd, $^{131}$I, 
$^{135}$Cs, $^{146}$Pm, $^{149}$Gd, $^{155}$Eu, $^{159}$Dy, $^{161}$Ho, $^{188}$W, and 
$^{189}$Ir \cite{Mus10b,Mus11,Haa13,Suh14}
(see also the recent review \cite{suh19} and an additional list of candidates in \cite{gamage:2019});
these decays are not observed yet
and their $Q_\beta$ values are known currently with a low accuracy ($\simeq 1$~keV or worse).
Precise measurements of the atomic mass differences and possibly of the energies of the
excited levels are needed in all these cases.
Because of the evident restriction: $m_\nu \leq Q_\beta$, if -- luckily --
an extremely low $Q_\beta^*$ value will be measured for one of these candidates, say, $Q_\beta^* = 1$~eV, 
one can soon obtain an interesting limit on neutrino mass: $m_\nu \leq 1$~eV.

\subsection{$^{123}$Te electron capture}
\label{te123}

$^{123}$Te (a naturally occurring isotope of Te) has an isotopic abundance of (0.89 $\pm$ 0.03) \% \cite{Meija:2016} and a $Q$-value of 51.91(7) 
keV \cite{Wang:2017}. 
Because of the large spin difference between the parent and the daughter nuclei, the electron capture occurs primarily from the L and M shells \cite{Bia97,ToI98} while the capture from the K shell is suppressed by about three orders of magnitude.
 
It is a second forbidden unique electron capture. It should occur to the ground state of $^{123}$Sb.
 This process, although expected to occur, has never been experimentally observed. 
  Measurements in an old work \cite{Watt:1962} concentrating on the detection of the 26.1 keV photons of the K X-rays from $^{123}$Sb resulted in a 
half-life $T_{1/2}$ = (1.24 $ \pm$ 0.10) $\times 10^{13}$ yr. However, in Ref. \cite{Ale96} it was found that the real value is 6 orders of magnitude 
larger: $T_{1/2}$ = (2.4 $\pm 0.9) \times$ 10$^{19}$ yr. Later, also this result was found incorrect, 
and only a limit was set for electron capture of $^{123}$Te from the K shell as $T_{1/2} > 5.0 \times 10^{19}$ yr \cite{Ale03}. The observation of Ref. \cite{Ale96} 
was explained with the electron capture in $^{121}$Te; this unstable isotope was created in the TeO$_{2}$ crystals, used in the measurements, through neutron capture by $^{120}$Te while the 
crystals were at sea level. The natural abundance of $^{120}$Te is very small, $\delta$ = 0.09\% \cite{Meija:2016}, 
and this is a good demonstration how even a tiny effect may mimic another rare 
effect. Taking into account the predicted K electron capture branching ratio, the corresponding lower limit on the $^{123}$Te electron capture half-life is $T_{1/2}$ $>$ 9.2 $\times 
10^{16}$ yr, which can be theoretically interpreted only on the basis of a strong suppression of the nuclear matrix elements.
 
 Considering the improved sensitivity of present TeO$_2$ thermal detectors, this limit could be 
 improved orders of magnitude by CUORE apparatus \cite{Ald18,Bellini:2018}.
Being the source part of the active volume, a signal corresponding to the total binding energy of the captured electron can be measured with almost 100\% efficiency.
Moreover, the excellent energy resolution is a useful tool to distinguish externally generated $\gamma$ rays from X-rays and electron cascades.
For the (suppressed) K-capture a line at 30.5 keV is expected, while for the most probable L3 capture (L1 and L2 intensities are about one and three orders of magnitude lower than L3 \cite{ToI98}) the line should appear at 4.1 keV. Finally, the M transitions are all beyond any possible detection capability given their too small energy release (of the order of keV). In addition, in a high granularity array like CUORE, 
background events have a high chance of generating energy depositions in multiple crystals, while a genuine electron capture process 
is most likely a single-crystal event. Other possible backgrounds can come from the electron capture of $^{121}$Te and $^{121m}$Te, 
both generated via neutron activation of the naturally occurring $^{120}$Te. The neutron flux in the underground location of CUORE, however, 
is small enough to assume the in situ activation to be negligible, while the activity generated between the growth of the crystals and the storage 
underground is expected to have already decayed by many orders of magnitude (and could anyway be tagged thanks to its well-known time-dependence).

The experimental measurement of the $^{123}$Te half-life has important theoretical implications: very low rates are predicted if a strong (up to six orders of magnitude \cite{Bia97}) 
suppression of the nuclear matrix element is generated by the cancellation
between particle--particle and particle--hole correlations. An experimental study of this effect would be a severe
test of the nuclear models that are used to calculate the matrix elements for rare electroweak decays \cite{Civ01}.

An alternative search for the second forbidden electron capture of $^{123}$Te was performed
with CdZnTe semiconductor detectors \cite{Mus03} setting a lower half-life limit (for $\varepsilon$ from the K shell) 
of $T_{1/2}$ $> 3.2 \times$ 10$^{16}$ yr (95\% C.L.). The study was performed within a pilot project for the COBRA experiment. 
A more sensitive experiment could only be done with a setup using a larger amount of CdZnTe detectors.
 
 Although the decay schemes for $^{123}$Te and $^{123}$Sb are known,
  the mass difference of $^{123}$Te and $^{123}$Sb, which governs the half-life of $^{123}$Te, 
  was not well known. It was recently measured \cite{Fil16} with a direct and high precision measurement. 
 Since the expected mass difference, i.e. $Q$-value, between the parent and the daughter nucleus is close to the energetic region of the total electron binding energies for Te, accurate and precise data are required. 
 All the attempts of determining the half-life concern the ground state of neutral $^{123}$Te. 
 However, in stars nuclides are ionized leading to a certain charge-state distribution, which depends on the temperature of the stellar interior. The decay energy for the electron capture from ionic ground states is expected to be smaller than that for neutral states. This substantially increases the half-life of ionized $^{123}$Te. 
In addition, it was noted that in hot stellar conditions the nuclei with populated excited states can undergo electron capture, whose strength depends on the excitation energy, the mass 
difference between the neutral parent and daughter atoms, the ionic charge state, and the astrophysical conditions \cite{Cam59,Arn72,Yok83}. Such a possibility was verified \cite{Tak87} for 
long-lived beta-emitters. The mass difference between ionic states depends on the mass difference of the neutral atoms and the excitation energy of the nuclear state, which can be populated in the stellar environment. Accurate knowledge of the former is important for the capture process, especially for small decay energies close to the binding energy of K-electrons equal to 30.49 keV for the electron capture in $^{123}$Te.

\begin{table*}[ht]
\caption{Recent results on $^{180m}$Ta half-lives in the electron capture and $\beta^-$ channels. They have been obtained
by $\gamma$-spectrometry using the different detectors reported in the first column. Only the results obtained after 1980 are 
included in the Table; for a complete list see Ref. \cite{aa37}. The two experiments in Ref. \cite{Norman:1981,Cumming:1985} have been 
performed using enriched Tantalum.}
\begin{center}
\begin{tabular}{lcccc}
 \hline \hline
 Detector               &    \multicolumn{3}{c}{Lower half-life limits}                                          &    Year, reference  \\
                        &  $\varepsilon$              &   $\beta^-$                 &   Total                    &                     \\
 \hline
 Ge(Li), enriched Ta    &  $5.6  \times 10^{13}$ yr   &  $5.6  \times 10^{13}$ yr   &  $2.8 \times 10^{13}$ yr   &  1981 \cite{Norman:1981}  \\
 HPGe, enriched Ta      &  $3.0  \times 10^{15}$ yr   &  $1.9  \times 10^{15}$ yr   &  $1.2 \times 10^{15}$ yr   &  1985 \cite{Cumming:1985}  \\
 HPGe                   &  $1.7  \times 10^{16}$ yr   &  $1.2  \times 10^{16}$ yr   &  $7.2 \times 10^{15}$ yr   &  2006 \cite{Hult:2006}  \\
 Sandwich HPGe          &  $4.45 \times 10^{16}$ yr   &  $3.65 \times 10^{16}$ yr   &  $2.0 \times 10^{16}$ yr   &  2009 \cite{Hult:2009}  \\
 Sandwich HPGe          &  $2.0  \times 10^{17}$ yr   &  $5.8  \times 10^{16}$ yr   &  $4.5 \times 10^{16}$ yr   &  2017 \cite{aa37}  \\
 HPGe                   &  $2.0  \times 10^{17}$ yr   &  $1.7  \times 10^{17}$ yr   &  $9.0 \times 10^{16}$ yr   &  2018 \cite{chan:2018} \\
\hline
\end{tabular}
\end{center}
\label{tab:180ta}
\end{table*}

The basic processes involved in heavy element nucleosynthesis are 
the slow (s) and rapid (r) neutron capture processes, and, to a lower extent, the p process.
 The s-process forms nuclides along the ``valley of stability'', over relatively long time periods during the helium burning phase of stellar evolution. Tellurium is unique in that, it is the only element with three s-only isotopes ($^{122,123,124}$Te) and they offer a unique opportunity to test s-process 
 systematics \cite{Lae94}. 

In particular an investigation of the decay-branches, attributed to the electron capture from excited states -- impossible under terrestrial conditions --
can shed light on the production and depletion of tellurium isotopes in stars.
The role of the electron-capture process from excited states of  $^{123}$Te in stellar conditions is discussed in Ref. \cite{Fil16}. Based on reliable data the impact on the effective decay rate
of $^{123}$Te in stellar conditions has been investigated and revealed dramatic changes in electron capture probabilities of this pure s-process nuclide, which exceeds the terrestrial value by many orders of magnitude. Thus the terrestrially nearly stable nuclide $^{123}$Te becomes radioactive in stars. This can lead to a depletion of $^{123}$Te in the s-process nucleosynthesis.

\subsection{$^{180m}$Ta $\beta$ decay}
\label{ta180m}

It is interesting to note that $^{180m}$Ta has some peculiar features. In fact,
to our knowledge it is: 
i) the only nuclide present in nature in the isomeric state; 
ii) one of the rarest naturally occurring isotopes;
iii) the longest lived metastable state;
iv) the heaviest nuclide among the nine odd-odd nuclei present in nature.

Tantalum is present in nature in two isotopes, the stable $^{181}$Ta
and the excited level, $^{180m}$Ta ($\delta = 0.012 \%$ \cite{Meija:2016}), at $E_{exc} = 77$ keV. 
The ground state of $^{180}$Ta is unstable, since it quickly decays with $T_{1/2} \simeq 8.1$ h 
by electron capture and $\beta^-$ with branching ratios 85(3)\% and 15(3)\%, respectively.

\begin{figure}[ht]
\begin{center}
\resizebox{0.45\textwidth}{!}{\includegraphics{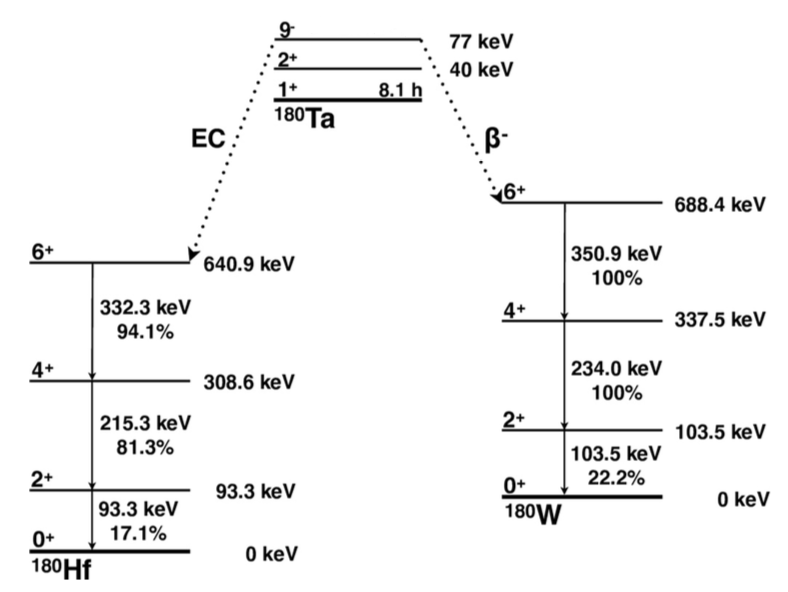}}
\end{center}
\caption{The decay schema of $^{180m}$Ta. Figure modified from Ref. \cite{aa37}.}
\label{fig:ta180_schema}
\end{figure}

$^{180m}$Ta has a high spin of $9^-$, while the ground state has a spin of $1^+$.
Due to the high spin difference, the isomeric transition to the ground state is very slow, and the decay of $^{180m}$Ta can also occur
by electron capture and $\beta^-$ decay into excited levels of $^{180}$Hf and $^{180}$W, respectively.
The decay scheme is shown in Fig. \ref{fig:ta180_schema}.
The transitions are towards the level with the lowest spin difference, that is the $6^+$ levels
of $^{180}$Hf and $^{180}$W. Afterward an electromagnetic cascade down to the ground state occurs.
The transitions can be classified as $\Delta J^{\Delta \pi} = 3^-$, so they are 3-fold forbidden non-unique (see Table \ref{tab:beta}).
The half-lives in relation to the isomeric transition (which occurs mainly by conversion electrons), the $\varepsilon$ and $\beta^-$ channels were calculated recently as:
$T_{1/2}$(IT) $= 8.0\times10^{18}$ yr,
$T_{1/2}(\varepsilon) = 1.4\times10^{20}$ yr,
$T_{1/2}(\beta^-) = 5.4\times10^{23}$ yr \cite{Eji17}, respectively.

The decay of $^{180m}$Ta (through electron capture and $\beta^-$) has not yet been observed; 
thus, $^{180m}$Ta is considered a quasi-stable nuclide. By the way, kinematically it can also decay
via $\alpha$ emission with $Q_\alpha = (2101.6\pm2.5)$ keV (see Table \ref{tab:alpha}), 
but this decay is strongly disfavoured compared with the $\varepsilon/\beta^-$ channels.

Several measurements have been performed in the past to look for electron capture and $\beta^-$ decay of $^{180m}$Ta; 
they are summarized in Table \ref{tab:180ta}.

Well established limits were set in the recent work of Ref. \cite{aa37} as: $T_{1/2}(\varepsilon) > 2.0 \times 10^{17}$ yr and 
$T_{1/2}(\beta^-) > 5.8 \times 10^{16}$ yr (90\% C.L.). The total half-life 
is $T_{1/2} > 4.5 \times 10^{16}$ yr (90\% C.L.) \cite{aa37}.

\begin{figure}[!t]
\begin{center}
\resizebox{0.32\textwidth}{!}{\includegraphics{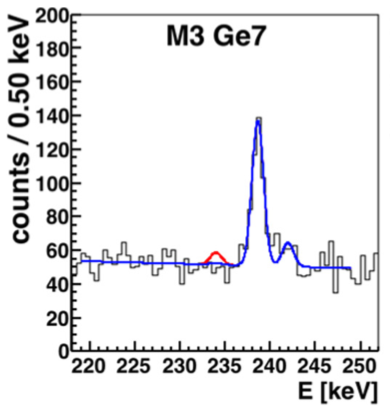}}
\resizebox{0.32\textwidth}{!}{\includegraphics{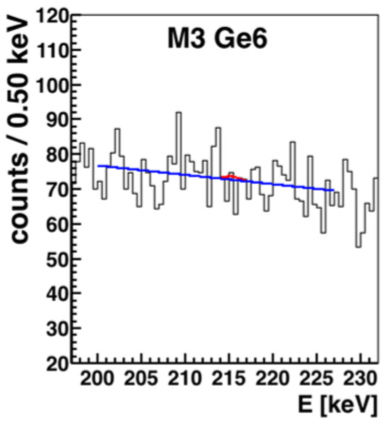}}
\resizebox{0.32\textwidth}{!}{\includegraphics{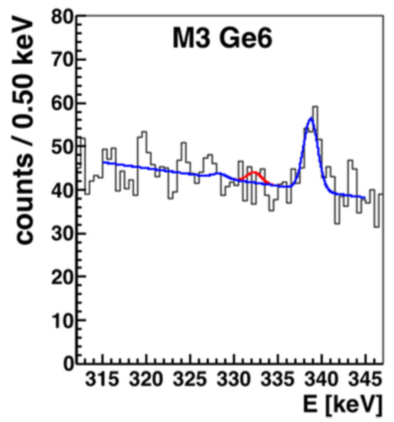}}
\end{center}
\caption{Three energy spectra measured by HPGe detectors used in the experimental set-up of Ref. \cite{aa37}.
{\it Top:} ROI for the $\beta^-$ channel of $^{180m}$Ta decay.
{\it Middle and bottom:} ROI for the electron capture channel.
The best fit is shown in blue (online) and the best fit with the signal peak set to the 90\% C.L. half-life limit 
is shown in red (online). Figures taken from Ref. \cite{aa37}.}
\label{fig:ta180_results}
\end{figure}

The measurements \cite{aa37} were performed in the HADES laboratory in Belgium,
where the atmospheric $\mu$ flux is reduced by a factor 5000 compared to the surface.
The used detectors were made of two coaxial HPGe detectors, Ge6 and Ge7, placed with
their end-caps faced each other (sandwich detector system). The sample, placed between the two HPGe detectors,
was made of six 2 mm thick disks with 100 mm diameter of high purity Tantalum of natural isotopic composition. 
The total mass was 1500.33 g, corresponding to 180 mg of $^{180m}$Ta. 

Some examples of the collected energy spectra accumulated in the total time of 176 days are reported in 
Fig. \ref{fig:ta180_results}. 
The $\beta^-$ channel was studied by the $\gamma$ peak at 234.0 keV;
two background peaks are also present in the fit region coming from contaminations of $^{212}$Pb and $^{224}$Ra.
As concerning the electron capture channel, it was studied by means of the $\gamma$ peaks at 215.4 keV and 332.3 keV;
also in this case two background $\gamma$ peaks from $^{228}$Ac are included in the fit. 
The other $\gamma$ peaks expected by $^{180m}$Ta decay were not used either for the presence of higher background or of lower detection efficiency.

Recently, the results from the Kamioka Underground Observatory (2700 m w.e.) were reported \cite{chan:2018}. 
In this experiment, a sample of Ta with mass of 863 g was measured with HPGe detector for 8597 h. Only limits were set as:
$T_{1/2}(\varepsilon) > 2.0\times10^{17}$ yr,
$T_{1/2}(\beta^-) > 1.7\times10^{17}$ yr,
and total $T_{1/2} > 9.0\times10^{16}$ yr (which are currently the best).

In conclusion, no signal was observed for either decay mode and new limits were set, as reported in Table \ref{tab:180ta}.
Improvements in the sensitivity of the experiments can be expected by using much lower background set-ups, in a deeper 
underground site and possibly with enriched samples.

\subsection{$^{222}$Rn $\beta$ decay}
\label{rn222}

The $^{222}$Rn isotope is known to undergo $\alpha$ decay in 100\% of the cases ($Q_\alpha = 5590$ keV, $T_{1/2} = 3.82$ d). 
However, it was recently noted \cite{Belli:2014} that single $\beta$ decay (1 FU) 
is also energetically possible with $Q_\beta = (24 \pm 21)$ keV\footnote{$^{222}$Rn
is an example of rare $\beta$ decay, despite its short half-life, because of a rather low branching ratio.}.
After the $\beta$ decay of $^{222}$Rn, one should observe a chain of $\alpha$ and $\beta$ decays 
which is different from that following its $\alpha$ decay. Looking for 
these chains in a BaF$_2$ crystal scintillator polluted by $^{226}$Ra, only a limit on the half-life 
of the $^{222}$Rn $\beta$ decay was set: $T_{1/2}(\beta) > 8.0$ yr \cite{Belli:2014}.
This search was performed using a 1.714 kg BaF$_2$ detector installed deep underground at LNGS in the low background 
DAMA/R\&D set-up. 

The $\beta$ decay to the ground state $^{222}$Rn(0$^+$) $\to$ $^{222}$Fr(2$^-$)
occurs with change in spin and parity: $\Delta J^{\Delta \pi}=2^-$, and thus it is
classified as first forbidden unique. 

The compilation of Log~$ft$ values in Ref. \cite{Sin98} gives the average value Log~$ft=9.5\pm0.8$
for all known 216 first forbidden unique $\beta$ decays.
Using the LOGFT tool at the National Nuclear Data Center \cite{logft}, the central value Log~$ft=9.5$ corresponds to a half-life
$4.8\times10^5$ yr for $Q_\beta=24$ keV; thus, considering the uncertainties in   
the $Q_\beta$ value, the half-life is 
$6.7\times10^4$ yr for $Q_\beta=45$ keV and
$2.4\times10^8$ yr for $Q_\beta=3$ keV.

The $\beta$ decay of $^{222}$Rn leads to a chain of subsequent decays:   
\begin{equation}\begin{split}
& ^{222}_{~86}\text{Rn}~\xrightarrow[24~\text{keV}]{\beta?~~\simeq4.8\times10^5~\text{y}}~
  ^{222}_{~87}\text{Fr}~\xrightarrow[2028~\text{keV}]{\beta~~14.2~\text{m}}~
  ^{222}_{~88}\text{Ra}~\xrightarrow[6679~\text{keV}]{\alpha~~38.0~\text{s}}~ \\
& ^{218}_{~86}\text{Rn}~\xrightarrow[7263~\text{keV}]{\alpha~~35~\text{ms}}~ 
  ^{214}_{~84}\text{Po}~\xrightarrow[7833~\text{keV}]{\alpha~~164.3~\mu\text{s}}~
  ^{210}_{~82}\text{Pb (22.3 y)}.
\end{split}\end{equation}

The first attempt to experimentally search for the $\beta$ decay
of $^{222}$Rn was performed in Ref. \cite{Belli:2014}; this is due, in particular, to the small half-life relatively to 
$\alpha$ decay
(3.82 d \cite{Sin11}), the low $Q_\beta$ value (typically below the energy threshold) and
the background level at low energies.
The live time was 101 h considering the  
quite-high activity of $^{226}$Ra (parent of $^{222}$Rn): 13.4 Bq in the used BaF$_2$ crystal.
The possibility of pulse-shape analysis to discriminate between $\alpha$ and $\beta/\gamma$
events,
and the knowledge of the expected energies and time differences between events
give the possibility to search for the peculiar chain in the accumulated data.
However, due to the high total event rate in the BaF$_2$ detector (75 counts/s), it was not feasible to look
for events with $T_{1/2}=14.2$ m ($^{222}$Fr $\beta$ decay) because of the many events of
other origin; at most, we should have restricted to the $\alpha$ decay of $^{222}$Ra with
$T_{1/2}=38.0$ s. Moreover, also the $\alpha$ decay of $^{214}$Po with $T_{1/2}=164.3~\mu$s could not be considered because of 
the 1.65 ms dead time in the measurements resulting in a very low efficiency
and in a low sensitivity. Thus, 
practically, only the decay chain: $^{222}$Fr $\to$ $^{222}$Ra $\to$ $^{218}$Rn
was considered by searching for the following sequence of events:
(1) an event in the energy window [30, 2207] keV ($^{222}$Fr $Q_\beta$¢ + FWHM$_\beta$) and with $\beta$¢ time shape;
(2) next event at [2109, 2623] keV ($^{222}$Ra $E_\alpha \pm $ FWHM$_\alpha$ in $\gamma$ scale), with $\alpha$ 
time shape and in the time interval [1.65 ms, 1.65 ms + $5 \times 38$ s];
(3) last event at [2398, 2946] keV ($^{218}$Rn $E_\alpha \pm$ FWHM$_\alpha$ in $\gamma$ scale), with $\alpha$
time shape and in the time interval [1.65 ms, 1.65 ms + $5 \times 35$ ms].

The energy spectrum of the last events in the chain is shown in Fig. \ref{fig222rn} - top;
it obviously contains also additional events of other origin due to the high counting rate
in the used BaF$_2$. 

\begin{figure}[htb]
\begin{center}
\resizebox{0.45\textwidth}{!}{ 
  \includegraphics{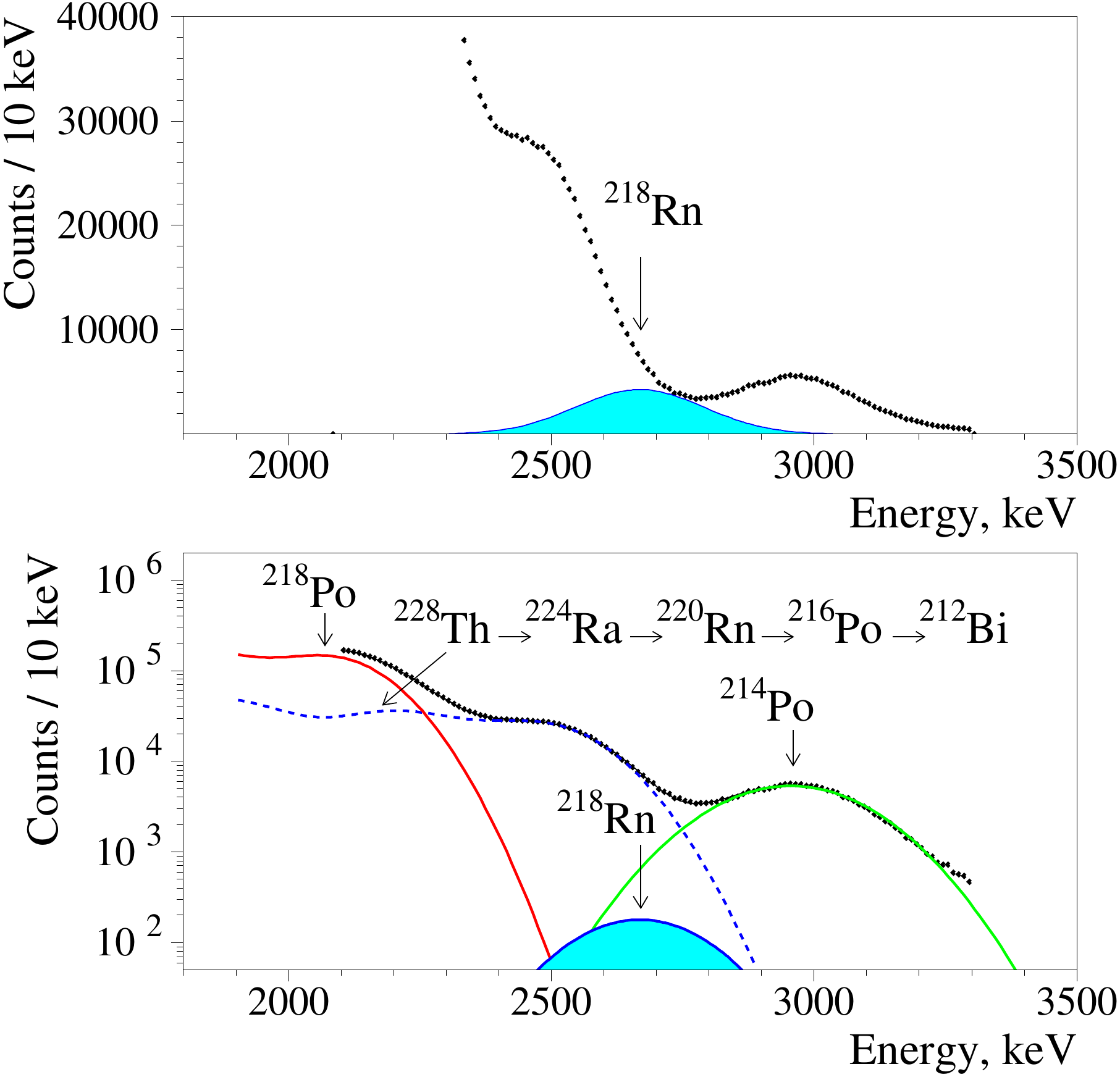}
}
\end{center}
\caption{(Color online) Top: Energy spectrum of possible events of $^{218}$Rn $\alpha$
decay in the chain $^{222}$Fr $\to$ $^{222}$Ra $\to$ $^{218}$Rn:
experimental data and maximal effect consistent with the data.
Bottom: Fit of the selected data by sum of the background model built from
$\alpha$ peaks of isotopes in U/Th chain together with the excluded $^{218}$Rn $\alpha$ peak.
Figure taken from Ref. \cite{Belli:2014}.}
\label{fig222rn}
\end{figure}

The limit on $T_{1/2}^\beta$ of $^{222}$Rn was conservatively calculated using the formula:
$$\lim T_{1/2}^\beta = \varepsilon \cdot t \cdot R^\alpha \cdot T_{1/2}^\alpha / \lim S,$$ 
\noindent and the values
$\lim S = 1.3\times10^5$,
$\varepsilon = 0.849$ (taking into account the time and energy intervals used and
the efficiency for the pulse-shape discrimination),
$t = 101$ h,
$R^\alpha = 13.4$ Bq (the same activity can be considered for $^{222}$Rn as for
$^{226}$Ra being $^{222}$Rn in equilibrium with $^{226}$Ra),
$T_{1/2}^\alpha = 3.8235$ d for $^{222}$Rn. The result was:
$T_{1/2}^\beta(^{222}$Rn$) > 122$ d at 90\% C.L.

A more realistic estimate was obtained by fitting
the spectrum with the sum of (1) some model that represents the background and (2)
a gaussian with known center and width that corresponds to the $^{218}$Rn peak searched for.
The isotopes in the $^{232}$Th and $^{238}$U chains randomly give contributions to the data
of Fig. \ref{fig222rn} due to the large time interval used in the selection ($\simeq5\times38$ s) and to the high
counting rate. Thus, one can build the background model from the $\alpha$ peaks of the
nuclei in the U/Th chains. The fit of the data in the energy interval
(2110 -- 3260) keV with this background model and the expected gaussian peak of $^{218}$Rn gave as area in the region of interest
$S = (3023 \pm 1476)$ counts
and, in accordance with the Feldman-Cousins procedure
\cite{Fel98}, the limit on the area was $\lim S = 5444$ counts at 90\% C.L., giving \cite{Belli:2014}:
$$T_{1/2}^\beta(^{222}\textrm{Rn}) > 8.0~\textrm{yr at 90\% C.L.}$$

\section{Prospects for future experiments and conclusions}
\label{prospects}

The most important requirements for the detectors used for the investigation of rare $\alpha$
and $\beta$ decays are: 1) high detection efficiency; 2) particle
discrimination ability; 3) low background; 4) high energy
resolution; 5) large amount of the isotope of interest. Moreover, the lowest possible energy 
threshold is requested for the $\beta$ decay studies.
It should be stressed that the characteristics 2 and 3 are, to
some extent, interrelated for $\alpha$ decay experiments: the particle
discrimination allows to suppress the background caused by $\gamma$
quanta ($\beta$ particles, cosmic muons) of $\alpha$ detectors.
However, the background caused by other $\alpha$ emitters should be
reduced by minimizing the naturally occurring $\alpha$ active
nuclides, and by applying $\alpha$ detectors with high
energy resolution. On the other hand, a low level of $\beta$ 
background is required in the $\beta$ decay experiments.

The development and use of scintillation detectors containing an isotope of interest look like
promising techniques to investigate rare $\alpha$ and $\beta$
decays, first of all thanks to their high detection efficiency,
to their particle discrimination ability and to their reasonable energy resolution.
For example, the $\alpha$ decay of the two naturally occurring thallium isotopes ($^{203}$Tl, $^{205}$Tl)
can be studied by massive ultra low background NaI(Tl) scintillators, which contain Tl as activator \cite{tretyak:2005}.
In addition, low temperature scintillating bolometers look like a
promising detection technique as well taking into account their better 
energy resolution, their low energy threshold and their particle
discrimination ability. A typical energy resolution of bolometers
is several keV and almost 100\% discrimination between $\alpha$
particles and $\beta$ particles ($\gamma$ quanta) could be achievable.
The recent observations of rare alpha decays (e.g., of $^{151}$Eu,
$^{180}$W, $^{209}$Bi) confirm this.

While low temperature scintillating bolometers look like a good
technique to investigate $\alpha$ and $\beta$ decays to the ground
states of daughter nuclei, HPGe $\gamma$ spectrometry is a favorable
technique to study the $\alpha$ and $\beta$ decays to excited levels
owing to their excellent energy resolution, the low background achieved
with the HPGe technique, and the possibility of long time operation
(see the review \cite{Laubenstein:2004} and the original works
\cite{Cattadori:2005,Belli:2011,Dombrowski:2011}).

The next issue that should be addressed is the use of
isotopically enriched materials for rare $\alpha$ and $\beta$ decay search.
In most of the cases the isotopes of interest have very low
isotopic abundances. Therefore, the production of crystals
scintillators or samples for HPGe $\gamma$ spectrometry
from isotopically enriched materials is 
requested for the next generation experiments aiming at the discovery of
$\alpha$ decay half-lives at the level of $T_{1/2} \sim
10^{19}-10^{26}$ yr. It should be noted that the production of
crystal scintillators from enriched isotopes is already well
developed for double beta decay experiments (see the review
\cite{Danevich:2012} and the original works
\cite{Belli:2010a,Barabash:2011,Alenkov:2011,Barabash:2014a,Grigorieva:2017,Armengaud:2017,Dafinei:2017a}).

An interesting experimental approach for the future could also be the study of the $\alpha$ decay rates 
as a function of the temperature.
There are some theoretical predictions that the $\alpha$ decay rate could change at low temperatures
(approximately of liquid helium) if atoms of radionuclides are incorporated into metal. The 
effect can be explained by the plasma model of Debye applied to the quasi-free electrons 
in the metallic samples ~\cite{Rai05,Ket25}. There is an indication of such an effect for alpha active $^{210}$Po \cite{Rai07,Spi07}. However, there 
are also theoretical considerations where the effect is considered to be very small, if present at all \cite{Zin07}. 
A  possible option of the experiment could be the measurement of the $\alpha$ decay of $^{190}$Pt 
to the excited level of daughter nuclide at room and at low temperature by using a low 
background HPGe detector. An advantage of such approach is the incorporation of $\alpha$ 
active platinum atoms in platinum metal, which exclude any surface location of radioactive 
atoms, or localization of radioactive atoms in dielectric (oxide, carbonate, etc.) micro 
inclusions, etc.
A better possibility to look for the dependence of alpha decay on temperature 
could be with a $^{226}$Ra source ($T_{1/2} = 1600$ yr). $^{226}$Ra decays with probability 
of $\approx 5.6$ \% to the 186.2 keV level of $^{222}$Rn, and this gamma peak is well-known
(usually it is present in all background spectra). Thus, using an external
target contaminated with $^{226}$Ra and changing its temperature, one
would be able to measure a weak change of the decay rate with a HPGe detector. 
This method avoids changing the temperature of the used detector.

\vspace{0.4cm}
In conclusion, in the recent past many interesting observations of rare $\alpha$ and $\beta$
decays, previously unseen due to their long half-lives ($10^{15} - 10^{20}$ yr) or small branching ratios ($10^{-3}-10^{-8}$)
have been achieved thanks to the improvements in
the experimental techniques, to the use of underground location and to
background suppression techniques.
The experimental status of such rare decays
has been described in this paper and future possibilities have also been underlined.


\begin{thebibliography}{999}

\bibitem{Bonet:1999}        R. Bonetti, A. Guglielmetti, in Heavy Elements and Related New Phenomena, edited by W. Greiner, R.K. Gupta, 
                            Vol. 2 (World Scientific, Singapore, 1999) p. 643.
\bibitem{Santhosh:2007}     K.P. Santhosh, I. Sukumaran, Eur. Phys. J. A \textbf{53}, 136 (2017).
\bibitem{Sandul:1984}       A. Sandulescu et al., Sov. J. Part. Nucl. \textbf{11}, 528 (1980). 
\bibitem{Rose:1984}         H.J. Rose, G.A. Jones, Nature \textbf{307}, 245 (1984).
\bibitem{Aleks:1984}        D.V. Aleksandrov et al., JETP Lett. \textbf{40}, 909 (1984).
\bibitem{Tretyak:2002}      V.I. Tretyak, Yu.G. Zdesenko, At. Data Nucl. Data Tables \textbf{80}, 83 (2002).
\bibitem{Saak:2013}         R. Saakyan, Annu. Rev. Nucl. Part. Sci. \textbf{63}, 503 (2013). 
\bibitem{Barabash:2015}     A.S. Barabash, Nucl. Phys. A \textbf{52}, 935 (2015).
\bibitem{Malam:2013}        J. Maalampi, J. Suhonen, Adv. High Energy Phys. \textbf{2013}, 505874 (2013).
\bibitem{Petters:1966}      B.G. Pettersson, in Alpha-, Beta-, and Gamma-Ray Spectroscopy, edited by K. Siegbahn (North-Holland, Amsterdam, 1966) p. 1569.
\bibitem{Ljub:1973}         A. Ljubicic, B.A. Logan, Phys. Rev. C \textbf{7}, 1541 (1973).
\bibitem{Arley:1938}        N. Arley, C.Møller, Kgl. Danske Videnskab. Selskab, Mat.-Phys. Medd. \textbf{15}, 9 (1938).
\bibitem{Tisza:1937}        L. Tisza, Phys. Z. Sowjetunion \textbf{11}, 245 (1937).
\bibitem{Greens:1956}       J.S. Greenberg, M. Deutsch, Phys. Rev. \textbf{102}, 415 (1956).
\bibitem{Bernabei:2013}     R. Bernabei et al., Eur. Phys. J. A \textbf{49}, 64 (2013).

\bibitem{rut99}             E. Rutherford, Philos. Mag. (Ser. 5) \textbf{47}, 109 (1899).

\bibitem{Gam1928}           G. Gamow, Z. Phys. \textbf{51}, 204 (1928).
\bibitem{Gur1928}           R.W. Gurney, E.U. Condon, Nature \textbf{122}, 439 (1928).
\bibitem{Gei1911}           H. Geiger, J.M. Nuttall, Phil. Mag. ser. 6 \textbf{22}, 613 (1911).
\bibitem{Ang2005}           G. Angloher et al., Astropart. Phys. \textbf{23}, 325 (2005).

\bibitem{den09}             V.Yu. Denisov, A.A. Khudenko, At. Data Nucl. Data Tables \textbf{95}, 815 (2009); erratum ibid. \textbf{97}, 187 (2011).
\bibitem{poe12}             D.N. Poenaru, R.A. Gherghescu, W. Greiner, J. Phys. G \textbf{39}, 015105 (2012).
\bibitem{qia14}             Y. Qian, Z. Ren, Phys. Lett. B \textbf{738}, 87 (2014).
\bibitem{Den15}             V.Yu. Denisov, O.I. Davidovskaya, I.Yu. Sedykh, Phys. Rev. C \textbf{92}, 014602 (2015).
\bibitem{san15}             K.P. Santhosh, I. Sukumaran, B. Priyanka, Nucl. Phys. A \textbf{935}, 28 (2015).
\bibitem{ash16}             N. Ashok, D.M. Joseph, A. Joseph, Mod. Phys. Lett. A \textbf{31}, 1650045 (2016). 
\bibitem{sah16}             B. Sahu, S. Bhoi, Phys. Rev. C \textbf{93}, 044301 (2016).
\bibitem{zha17}             S. Zhang et al., Phys. Rev. C \textbf{95}, 014311 (2017).
\bibitem{akr17}             D.T. Akrawy, D.N. Poenaru, J. Phys. G \textbf{44}, 105105 (2017).
\bibitem{qia11}             Y. Qian and Z. Ren, Phys. Rev. C \textbf{84}, 064307 (2011).
\bibitem{qia11b}            Y. Qian, Z. Ren, D. Ni, J. Phys. G: Nucl. Part. Phys. \textbf{38}, 015102 (2011).
\bibitem{hof15}             S. Hofmann, J. Phys. G \textbf{42}, 114001 (2015).
\bibitem{oga17}             Yu.Ts. Oganessian, A. Sobiczewski, G.M. Ter-Akopian, Phys. Scripta \textbf{92}, 023003 (2017).


\bibitem{Buck:1991}     B. Buck, A.C. Merchant, S.M. Perez, Mod. Phys. Lett. A \textbf{06}, 2453 (1991).
\bibitem{Ni:2010}       D. Ni, Z. Ren, Phys. Rev. C \textbf{81}, 064318 (2010).
\bibitem{Ni:2012}       D. Ni, Z. Ren, Nucl. Phys. A \textbf{893}, 13 (2012).
\bibitem{Chow:2008}     P.R. Chowdhury, C. Samanta, D.N. Basu, Phys. Rev. C \textbf{77}, 044603 (2008).
\bibitem{Sama:2007}     C. Samanta, P.R. Chowdhury, D.N. Basu, Nucl. Phys. A \textbf{789}, 142 (2007).
\bibitem{Wang:2010}     Y.Z. Wang et al., Phys. Rev. C \textbf{81}, 067301 (2010).
\bibitem{Denis:2005}    V.Yu. Denisov, H. Ikezoe, Phys. Rev. C \textbf{72}, 064613 (2005).
\bibitem{Denis:2010}    V.Yu. Denisov, A.A. Khudenko, Phys. Rev. C \textbf{81}, 034613 (2010).
\bibitem{Cob:2012}      A. Coban, O. Bayrak, A. Soylu, I. Boztosun, Phys. Rev. C \textbf{85}, 044324 (2012).
\bibitem{Pahl:2013}     M.R. Pahlavani, S.A. Alavi, N. Tahanipour, Mod. Phys. Lett. A \textbf{28}, 1350065 (2013).
\bibitem{Ghod:2016}     O.N. Ghodsi, E. Gholami, Phys. Rev. C \textbf{93}, 054620 (2016).
\bibitem{Ism:2017}      M. Ismail, W.M. Seif, A. Adel, A. Abdurrahman, Nucl. Phys. A \textbf{958}, 202 (2017).
\bibitem{Bohr:1998}     A. Bohr, B. Mottelson, {\it Nuclear Structure}, vol. 2: {\it Nuclear Deformations}, World Scientific, Singapore, 1998.
\bibitem{Saxe:1986}     A. Saxena, V.S. Ramamurthy, Pramana J. Phys. \textbf{27}, 15 (1986).
\bibitem{Ishk:2005}     B.S. Ishkhanov, V.N. Orlin, Phys. At. Nucl. \textbf{68}, 1352 (2005).
\bibitem{Scam:2013}     G. Scamps et al., Phys. Rev. C \textbf{88}, 064327 (2013).
\bibitem{Adam:2014}     G.G. Adamian, N.V. Antonenko, L.A. Malov, G. Scamps, D. Lacroix, Phys. Rev. C \textbf{90}, 034322 (2014).
\bibitem{Dahmar:2017}   S. Dahmardeh et al., Nucl. Phys. A \textbf{963}, 68 (2017).

\bibitem{Qi2019}        C. Qi, R. Liotta, R. Wyss, Prog. Part. Nucl. Phys. \textbf{105}, 214 (2019).

\bibitem{bel18}         P. Belli et al., {\it Search for $\alpha$ decay of naturally occurring osmium isotopes to excited levels of daughter nuclei}, to be published.
\bibitem{Buc91}         B.~Buck, A.C.~Merchant, S.M.~Perez, J. Phys. G \textbf{17}, 1223 (1991).
\bibitem{Buc92}         B.~Buck, A.C.~Merchant, S.M.~Perez, J. Phys. G \textbf{18}, 143 (1992).
\bibitem{Poe83}         D.N. Poenaru, M. Ivascu, J. Physique \textbf{44}, 791 (1983).

\bibitem{Hey99}         K.~Heyde, {\it Basic Ideas and Concepts in Nuclear Physics}, 2nd ed., IoP Bristol, (1999).

\bibitem{geoneu}        G. Fiorentini, M. Lissia, F. Mantovani, Phys. Rep. \textbf{453}, 117-172 (2007).

\bibitem{Jenkner:1949}  K.~Jenkner, E.~Broda, Nature \textbf{164}, 412 (1949).
\bibitem{Faraggi:1951}  H.~Faraggi, A.~Berthelot, C. R. Acad. Sci. \textbf{232}, 2093 (1951).
\bibitem{Riezler:1952}  W.~Von~Riezler, W.~Porschen, Z. Naturforsch. A \textbf{7}, 634 (1952).

\bibitem{Meija:2016}    J.~Meija et al., Pure Appl. Chem. \textbf{88}, 293 (2016).
\bibitem{Wang:2017}     M.~Wang et al., Chin. Phys. C \textbf{41}, 030003 (2017).

\bibitem{bel03}         P. Belli et al., Nucl. Instrum. Meth. A \textbf{498}, 352 (2003).

\bibitem{Sonz:2001}     A.A. Sonzogni, Nuclear  Data  Sheets \textbf{93},  599  (2001).
\bibitem{nd144_04}      A. Isola, M. Nurmia, Z. Naturforsch. \textbf{20a}, 541 (1965).
\bibitem{Stengl:2015}   C. Stengl, H. Wilsenach, K. Zuber, Int. J. Mod. Phys. E \textbf{24}, 1550043 (2015).
\bibitem{wil17}         H. Wilsenach et al., Phys. Rev. C \textbf{95}, 034618 (2017).

\bibitem{cas16}         N. Casali et al., J. Low Temp. Phys. \textbf{184}, 952 (2016).
\bibitem{aud17}         G. Audi et al., Chinese Phys. C \textbf{41}, 030001 (2017).
\bibitem{cas14}         N. Casali et al., J. Phys. G \textbf{41}, 075101 (2014).
\bibitem{dan12}         F.A. Danevich et al., Eur. Phys. J. A \textbf{48}, 157 (2012).

\bibitem{Macfarlane:1961}   R.D.~Macfarlane, T.P.~Kohman, Phys. Rev. \textbf{121}, 1758 (1961).
\bibitem{Riezler:1958}      W. Von Riezler, G. Kauw, Z. Naturforsch. A \textbf{13a}, 904 (1958).
\bibitem{Porschen:1956}     W.~Porschen, W.~Riezler, Z. Naturforsch. A \textbf{11a}, 143 (1956).
\bibitem{Munster:2014a}     A.~M\"{u}nster et al., JCAP \textbf{05}, 018 (2014).
\bibitem{Cozzini:2004a}     C.~Cozzini et al., Phys. Rev. C \textbf{70}, 064606 (2004).

\bibitem{pet14}         S.T.M. Peters et al., Earth and Planetary Sci. Lett. \textbf{391}, 69 (2014).
\bibitem{vio75}         V.E. Viola et al., J. Inorg. Nucl. Chem. \textbf{37}, 11 (1975).

\bibitem{Brau17}        M. Braun et al., Phys. Lett. B \textbf{768}, 317 (2017).
\bibitem{Belli:2011}    P.~Belli et al., Phys. Rev. C \textbf{83}, 034603 (2011).
\bibitem{Kau66}         G. Kauw, Forschungsber. Landes Nordrhein-Westfalen (1966) (No. 1640).
\bibitem{bee13}         J.W. Beeman et al., Eur. Phys. J. A \textbf{49}, 50 (2013).
\bibitem{Beeman:2012a}  J.W.~Beeman et al., Phys. Rev. Lett. \textbf{108}, 062501 (2012).

\bibitem{Hincks:1958}        E.P.~Hincks, C.H.~Millar, G.C.~Hanna, Can. J. Phys. \textbf{36}, 231 (1958).
\bibitem{Carvalho:1972}      H.G. de Carvalho, M. de Araujo Penna, Lett. Nuovo Cim. \textbf{3}, 720 (1972).
\bibitem{Norman:2000}        E.B.~Norman et al., Bull. Am. Phys. Soc. \textbf{45}, 30 (2000).
\bibitem{Marcillac:2003a}    P.~de~Marcillac et al., Nature \textbf{422}, 876 (2003).

\bibitem{Tretyak:2010}       V.I. Tretyak, Astropart. Phys. \textbf{33}, 40 (2010).

\bibitem{Coo04}           D. Cook et al., Geochim. Cosmochim. Acta \textbf{68} (6), 1413 (2004).
\bibitem{Bra06}           A. Brandon et al., Geochim. Cosmochim. Acta \textbf{70} (8), 2093 (2006).
\bibitem{Hof21}           G. Hoffman, Z. Phys. \textbf{7}, 254 (1921).

\bibitem{Tav06}           O.P. Tavares, M.L. Terranova, E.L. Medeiros, Nucl. Instrum. Methods B \textbf{243}, 256 (2006).
\bibitem{Bag03}           C.M. Baglin, Nucl. Data Sheets \textbf{99}, 1 (2003).
\bibitem{ToI98}           R.B. Firestone et al., Table of Isotopes, 8th edition (John Wiley \& Sons, New York, 1996) and CD update (1998).
\bibitem{EGS4}            W.R. Nelson et al., SLAC-Report-265, Stanford, 43 (1985). 
\bibitem{GEANT4}          S. Agostinelli et al., Nucl. Instrum. Methods A \textbf{506}, 250 (2003).
                          J. Allison et al., IEEE Trans. Nucl. Sci.  \textbf{53}, 270 (2006);
                          J. Allison et al., Nucl. Instrum. Meth. A \textbf{835}, 186 (2016).

\bibitem{Boh2005}         J.K. Bohlke et al., J. Phys. Chem. Ref. Data \textbf{34}, 57 (2005).
\bibitem{WWW}             WWW Table of Radioactive Isotopes, \\
                          http://\-nu\-clear\-data.nu\-clear.lu.se/nu\-clear\-data/toi/radSear\-ch.asp.
\bibitem{Mar92}           C.J. Martoff, P.D. Lewin, Comp. Phys. Comm. \textbf{72}, 96 (1992).
\bibitem{Fel98}           G.J. Feldman, R.D. Cousins, Phys. Rev. D \textbf{57}, 3873 (1998).

\bibitem{Tav11}           O.A.P. Tavares, E.I. Medeiros, Phys. Scripta \textbf{84}, 045202 (2011).
\bibitem{Ni011}           D. Ni, Z. Ren, Phys. Rev. C \textbf{84}, 037301 (2011).

\bibitem{bag10}           C.M. Baglin, Nucl. Data Sheets \textbf{111}, 275 (2010).
\bibitem{spe76}           R.F. Sperlein, R.L. Wolke, J. Inorg. Nucl. Chem. \textbf{38}, 27 (1976).

\bibitem{Med06}           E.L. Medeiros et al., J. Phys. G \textbf{32}, B23 (2006).
\bibitem{Gan09}           G. Gangopadhyay, J. Phys. G \textbf{36}, 095105 (2009).
\bibitem{Ism17}           M. Ismail et al., Int. J. Mod. Phys. E \textbf{26}, 1750026 (2017).
\bibitem{Wu18}            S.X. Wu, Y.B. Qian, Z.Z. Ren, Phys. Rev. C \textbf{97}, 054316 (2018).

\bibitem{bel13}           P. Belli et al., Eur. Phys. J. A \textbf{49}, 24 (2013).

\bibitem{Porschen:1953}   W.~Porschen, W.~Riezler, Z. Naturforsch. \textbf{8a}, 502 (1953).
\bibitem{Danevich:2003a}  F.A.~Danevich et al., Phys. Rev. C \textbf{67}, 014310 (2003).

\bibitem{Beard:1960a}     G.B.~Beard, W.H.~Kelly, Nucl. Phys. \textbf{16}, 591 (1960).
\bibitem{Georgadze:1995a} A.Sh. Georgadze et al., JETP Lett. \textbf{61}, 882 (1995).
\bibitem{Zdesenko:2005a}  Yu.G.~Zdesenko et al., Nucl. Instr. Meth. A \textbf{538}, 657 (2005).
\bibitem{Belli:2011a}     P.~Belli et al., Nucl. Instr. Meth. A \textbf{626-627}, 31 (2011).
\bibitem{Belli:2011b}     P.~Belli et al., J. Phys. G \textbf{38}, 115107 (2011).
\bibitem{Danevich:2018a}  F.A.~Danevich, V.I.~Tretyak, Int. J. Mod. Phys. A \textbf{33}, 1843007 (2018).
\bibitem{Danevich:2005a}  F.A.~Danevich et al., Nucl. Instr. Meth. A \textbf{544}, 553 (2005).

\bibitem{Ach09}           E. Achterberg, O.A. Capurro, G.V. Marti, Nucl. Data Sheets \textbf{110}, 1473 (2009).
\bibitem{Kli80}           J. Van Klinken et al., Nucl. Phys. A \textbf{339}, 189 (1980).
\bibitem{Karamian:2007}   S.A.~Karamian et al., Phys. Rev. C \textbf{75}, 057301 (2007).
\bibitem{Sch04}           B. Schwarzschild, Phys. Today \textbf{57(5)}, 21 (2004).
\bibitem{Tka05}           E.V. Tkalya, Phys. Usp. \textbf{48}, 525 (2005).
\bibitem{Col99}           C.B. Collins et al., Phys. Rev. Lett. \textbf{82}, 695 (1999).
\bibitem{Car04}           J.J. Carroll, Laser. Phys. Lett. \textbf{1}, 275 (2004).
\bibitem{Har08}           E.P. Hartouni et al., preprint LLNL-TR-407631 (2008).
\bibitem{Kir15}           V.I. Kirischuk et al., Phys. Lett. B \textbf{750}, 89 (2015).
\bibitem{Col05}           C.B. Collins et al., Laser. Phys. Lett. \textbf{2}, 162 (2005).
\bibitem{Bro99}           E. Browne, H. Junde, Nucl. Data Sheets \textbf{87}, 15 (1999).

\bibitem{bel07}           P. Belli et al., Nucl. Phys. A \textbf{789}, 15 (2007).
\bibitem{Poe85}           D.N. Poenaru et al., Phys. Rev. C \textbf{32}, 2198 (1985).
\bibitem{Derm92}          E. Der Mateosian, L.K. Peker, Nucl. Data Sheets \textbf{66}, 705 (1992).
\bibitem{Bell07}          P.~Belli et al., Nucl. Instr. Meth. A \textbf{607}, 573 (2009). 

\bibitem{Kvi66}           J. Kvitek, Yu.P. Popov, Phys. Lett. \textbf{22}, 186 (1966).
\bibitem{Oak67}           N.S. Oakey, R.D. Macfarlane, Phys. Lett. B  \textbf{24}, 142 (1967).
\bibitem{Kvi70}           J. Kvitek, Yu.P. Popov, Nucl. Phys. A \textbf{154}, 177 (1970).
\bibitem{Pop72}           Yu.P. Popov et al., Nucl. Phys. A  \textbf{188}, 212 (1972).
\bibitem{Fau09}           G. Faure, T. Mensing, {\it Isotopes: Principles and Applications} (Wiley, India, 2009). 
\bibitem{Hev33}           G. Hevesy, M. Pahl, Nature  \textbf{131}, 434 (1933); 
                          G. Hevesy, M. Pahl, R. Hosemann, Z. Phys. \textbf{83}, 43 (1933).
\bibitem{Cur34}           M. Curie, F. Joliot, Compt. Rend. Acad. Sci. \textbf{198}, 360 (1934).
\bibitem{Har16}           A. Hartmann et al., Nucl. Instr. Meth. A \textbf{814}, 12 (2016).
\bibitem{Nic09}           N. Nica, Nucl. Data Sheets \textbf{110}, 749 (2009).
\bibitem{Raj92}           M. Rajput, T.M. Mahon, Nucl. Instr. Meth. A \textbf{312}, 289 (1992).
\bibitem{Tav18}           O.A.P. Tavares, M.L. Terranova, Applied Radiation and Isotopes \textbf{139}, 26 (2018).
\bibitem{Bro12}           E. Browne, J. Tuli, Nucl. Data Sheets \textbf{113}, 715 (2012).
\bibitem{Koh09}           M. K\"{o}hler et al., Appl. Radiat. Isot.  \textbf{67}, 736 (2009).
\bibitem{kor68}           V.A. Korolev et al., Yad. Fiz. \textbf{8}, 227 (1968); Sov. J. Nucl. Phys. \textbf{8}, 131 (1969).
\bibitem{gup70}           M.C. Gupta, R.D. MacFarlane, J. Inorg. Nucl. Chem. \textbf{32}, 3425 (1970).
\bibitem{Roy00}           G. Royer, J. Phys. G \textbf{26}, 1149 (2000).
\bibitem{kin11}           KINR UMADAC Code: http://www.nndc.bnl.gov/codes/UMADAC/

\bibitem{Barabash:2018}   A.S. Barabash et al., Nucl. Phys. At. Energy \textbf{19}, 95 (2018).
\bibitem{bellotti:1982}   E. Bellotti et al., Nuovo Cimento A \textbf{33}, 273 (1982).
\bibitem{nd144_01}        B. Al-Bataina, J. Janecke, Radiochim. Acta \textbf{42}, 159 (1987).
\bibitem{Riezler:1959}    W. Von Riezler, G. Kauw, Z. Naturforsch. A \textbf{14a}, 196 (1959).

\bibitem{ber97}  R. Bernabei et al., Nuovo Cimento \textbf{110}, 189 (1997).
\bibitem{dan01}  F.A. Danevich et al., Nucl. Phys. A \textbf{694}, 375 (2001).
\bibitem{bel09}  P. Belli et al., Nucl. Phys. A \textbf{824}, 101 (2009).
\bibitem{bel14}  P. Belli et al., Nucl. Phys. A \textbf{930}, 195 (2014).
\bibitem{bel17}  P. Belli et al., Eur. Phys. J. A \textbf{53}, 172 (2017).

\bibitem{alb88}  B. Al-Bataina, J. Janecke, Phys. Rev. C \textbf{37}, 1667 (1988).

\bibitem{Dy}        P. Belli et al., Nucl. Phys. A \textbf{859}, 126 (2011).
\bibitem{Art96}     A. Artna-Cohen, Nucl. Data Sheets \textbf{79}, 1 (1996).
\bibitem{Rei09}     C.W. Reich, Nucl. Data Sheets \textbf{110}, 2257 (2009).
\bibitem{Rei03}     C.W. Reich, Nucl. Data Sheets \textbf{99}, 753 (2003).   
\bibitem{Hel04a}    R.G. Helmer, Nucl. Data Sheets \textbf{103}, 565 (2004).
\bibitem{Hel04}     R.G. Helmer, Nucl. Data Sheets \textbf{101}, 325 (2004).

\bibitem{Dick:05}         A.P. Dickin, {\it Radiogenic Isotope Geology}, 2nd edition (2005), Cambridge University Press, Cambridge.
\bibitem{suh19}           H. Ejiri, J. Suhonen, K. Zuber, Phys. Rep. \textbf{797}, 1 (2019).
\bibitem{Suhonen:2017b}   J.T.~Suhonen, Front. in Phys. \textbf{5}, 55 (2017).
\bibitem{aa02}            G. Bellini et al., Nature \textbf{512}, 383 (2014).
\bibitem{aa03}            M. Agostini et al., Nature \textbf{544}, 47 (2017).
\bibitem{aa04}            P. Agnes et al., Phys. Rev. D \textbf{93}, 081101 (2016).
\bibitem{kosten:2018}     J. Kostensalo, J. Suhonen, Int. J. Mod. Phys. A \textbf{33(9)}, 1843008 (2018).
\bibitem{tretyak:2017}    V.I. Tretyak, AIP Conf. Proc. \textbf{1894}, 020026 (2017).
\bibitem{aa05}            E.E. Berlovich, Izv. AN SSSR, ser. fiz. \textbf{16}, 314 (1952) (in Russian).
\bibitem{aa06}            S. Kageyama, J. Phys. Soc. Japan \textbf{8}, 689 (1953).
\bibitem{aa07}            R.A. Ricci, G. Trivero, Nuov. Cim. \textbf{2}, 745 (1955).

\bibitem{Suh07}           J.~Suhonen, {\it From Nucleons to Nucleus}, Springer (2007).

\bibitem{Bak02}           A. Bakalyarov et al., JETP Lett. \textbf{76}, 545 (2002).
\bibitem{Aun99}           M. Aunola, J. Suhonen, T. Siiskonen, Europhys. Lett. \textbf{46}, 577 (1999).
\bibitem{Haa14}           M. Haaranen et al., Phys. Rev. C \textbf{89}, 034315 (2014).
\bibitem{Kos17}           J. Kostensalo, J. Suhonen, Phys. Rev. C \textbf{96}, 024317 (2017).
\bibitem{Lau18}           M. Laubenstein et al., Phys. Rev. C \textbf{99}, 045501 (2019).
\bibitem{Nis79}           M. Nishimura, Z. Phys. A \textbf{289}, 307 (1979).
\bibitem{Haa14b}          M. Haaranen et al., Phys. Rev. C \textbf{90}, 044314 (2014).

\bibitem{Sas69}           K.S.R. Sastry, Phys. Lett. B \textbf{28}, 462 (1969).
\bibitem{Kop72}           I.V. Kopytin, Y.P. Suslov, Sov. J. Nucl. Phys. \textbf{14}, 344 (1972).
\bibitem{Szy76}           L. Szybisz, Nucl. Phys. A \textbf{267}, 246 (1976).

\bibitem{Kos17a}          J.~Kostensalo, M.~Haaranen, J.~Suhonen, Phys. Rev. C \textbf{95}, 044313 (2017).
\bibitem{aa19}            M. Arpesella et al., Europhys. Lett. \textbf{27}, 29 (1994).

\bibitem{Hei07}           H.~Heiskanen, M.T.~Mustonen, J.~Suhonen, J. Phys. G \textbf{34}, 837 (2007).

\bibitem{Belli:2007a}     P.~Belli et al., Phys. Rev. C \textbf{76}, 064603 (2007).

\bibitem{Mus07}           M. Mustonen, J. Suhonen, Phys. Lett. B \textbf{657}, 38 (2007).

\bibitem{Pfe78}           L. Pfeiffer et al., Phys. Rev. Lett. \textbf{41}, 63 (1978).
\bibitem{Pfe79}           L. Pfeiffer et al., Phys. Rev. C \textbf{19}, 1035 (1979).
\bibitem{Wie09}           E. Wieslander et al., Phys. Rev. Lett. \textbf{103}, 122501 (2009).
\bibitem{And11}           E. Andreotti et al., Phys. Rev. C \textbf{84}, 044605 (2011).

\bibitem{Kos17c}          J. Kostensalo, J. Suhonen, Phys. Rev. C \textbf{95}, 014322 (2017). 

\bibitem{Hult:2014}       M. Hult et al., Appl. Rad. Isot. \textbf{87}, 112 (2014).
\bibitem{chan:2018}       W.M. Chan et al., AIP Conf. Proc. \textbf{1921}, 030004 (2018).

\bibitem{Eji17}           H. Ejiri, T. Shima, J. Phys. G \textbf{44}, 065101 (2017).

\bibitem{Basu:2009} M.S. Basunia, Nuclear Data Sheets \textbf{110}, 999 (2009).
\bibitem{Ale03}     A. Alessandrello et al., Phys. Rev. C \textbf{67}, 014323 (2003).
\bibitem{Civ01}     O. Civitarese, J. Suhonen, Phys. Rev. C  \textbf{64}, 064312 (2001).
\bibitem{Arn16}     R. Arnold et al., Phys. Rev. D \textbf{93}, 112008 (2016).
\bibitem{aa20}      F. Nova et al., AIP Conf. Proc. \textbf{1560}, 184 (2013).
\bibitem{aa08}      H. Behrens, J. Janecke, \textit{Numerical Tables for Beta-Decay and Electron Capture}, Berlin, Springer-Verlag, 1969.
\bibitem{aa09}      H. Paul, Nucl. Data Tables A \textbf{2}, 281 (1966).
\bibitem{aa10}      H. Daniel, Rev. Mod. Phys. \textbf{40}, 659 (1968).
\bibitem{aa11}      H. Behrens, L. Szybisz, Phys. Data \textbf{6-1}, 1 (1976).
\bibitem{aa12}      X. Mougeot, Phys. Rev. C \textbf{91}, 055504 (2015).
\bibitem{aa13}      W.H. Kelly et al., Nucl. Phys. A \textbf{11}, 492 (1959).
\bibitem{aa14}      S.T. Hsue et al., Nucl. Phys. A \textbf{86}, 47 (1966).
\bibitem{Bern:2005} R. Bernabei et al., Nucl. Instrum. and Meth. A \textbf{555}, 270 (2005).

\bibitem{Ber02}           R. Bernabei et al., Nucl. Phys. A \textbf{705}, 29 (2002).
\bibitem{War85}           E.K. Warburton, Phys. Rev. C \textbf{31}, 1896 (1985).
\bibitem{Burrow:2006}     T.W. Burrows, Nucl. Data Sheets \textbf{107}, 1747 (2006).

\bibitem{Ele11}             Z. Elekes, J. Timar, B. Singh, Nucl. Data Sheets \textbf{112}, 1 (2011).
\bibitem{Lel49}             W.T. Leland, Phys. Rev. \textbf{76}, 1722 (1949).
\bibitem{Glo57}             R.N. Glover, D.E. Watt, Philos. Mag. \textbf{2}, 697 (1957).
\bibitem{Bau58}             E.R. Bauminger, S.G. Cohen, Phys. Rev. \textbf{110}, 953 (1958).
\bibitem{Wat62}             D.E. Watt, R.L.G. Keith, Nucl. Phys. \textbf{29}, 648 (1962).
\bibitem{Sim89}             J.J. Simpson, P. Moorhouse,  P. Jagam, Phys. Rev. C \textbf{39}, 2367 (1989).
\bibitem{Hei55}             J. von Heintze, Z. Naturforschung \textbf{10}, 77 (1955).
\bibitem{McN61}             A. McNair, Philos. Mag. \textbf{6}, 559 (1961).
\bibitem{Son66}             Ch. Sonntag, K.O. Munnich, Z. Phys. \textbf{197}, 300 (1966).
\bibitem{Pap77}             A. Pape, S.M. Refaei, J.C. Sens, Phys. Rev. C \textbf{15}, 1937 (1977).
\bibitem{Alb84}             D.E. Alburger, E.K. Warburton, J.B. Cumming, Phys. Rev. C \textbf{29}, 2294 (1984).
\bibitem{Sim85}             J.J. Simpson, P. Jagam,  A.A. Pilt, Phys. Rev. C \textbf{31}, 575 (1985).
\bibitem{Dombrowski:2011}   H.~Dombrowski, S.~Neumaier, K.~Zuber, Phys. Rev. C \textbf{83}, 054322 (2011).
\bibitem{Pat18}             L. Pattavina et al., Eur. Phys. J. A \textbf{54}, 79 (2018).

\bibitem{aa19_2}    M. Alanssari et al., Phys. Rev. Lett. \textbf{116}, 072501 (2016).
\bibitem{aa19_3}    J. \"{A}yst\"{o}, Nucl. Phys. A \textbf{693}, 477 (2001).
\bibitem{aa19_4}    I.D. Moore et al., Nucl. Instr. Meth. B \textbf{317}, 208 (2013).

\bibitem{Martell:1950}    E.A.~Martell, W.F.~Libby, Phys. Rev. \textbf{80}, 977 (1950).
\bibitem{Watt:1962}       D.E.~Watt, R.N.~Glove, Philosophical Magazine \textbf{7}, 105 (1962).
\bibitem{Greth:1970}      W.E.~Greth, S.~Gangadharan, R.L.~Wolke, J. Inorg. Nucl. Chem. \textbf{32}, 2113 (1970).
\bibitem{Mitchell:1988}   L.W.~Mitchell, P.H.~Fisher, Phys. Rev. C \textbf{38}, 895 (1988).
\bibitem{Alessandrello:1994a}  A.~Alessandrello et al., Nucl. Phys. B Proc. Suppl. \textbf{35}, 394 (1994).
\bibitem{Alessandrello:1994b}  A.~Alessandrello et al., Nucl. Instrum. Methods A \textbf{344}, 243 (1994).
\bibitem{Danevich:1996a}       F.A.~Danevich et al., Phys. At. Nucl. \textbf{59}, 1 (1996).
\bibitem{Goessling:2005}       G.~Goessling et al., Phys. Rev. C \textbf{72}, 064328 (2005).
\bibitem{Haaranen:2016}    M.~Haaranen, P.C.~Srivastava, J.~Suhonen, Phys. Rev. C \textbf{93}, 034308 (2016).
\bibitem{Haaranen:2017}    M.~Haaranen, J.~Kotila, J.~Suhonen, Phys. Rev. C \textbf{95}, 024327 (2017).
\bibitem{Barea:2013}       J.~Barea, J.~Kotila, F.~Iachello, Phys. Rev. C \textbf{87}, 014315 (2013).
\bibitem{DellOro:2014}     S.~Dell'Oro, S.~Marcocci, F.~Vissani, Phys. Rev. D \textbf{90}, 033005 (2014).
\bibitem{Suhonen:2017a}    J.T.~Suhonen, Phys. Rev. C \textbf{96}, 055501 (2017).
\bibitem{Belli:2014}       P.~Belli et al., Eur. Phys. J. A \textbf{50}, 134 (2014).
\bibitem{Dorenbos:1995}    P.~Dorenbos, J.T.M. de Haas, C.W.E.~van~Eijk, IEEE Trans. Nucl. Sci. \textbf{42}, 2190 (1995).
\bibitem{Bizzeti:2012}     P.G.~Bizzeti et al., Nucl. Instrum. Meth. A \textbf{696}, 144 (2012).

\bibitem{Coh51}            S.G. Cohen, Nature \textbf{167}, 779 (1951).
\bibitem{Bea61}            G.B. Beard, W.H. Kelly, Phys. Rev. \textbf{122}, 1576 (1961).
\bibitem{Led18}            A. Leder, talk at the APS DNP meeting, April 14-17, 2018.

\bibitem{Bla12}             J. Blachot, Nucl. Data Sheets \textbf{113}, 2391 (2012).
\bibitem{Cattadori:2005}    C.M.~Cattadori et al., Nucl. Phys. A \textbf{748}, 333 (2005).
\bibitem{Cat07}             C.M. Cattadori et al., Phys. At. Nucl. \textbf{70}, 127 (2007).
\bibitem{Aud03}             G.~Audi, A.~H.~Wapstra, C.~Thibault, Nucl. Phys. A \textbf{729}, 337 (2003).
\bibitem{Mou09}             B.J. Mount et al., Phys. Rev. Lett. \textbf{103}, 122502 (2009).
\bibitem{Zhe18}             V.A. Zheltonozhsky et al., Europhys. Lett. \textbf{121}, 12001 (2018).
\bibitem{Urb16}             W. Urban et al., Phys. Rev. C \textbf{94}, 011302 (2016).
\bibitem{Mus10a}            M.T. Mustonen, J. Suhonen, J. Phys. G \textbf{37}, 064008 (2010).
\bibitem{Dvo11}             R. Dvornicky, F. Simkovic, AIP Conf. Proc. \textbf{1417}, 33 (2011).
\bibitem{Mus10b}            M.T. Mustonen, J. Suhonen, AIP Conf. Proc. \textbf{1304}, 401 (2010).
\bibitem{Mus11}             M.T. Mustonen, J. Suhonen, Phys. Lett. B \textbf{703}, 370 (2011).  
\bibitem{Haa13}             M. Haaranen, J. Suhonen, Eur. Phys. J. A \textbf{49}, 93 (2013). 
\bibitem{Suh14}             J. Suhonen, Phys. Scr. \textbf{89}, 054032 (2014).  

\bibitem{gamage:2019}       N.D. Gamage et al., Hyperfine Inter. \textbf{240}, 43 (2019).

\bibitem{Bia97}           M. Bianchetti et al., Phys. Rev. C \textbf{56}, 1675 (1997).
\bibitem{Ale96}           A. Alessandrello et al., Phys. Rev. Lett. \textbf{77}, 3319 (1996).
\bibitem{Ald18}           C. Alduino et al., Int. J. Mod. Phys. A \textbf{33}, 1843002 (2018).
\bibitem{Bellini:2018}    F. Bellini, Int. J. Mod. Phys. A  \textbf{33}, 1843003 (2018).
\bibitem{Mus03}           D. M\"{u}nstermann, K. Zuber, J. Phys. G: Nucl. Part. Phys. \textbf{29}, B1 (2003).
\bibitem{Fil16}           P. Filianin et al., Phys. Lett. B \textbf{758}, 407 (2016).
\bibitem{Cam59}           A. Cameron, Astrophys. J. \textbf{130}, 452 (1959).
\bibitem{Arn72}           M. Arnould, Astron. Astrophys. \textbf{21}, 401 (1972).
\bibitem{Yok83}           K. Yokoi, K. Takahashi, M. Arnold, Astron. Astrophys. \textbf{117}, 65 (1983).
\bibitem{Tak87}           K. Takahashi, K. Yokoi, At. Data Nucl. Data Tables \textbf{36}, 375 (1987).
\bibitem{Lae94}           J.R. De Laeter, C.L. Smith, K.J.R. Rosman, Phys. Rev. C \textbf{49}, 1227 (1994).

\bibitem{aa37}            B. Lehnert et al., Phys. Rev. C \textbf{95}, 044306 (2017).

\bibitem{Norman:1981}       E.B. Norman, Phys. Rev. C \textbf{24}, 2334 (1981).
\bibitem{Cumming:1985}      J.B. Cumming, D.E. Alburger, Phys. Rev. C \textbf{31}, 1494 (1985).
\bibitem{Hult:2006}         M. Hult et al., Phys. Rev. C \textbf{74}, 054311 (2006).
\bibitem{Hult:2009}         M. Hult et al., Appl. Rad. Isot. \textbf{67}, 918 (2009).

\bibitem{Sin98}     B. Singh et al., Nucl. Data Sheets \textbf{84}, 487 (1998).
\bibitem{logft}     National Nuclear Data Center, http://www.nndc.bnl.gov/logft/.
\bibitem{Sin11}     S. Singh, A.K. Jain, J.K. Tuli, Nucl. Data Sheets \textbf{112}, 2851 (2011).

\bibitem{tretyak:2005}         V.I. Tretyak, DAMA internal note (November 2005).
\bibitem{Laubenstein:2004}     M.~Laubenstein et al., Appl. Radiat. Isot. \textbf{61}, 167 (2004).
\bibitem{Danevich:2012}        F.A.~Danevich, IEEE Trans. Nucl. Sci. \textbf{59}, 2207 (2012).
\bibitem{Belli:2010a}          P.~Belli et al., Nucl. Instr. Meth. A \textbf{615}, 301 (2010).
\bibitem{Barabash:2011}        A.S.~Barabash et al., J. Instrum. \textbf{6}, P08011 (2011).
\bibitem{Alenkov:2011}         V.V.~Alenkov et al., Cryst. Res. Technol. \textbf{46}, 1223 (2011).
\bibitem{Barabash:2014a}       A.S.~Barabash et al., Eur. Phys. J. C \textbf{74}, 3133 (2014).
\bibitem{Grigorieva:2017}      V.~Grigorieva et al., J. Mat. Sci. Eng. B \textbf{7}, 63 (2017).
\bibitem{Armengaud:2017}       E.~Armengaud et al., Eur. Phys. J. C \textbf{77}, 785 (2017).
\bibitem{Dafinei:2017a}        I.~Dafinei et al., J. Cryst. Growth \textbf{475}, 158 (2017).

\bibitem{Rai05}           F. Raiola et al., J. Phys. G \textbf{31}, 1141 (2005).
\bibitem{Ket25}           K.U. Kettner et al., J. Phys. G \textbf{32}, 489 (2006).
\bibitem{Rai07}           F. Raiola et al., Eur. Phys. J. A \textbf{32}, 51 (2007).
\bibitem{Spi07}           T. Spillane et al., Eur. Phys. J. A \textbf{31}, 203 (2007).
\bibitem{Zin07}           N.T. Zinner, Nucl. Phys. A \textbf{781}, 81 (2007).

\end{thebibliography}
\end{document}